\documentclass[preprint]{aastex}
\usepackage{graphics}
\usepackage{epsfig}

\newcommand{\etal}{et\,al.}

\newcommand{\ltsimeq}{\la}

\newcommand{\msun}{M$_{\odot}$}

\newcommand{\HI}{H{\sc i}}

\received{}
\revised{}
\accepted{}
\shortauthors{McQuinn et al.}
\shorttitle{The Star Formation Histories of 18 Starburst Galaxies}

\begin{document}
\title{The Nature of Starbursts: I. The Star Formation Histories of Eighteen Nearby Starburst Dwarf Galaxies\footnote{Based on observations made with the NASA/ESA Hubble Space Telescope, obtained from the Data Archive at the Space Telescope Science Institute, which is operated by the Association of Universities for Research in Astronomy, Inc., under NASA contract NAS 5-26555.}}
\author{Kristen B.~W. McQuinn\altaffilmark{1}, 
Evan D. Skillman\altaffilmark{1},
John M. Cannon\altaffilmark{2},
Julianne Dalcanton\altaffilmark{3},
Andrew Dolphin\altaffilmark{4},
Sebastian Hidalgo-Rodr\'{i}guez\altaffilmark{5},
Jon Holtzman\altaffilmark{6},
David Stark\altaffilmark{1},
Daniel Weisz\altaffilmark{1},
Benjamin Williams\altaffilmark{3}
}

\altaffiltext{1}{Department of Astronomy, School of Physics and
Astronomy, 116 Church Street, S.E., University of Minnesota,
Minneapolis, MN 55455, \ {\it kmcquinn@astro.umn.edu}} 
\altaffiltext{2}{Department of Physics and Astronomy, 
Macalester College, 1600 Grand Avenue, Saint Paul, MN 55105}
\altaffiltext{3}{Department of Astronomy, Box 351580, University 
of Washington, Seattle, WA 98195}
\altaffiltext{4}{Raytheon Company, 1151 E. Hermans Road, Tucson, AZ 85706}
\altaffiltext{5}{Instituto de Astrof\'{i}sica de Canarias, V\'{i}a L\'{a}ctea s/n.~E38200, La Laguna, Tenerife, Canary Islands, Spain}
\altaffiltext{6}{Department of Astronomy, New Mexico State University, Box 30001-Department 4500, 1320 Frenger Street, Las Cruces, NM 88003}

\begin{abstract}

We use archival Hubble Space Telescope observations of resolved stellar populations to derive the star formation histories (SFHs) of eighteen nearby starburst dwarf galaxies. In this first paper we present the observations, color-magnitude diagrams, and the SFHs of the eighteen starburst galaxies, based on a homogeneous approach to the data reduction, differential extinction, and treatment of photometric completeness. We adopt a star formation rate (SFR) threshold normalized to the average SFR of the individual system as a metric for classifying starbursts in SFHs derived from resolved stellar populations. This choice facilitates finding not only currently bursting galaxies but also ``fossil" bursts increasing the sample size of starburst galaxies in the nearby (D$<8$ Mpc) universe. Thirteen of the eighteen galaxies are experiencing ongoing bursts and five galaxies show fossil bursts. From our reconstructed SFHs, it is evident that the elevated SFRs of a burst are sustained for hundreds of Myr with variations on small timescales. A long $>100$ Myr temporal baseline is thus fundamental to any starburst definition or identification method. The longer lived bursts rule out rapid ``self-quenching'' of starbursts on global scales. The bursting galaxies' gas consumption timescales are shorter than the Hubble time for all but one galaxy confirming the short-lived nature of starbursts based on fuel limitations. Additionally, we find the strength of the H$\alpha$ emission usually correlates with the CMD based SFR during the last $4-10$ Myr. However, in four cases, the H$\alpha$ emission is significantly less than what is expected for models of starbursts; the discrepancy is due to the SFR changing on timescales of a few Myr. The inherently short timescale of the H$\alpha$ emission limits identifying galaxies as starbursts based on the current characteristics which may or may not be representative of the recent SFH of a galaxy. 
\end{abstract} 


\section{A Perspective on the Starburst Phenomenon in Dwarf Galaxies \label{starburst} }

A starburst is defined as a short-lived intense period of star formation (SF) that is unsustainable over the Hubble time due to the limited supply of gas within a galaxy. Starburst galaxies were first classified by \citet{Searle1972} and \citet{ Searle1973}, based on the blue colors produced by the massive stars formed during the burst. In the local universe, starbursts create approximately $10\%$ of the radiant energy and $20\%$ of the massive stars \citep{Heckman1998, Brinchmann2004}. At \textit{z} = 1, starburst characteristics are found in $15\%$ of galaxies \citep{OConnell2005}, presumably attributable to the greater amounts of gas typically present in young galaxies and increased galactic interactions \citep[e.g.,][]{Condon1982, Kennicutt1987, Telesco1988}.  

The starburst's impact on a galaxy and the surrounding intergalactic medium (IGM) is primarily due to the consumption of gas that fuels the burst and the feedback from massive stars formed in the burst. For example, one of the important mechanisms is the ionizing radiation produced and the mass loss experienced during the lifetime of the massive stellar population and the resulting supernovae (SNe). The mass loss from massive stars alters the chemical composition of the interstellar medium (ISM) and future stellar populations. In addition, the flux of energetic photons escaping the galaxy changes the ionization state of the surrounding IGM. It is possible for starbursts to generate the energy and material to fuel galactic ``super-winds'' which can escape from the low potential wells of dwarf galaxies enriching the metallicity of the IGM \citep{Skillman1997}. This has been suggested as an explanation for the low metallicity measurements of many local dwarf galaxies \citep[e.g.,][]{Spaans1997, Romano2006}. Yet, without understanding the SFH of a galaxy, it is difficult to constrain the impact of a starburst. 

Starbursts have been observed in gas rich galaxies such as disk galaxies and low mass dwarfs \citep[e.g.,][]{Heckman1998}. It is unclear whether most dwarf galaxies experience starbursts during their evolution or whether only a subgroup of dwarf galaxies exhibit bursts \citep[e.g.,][]{Marlowe1997, Mayer2001, vanZee2001b, Pasetto2003}. It is possible that for the dwarf galaxies with starbursts there may be a duty cycle to the bursts \citep[][and references therein]{Marlowe1999, Lee2009} making this a repetitive mode of star formation in these galaxies. 

The ionizing radiation flux created during a starburst makes bursting galaxies easily studied at optical, ultraviolet (UV) , far infrared, and H$\alpha$ rest wavelengths. Numerous studies over the past decade have used these different wavelength observations to study the properties of both nearby and distant actively bursting systems \citep[e.g.,][among many others]{Colless1994, Babul1996, Meurer1997, Heckman2005, Oey2007, Lee2009}. The goals of these studies have been many, including characterizing the different properties of a currently observable starburst and quantifying the feedback of starbursts to the different components of the ISM and IGM. 

Most studies have been linked to either instantaneous properties of starbursts (i.e., UV and H$\alpha$ studies) or statistical properties of galaxy samples. The latter are typically based on broadband colors \citep[e.g.,][]{Madau1996}, spectral indices \citep[e.g.,][]{Kauffmann2003} or full spectro-photometric modeling \citep[e.g.,][]{Bruzual2003} on integrated galaxy light. However, with resolved stellar populations, we can derive complete SFHs allowing us to study the evolution of starbursts over hundreds of Myr within individual galaxies. Among the many relevant studies on resolved stellar populations in individual galaxies are those of NGC~1569 by \citet{Angeretti2005} who estimate the SFH over the last $1-2$ Gyr using synthetic color-magnitude diagrams (CMDs) and by \citet{Grocholski2008} who estimate the SF activity by overlaying metallicity specific age isochrones on the observed CMD of NGC~1569. A similar approach was taken by \citet{Cannon2003} on NGC~625 and \citet{Annibali2008} on NGC~4449 where SFRs were estimated by overlaying age isochrones. In the case of \citet{Cannon2003}, the recent SFR estimates were refined in increments of 100 Myr using a model of blue helium burning (BHeB) stars. 

The high-resolution data used in such studies presents an opportunity to reconstruct the detailed SFHs over a longer temporal baseline for a sample of starburst dwarf galaxies in a homogeneous and consistent way. To that end, we have gathered the observations of eighteen galaxies from the HST public archive and undertaken a detailed analysis of star formation on these systems. Using information from the CMDs and stellar evolutionary models, we derived SFRs for these galaxies as a function of time taking into account both systematic and statistical uncertainties. Our emphasis on uniformity allows us to consider the strength and importance of a starburst event in the context of the host galaxy's star formation in past epochs, and facilitates a direct comparison between galaxies while minimizing systematic uncertainties.

In this first paper, Paper I, we present the observations and data reduction (\S\ref{obs}) and CMDs and SFHs (\S\ref{sfh}) for eighteen dwarf starbursts. In \S\ref{classify} we discuss the classification of the galaxies as starbursts using a gas consumption timescale, compare our SFRs with existing measurements of H$\alpha$ emission, and present a SFR normalized to the average SFR of each galaxy. The hypothesis that SF in a burst is ``self-quenching'' is discussed in \S\ref{quench} and a summary of our conclusions in \S\ref{conclusions}. In Paper II \citep{McQuinn2010} we will present additional analysis on these starbursts, and derive the burst duration for each system. Our third paper in this series (McQuinn et al. in prep) will explore the spatial structure within the starbursts. 

\section {Galaxy Observations and Photometric Reduction \label{obs}}

\subsection{The Galaxy Sample \label{galaxies}}

Our sample of galaxies was selected from observations in the HST public archive and is listed in Table~\ref{tab:galaxies}. The majority of the galaxies were previously identified as starbursts in the literature using a number of methods. The different methods all rely on evidence of recent star formation such as the presence of high surface brightness regions, high H$\alpha$ emission, or the blue color typical of massive, young stars (see Table~\ref{tab:galaxies} for citations and identification methods used). We explore a new method for identifying galaxies as starbursts based on the presence and distribution of young main sequence and helium burning stars in the optical CMDs. The higher concentration in dwarf galaxies of these young stars is a sign of recent SF activity that is likely more intense than the average SF experienced in low-mass systems. Using the characteristics of a CMD also has the potential to identify older or fossil bursts which may or may not be producing significant H$\alpha$ emission at the current epoch. The identification of fossil starbursts potentially greatly increases the number of starbursts in the nearby universe that can be identified and studied. Using this method, we added four additional galaxies to our sample as possible starbursts (Antlia dwarf, ESO~154$-$023, NGC~784, and UGC~9128). The classification of these systems as starbursts is explored and discussed below in \S\ref{classify}.  

The dwarf galaxies in the sample cover a range of brightness, morphologies (i.e., SBdm, IAB(s)m, pec, IBm), and spatial extent, facilitating an exploration of the starburst phenomenon in a range of dwarf galaxy properties. While the majority of the galaxies have high Galactic latitudes and low foreground extinction \citep[i.e., A$_{R} < 0.2$;][]{Schlegel1998}, two of the systems, NGC~1569 and NGC~6822, have high foreground extinction ($A_{R}$ of 1.9 and 0.6 mag respectively). If we assume that the foreground extinction is uniform across an individual galaxy, the extinction can be easily taken into account. However, it is more likely that the foreground extinction has variations and is not a uniform gray screen \citep[see \S\ref{sfh_method} and][]{McQuinn2009}. We chose to include these well-studied systems in the sample with the caveat that the high extinction values increase the uncertainties on the SFRs. 

Our starburst sample was selected from observations in the HST archive based on three photometric properties. First, we require both V and I band images of a galaxy in the archival observations. Second, we set a minimum photometric depth of the I band observations of $\sim$2 mag below the tip of the red giant branch; a requirement for accurately constraining the recent SFH of a galaxy \citep{Dolphin2002, Dohm-Palmer2002}. Third, the galaxies were required to lie close enough such that their stellar populations were resolved by HST imaging instruments. These properties were chosen to ensure robust reconstruction of the SFH. Seventeen of the eighteen dwarf galaxies in our sample meet these criteria. We include an additional, galaxy to the sample, SBS $1415+437$, even though the photometry stops somewhat short ($\sim~0.5$ mag) of reaching our required photometric depth despite long integration times (P.I. Aloisi, Program I.D. 9361). While it does not strictly meet our photometric requirements, SBS~1415$+437$ is a well-studied starburst with an interesting recent SFH. We include it in our sample for comparison purposes and although the ancient SFH derived for this system is uncertain, the recent SFRs are still well constrained. Four other data sets were considered that were at least one magnitude above our photometric depth cut-off (UGCA~290 P.I. Schulte-Ladbeck, Program I.D. 8122; NGC~1705 P.I. Tosi, Program I.D. 7506; NGC~3077 P.I. Seitzer, Program I.D. 8601; I~Zw~18 P.I. Thuan, Program I.D. 9400). Testing showed that the recent SFHs on these galaxies were not properly constrained by the shallower photometry and are therefore not included in this work. 

\begin{figure}
\centering
\begin{tabular}{cc}
\epsfig{file=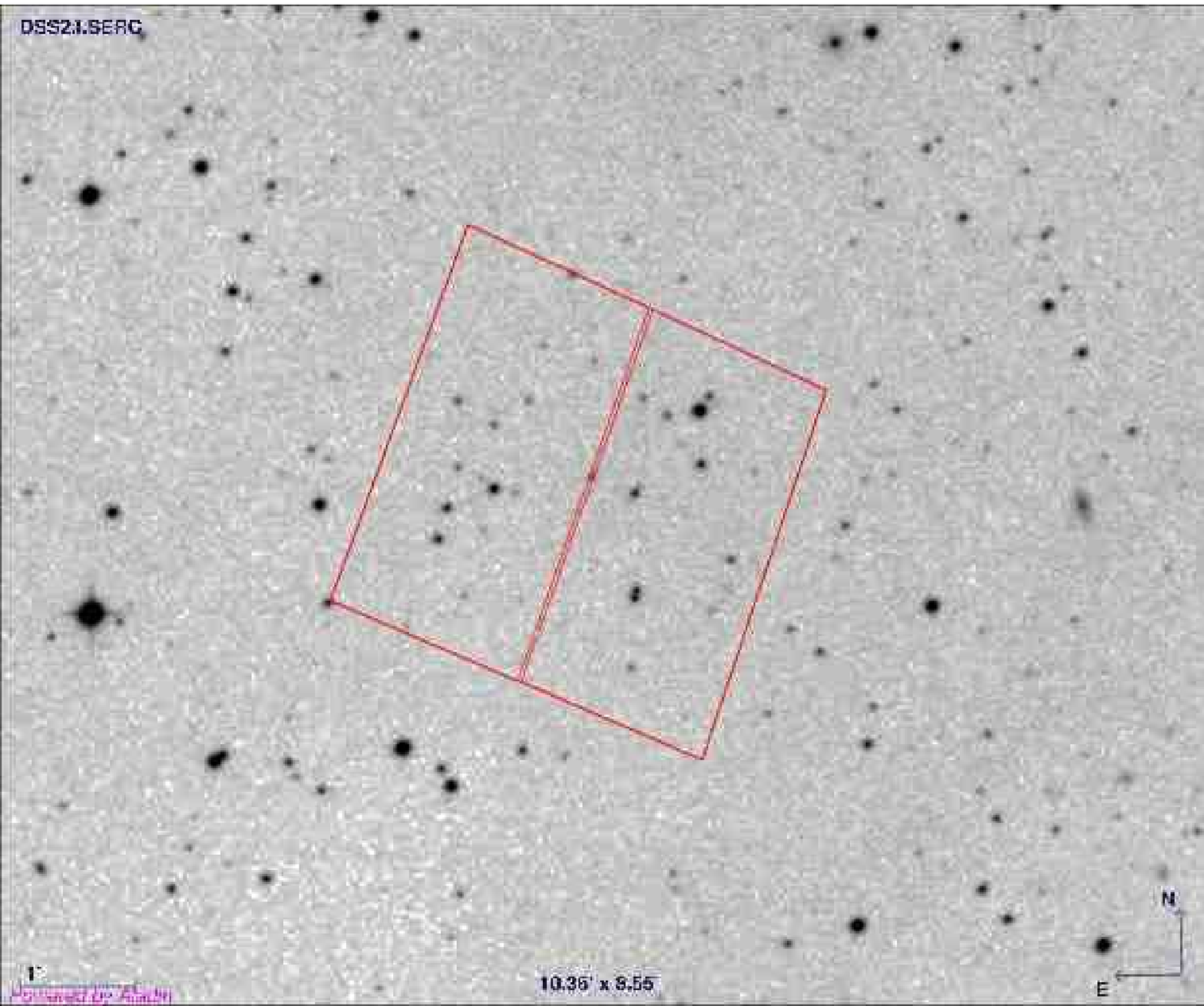,width=0.5\linewidth,clip=} & 
\epsfig{file=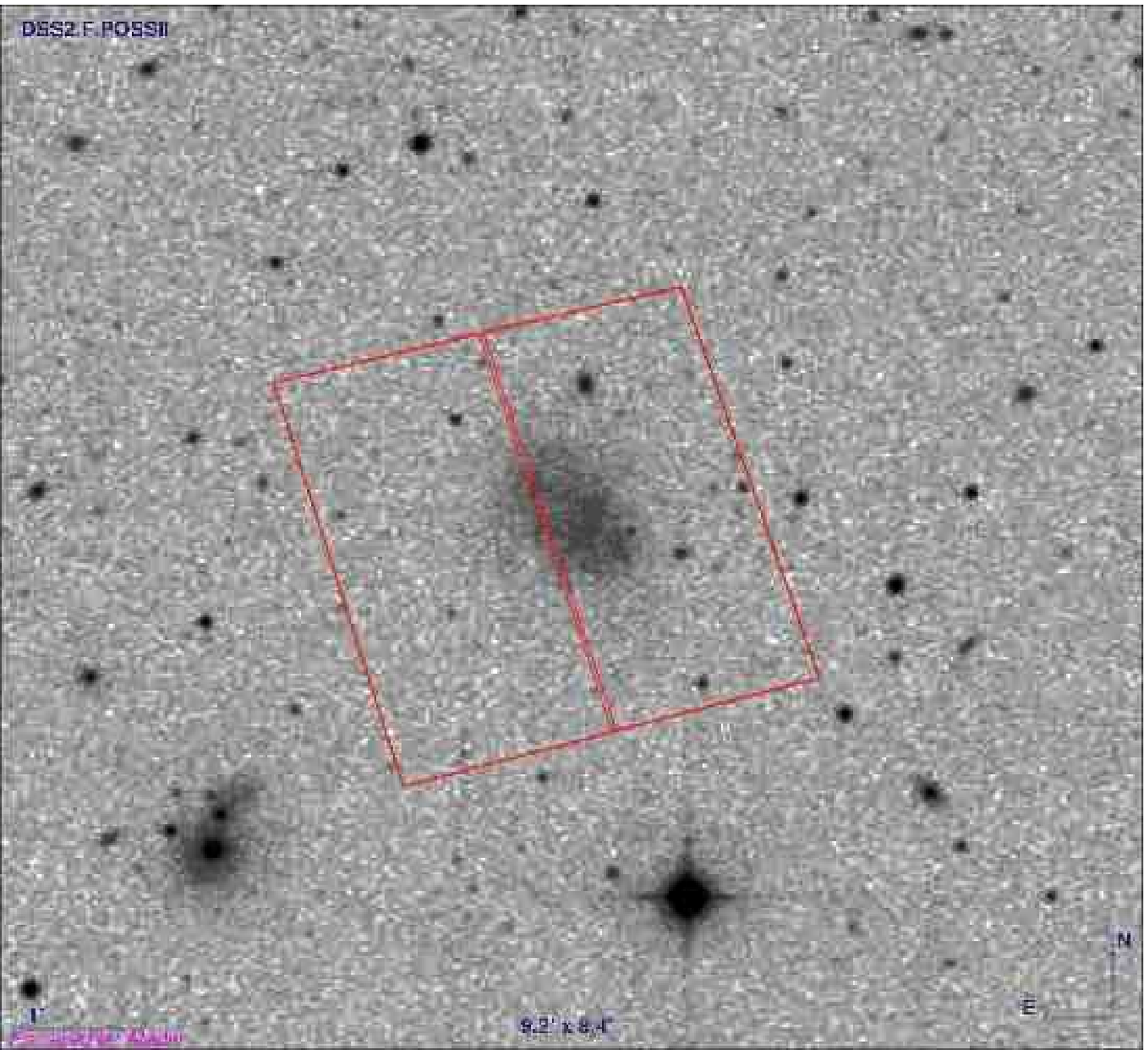,width=0.5\linewidth,clip=} \\
\epsfig{file=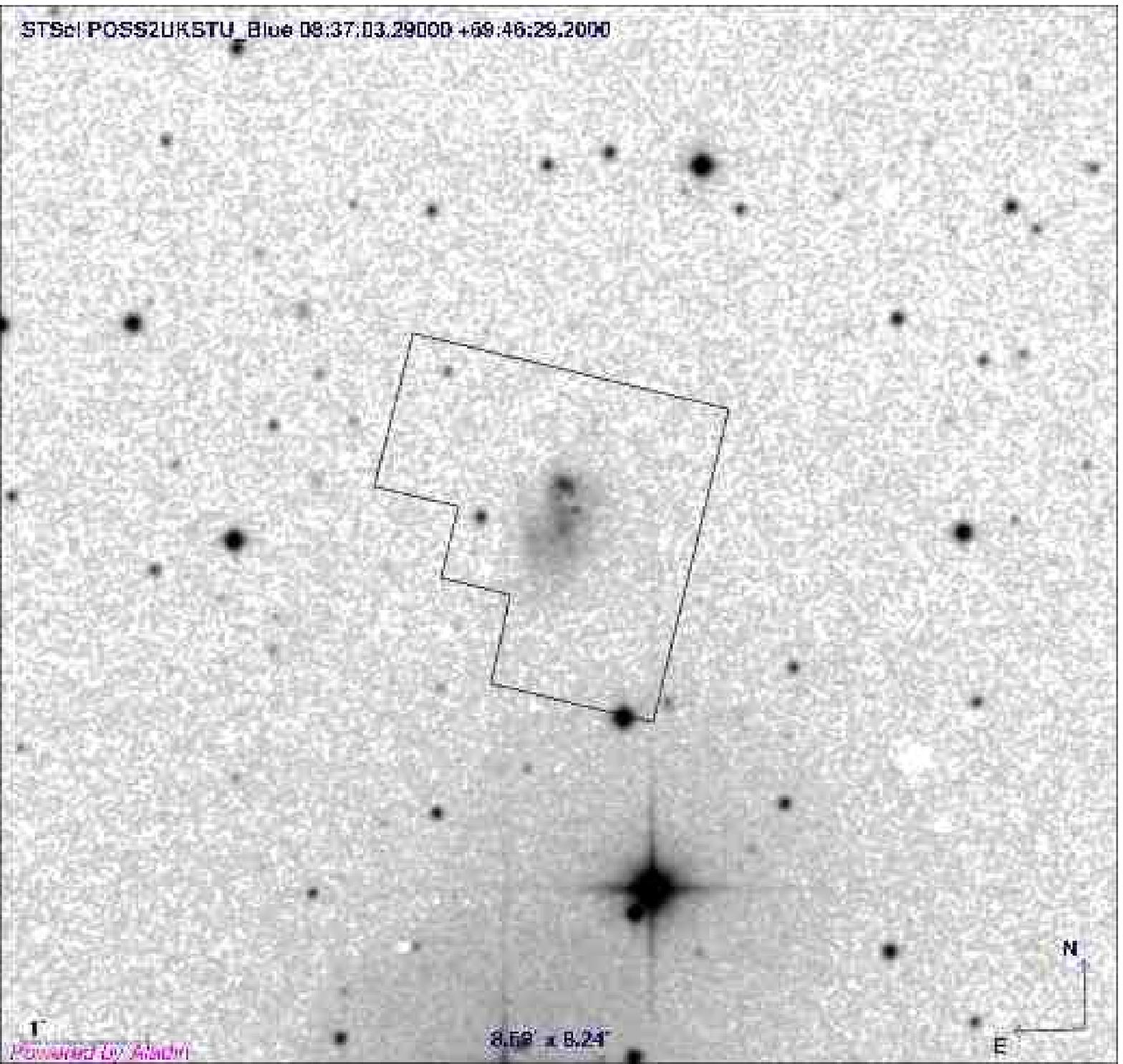,width=0.5\linewidth,clip=} &
\epsfig{file=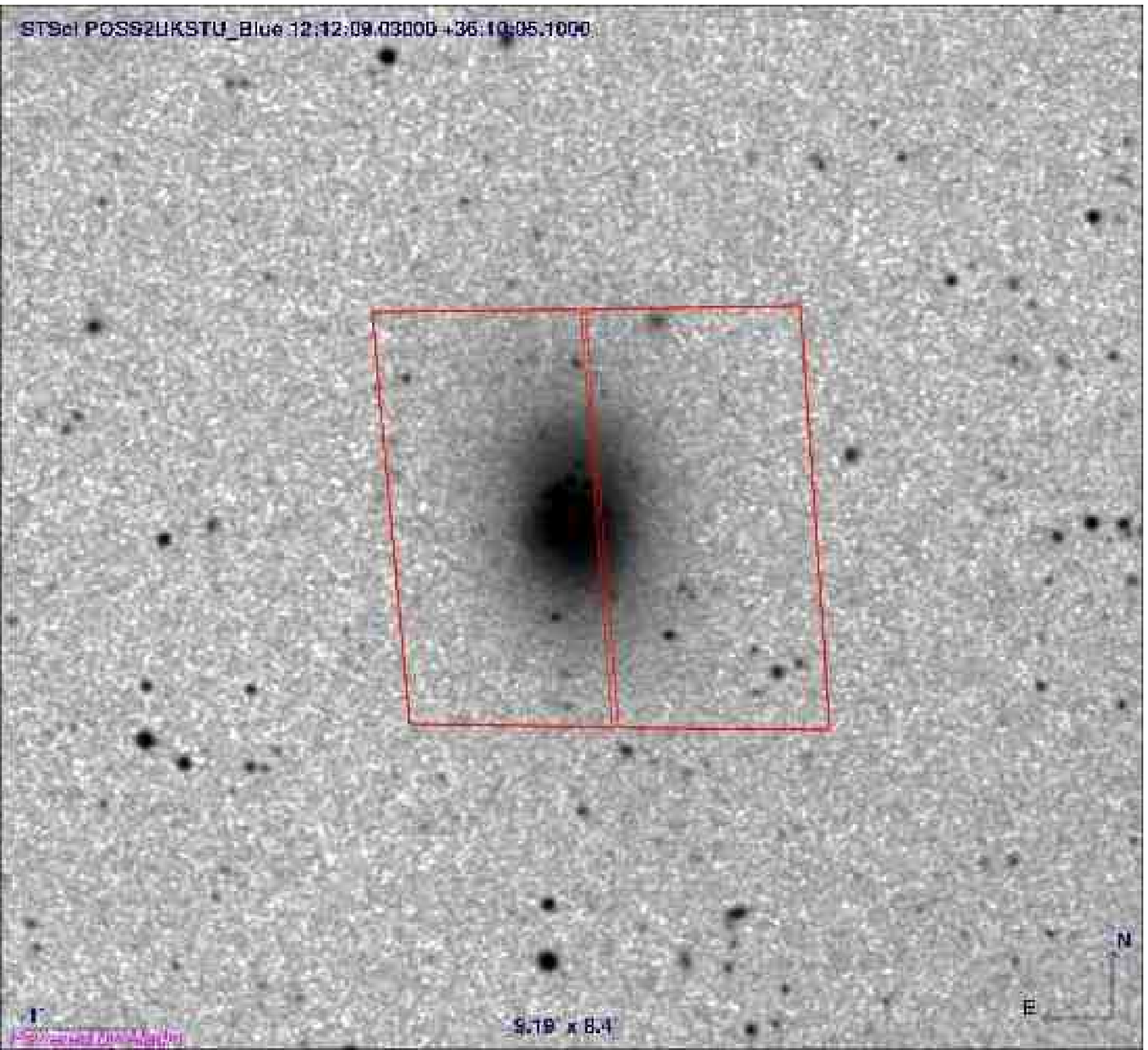,width=0.5\linewidth,clip=}
\end{tabular}

\caption{The footprints of the HST archival observation fields of view used to derive the SFHs shown on DSS V band images:~Antlia,~UGC~9128,~UGC~4483,~NGC~4163 \citep{Bonnarel2000}.}
\label{fig:images}
\end{figure}

The field of view of the HST observations are represented as footprints on DSS images \citep{Bonnarel2000} in Figure~\ref{fig:images}. Regions of active and quiescent star formation can be seen in each of the images in these inhomogeneous systems. The galaxies span a large range both in angular size (from NGC~6822 at 15.5$\arcmin$ to NGC~6789 at 1.3$\arcmin$) and in distance (ranging from Antlia at 1.3 Mpc to NGC~784 at 5.9 Mpc with the outlier SBS~1415$+$437 at 13.6 Mpc). The HST imaging instruments' field of view covers a large fraction of most galaxies in the sample, with the exception of NGC~6822.  The HST areal coverage thus includes a sufficient fraction of the stellar populations to profile the SFH although the SFRs derived for NGC~6822 will be lower limits, as noted below. The exposure times and basic data on each galaxy in the sample are listed in Table~\ref{tab:galaxies}.

\subsection {Photometric Processing \label{photometry} }

Twelve of the eighteen galaxies were observed using the Advanced Camera for Surveys (ACS) Wide Field Channel (WFC) on HST. The observations for three of the galaxies were snapshots taken on a single HST orbit (CRSPLIT = 2) (i.e., NGC~4068, NGC~4163, and IC~4662), while the observations for the remaining objects were obtained across multiple HST orbits. Photometry was performed on the pipeline processed, cosmic ray cleaned images using the ACS module of the DOLPHOT photometry package \citep{Dolphin2000} for the ACS images that were cosmic-ray split. The other six galaxies observed with the Wide Field Planetary Camera 2 (WFPC2) instrument were reduced using the pipeline developed by \citet{Holtzman2006}. Briefly, the WFPC2 images were pre-processed with the standard WFPC2 tasks \textit{mask}, which flags bad pixels as saturated using data quality files and \textit{crmask} which cleans the images. Photometry was performed as part of the pipeline using the HSTphot photometry package \citep{Dolphin2000} in the PSF mode. Four galaxies, NGC~2366, NGC~4449, NGC~5253, and NGC~6822, were observed with multiple pointings. In these cases, the photometry measurements from each image were combined by geometrically modeling any overlapping regions in the images and selecting the photometry from only one of the images in the overlapping region. 

The definitions and derivations of the parameters from the two photometry methods, DOLPHOT and HSTphot, used in the selecting high fidelity point sources are identical ensuring that our final photometry lists from the two HST instruments and photometric packages were homogeneously processed and filtered. The photometry output of DOLPHOT and HSTphot includes a characterization of the quality of each point source including the sharpness of the source, the degree of crowding, and the quality of the fit, etc; a detailed explanation of these parameters can be found in \citet{Dolphin2000}. We selected point sources from the photometry that were characterized as well-recovered stars with a minimum signal-to-noise ratio of 5. In addition to these criteria, the sum of the squared sharpness values for the \textit{two} wavelength observations were required to be $\leq 0.39$ for an individual point source. We rejected cuts based on crowding, because our tests showed this additional filtering preferentially eliminated bright point sources in areas of active star formation $-$ one of the areas of interest for studying starbursts. Note that while crowding is typically not as significant in low surface brightness galaxies, in regions where crowding is high, the derived SFR may still be a lower limit. 

\begin{figure}
\plotone{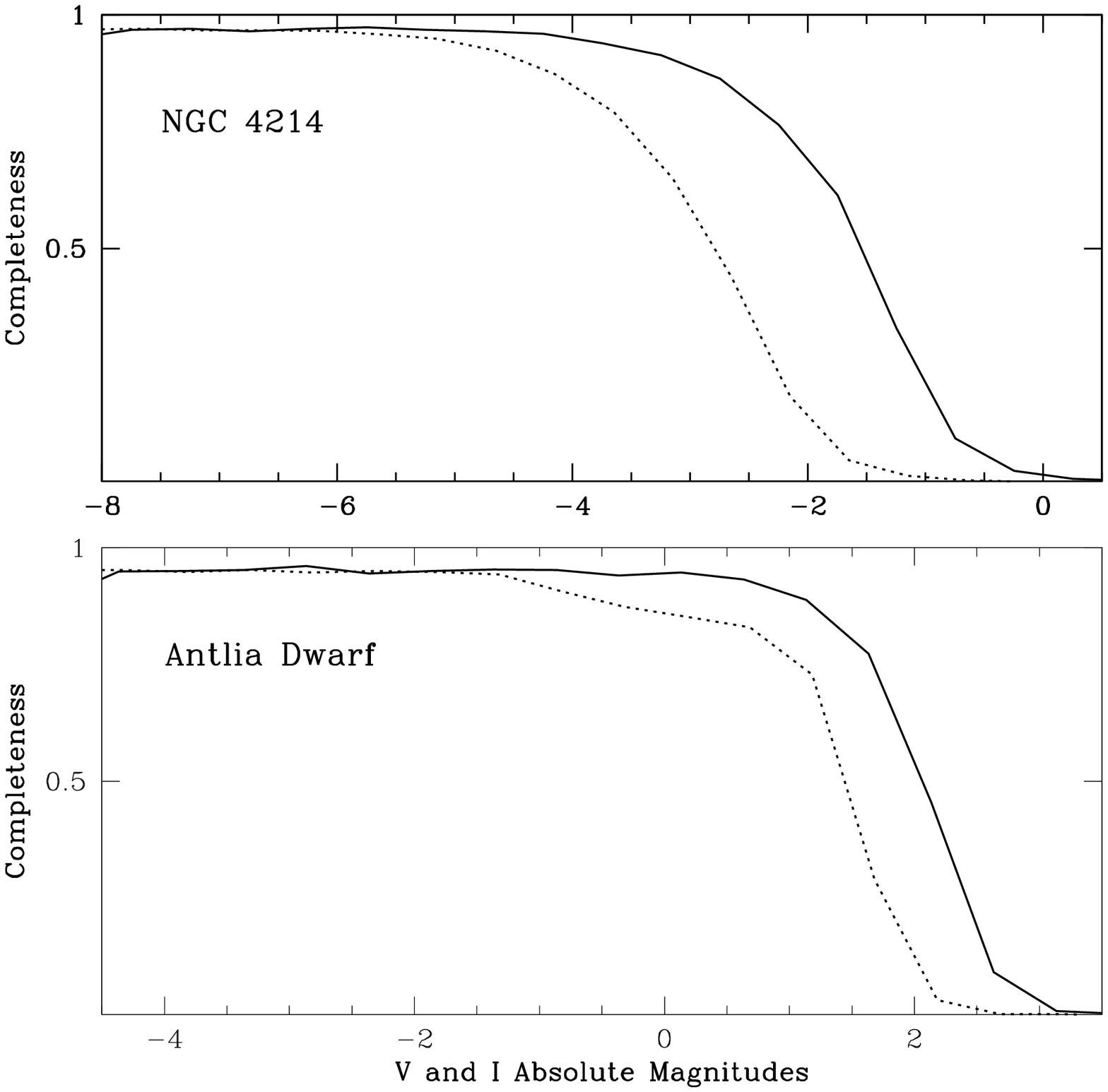}
\caption{The completeness functions for the V band (solid lines) and I band (dotted lines) for the shallowest photometry in our study (top panel: NGC~4214) and the deepest photometry in our study (bottom panel: Antlia). These completeness limits bracket the range in photometry for all eighteen galaxies with the exception of SBS~$1415+437$ whose completeness is lower than NGC~4214 as noted in the text (\S\ref{galaxies}).}
\label{fig:complete}
\end{figure}

Artificial star tests were performed to quantify the completeness and systematic uncertainty of the photometry in each of the images. The artificial stars were evenly distributed across the field of view at the range of magnitudes and colors that bracketed the measured photometry. We present the percentage of stars recovered across each magnitude bin for one of the shallowest observations (NGC~4214) and for one of the deepest observations (Antlia dwarf galaxy) in our sample in Figure~\ref{fig:complete}.\footnote{Note that SBS$1415+437$ has a lower completeness limit than NGC~4214 despite long integration times, but is not representative of our galaxy sample.} The artificial star lists were filtered according to the same parameters as the photometry lists. Point sources not meeting the criteria were retained as unrecovered stars.

\section{Star Formation Histories \label{sfh}}

\subsection{Connecting CMDs and Stellar Evolutionary Populations to SFRs\label{cmd}}

\begin{figure}
\centering
\begin{tabular}{cc}
\epsfig{file=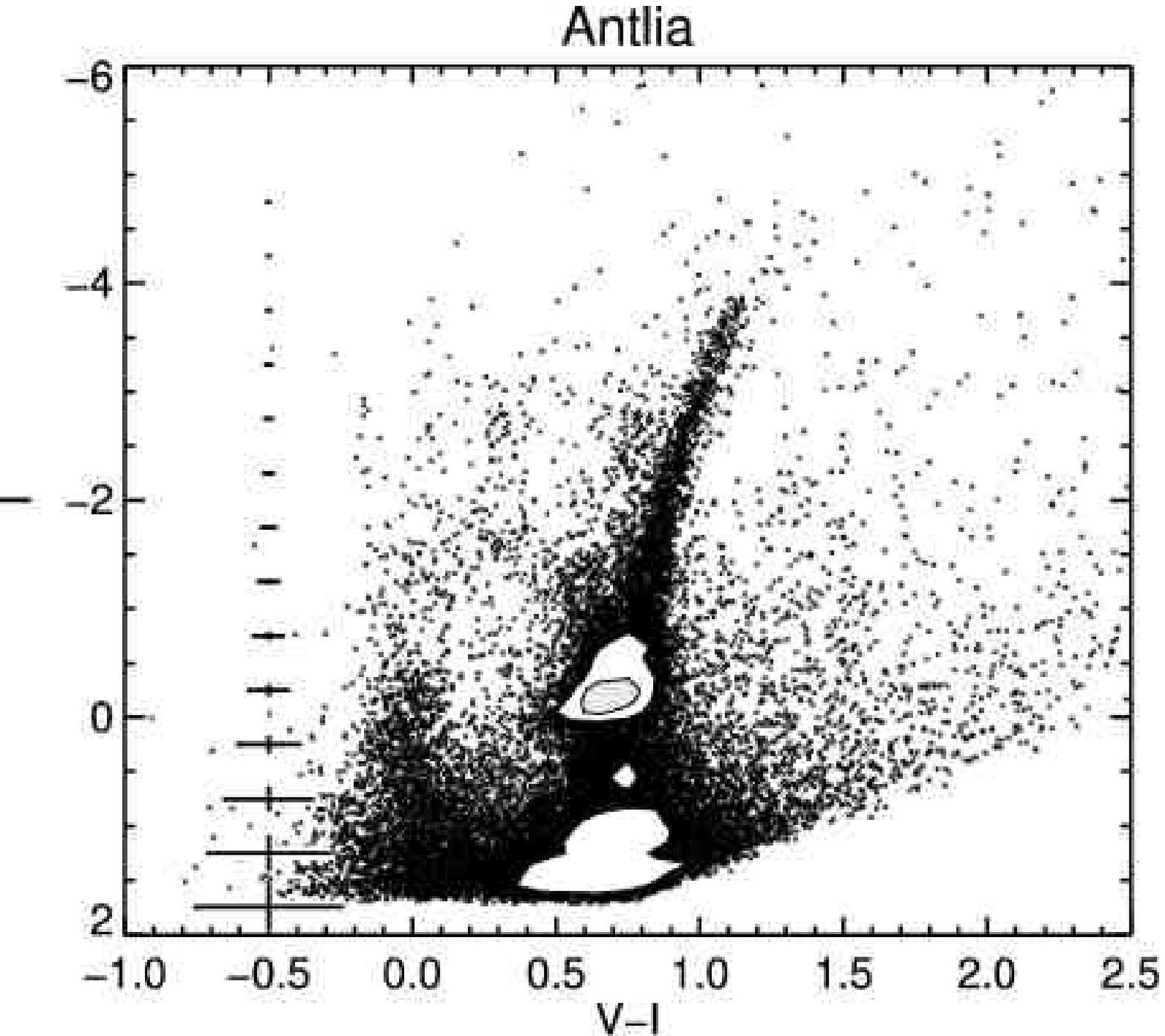,width=0.5\linewidth,clip=} & 
\epsfig{file=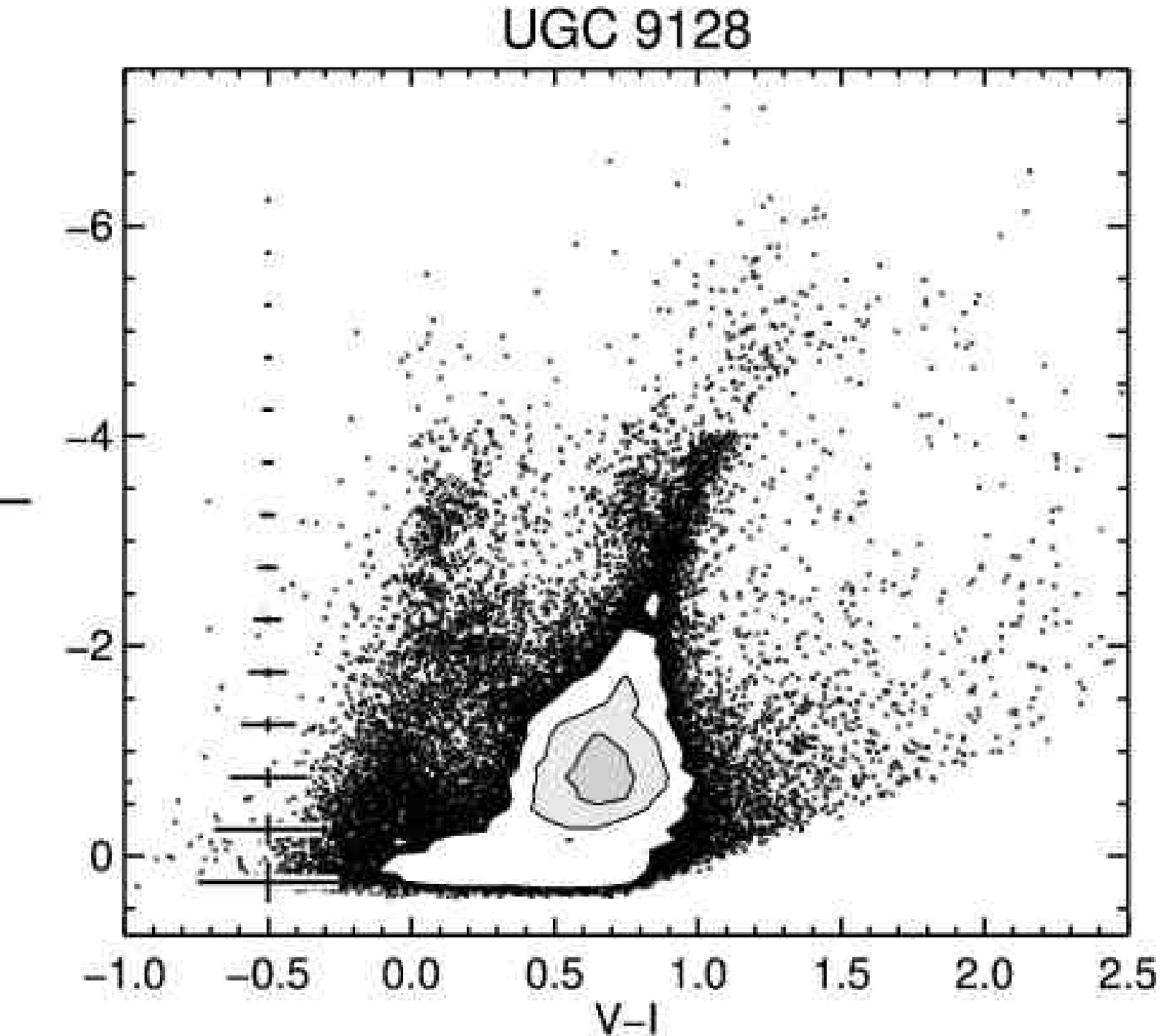,width=0.5\linewidth,clip=} \\
\epsfig{file=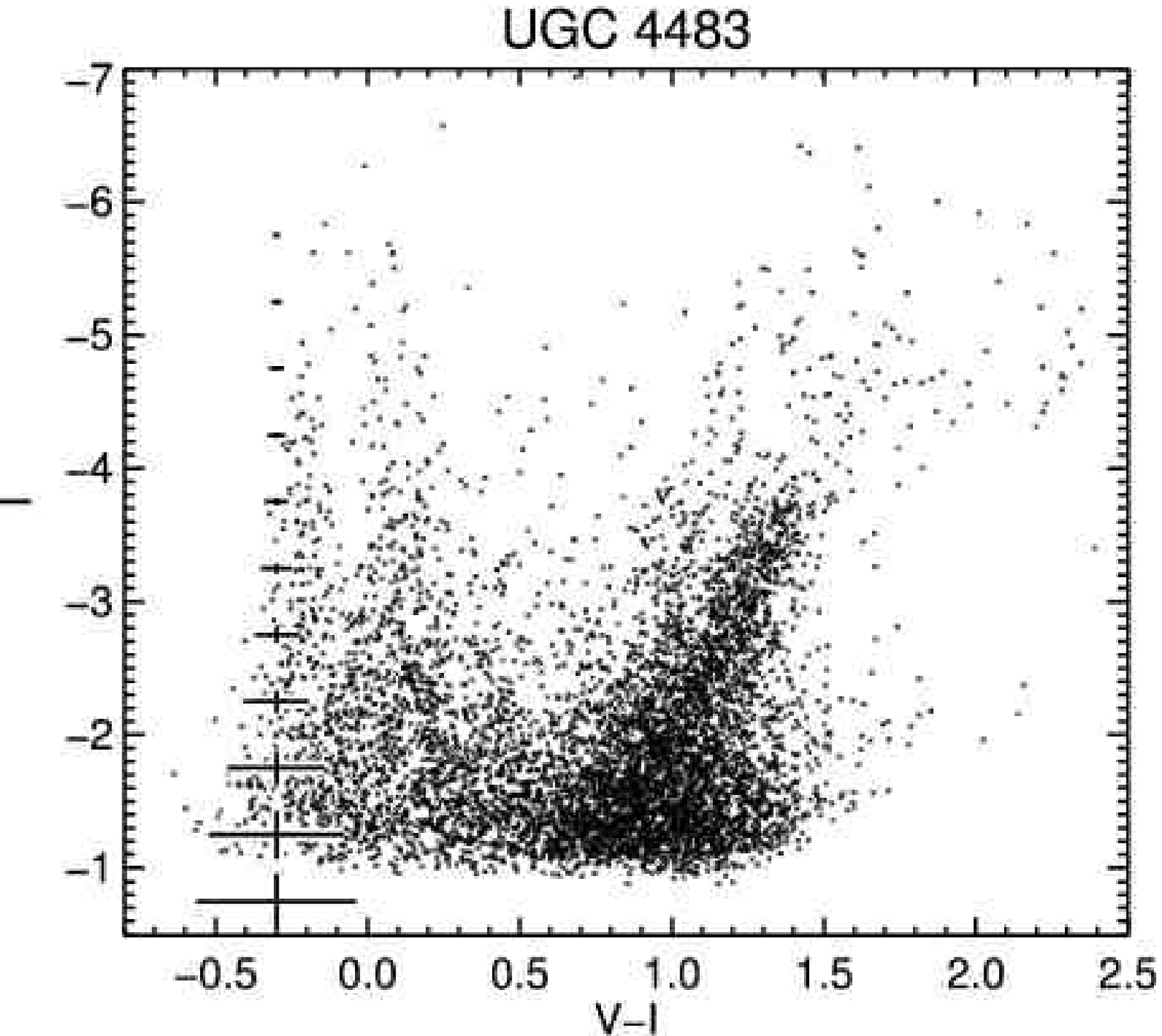,width=0.5\linewidth,clip=} &
\epsfig{file=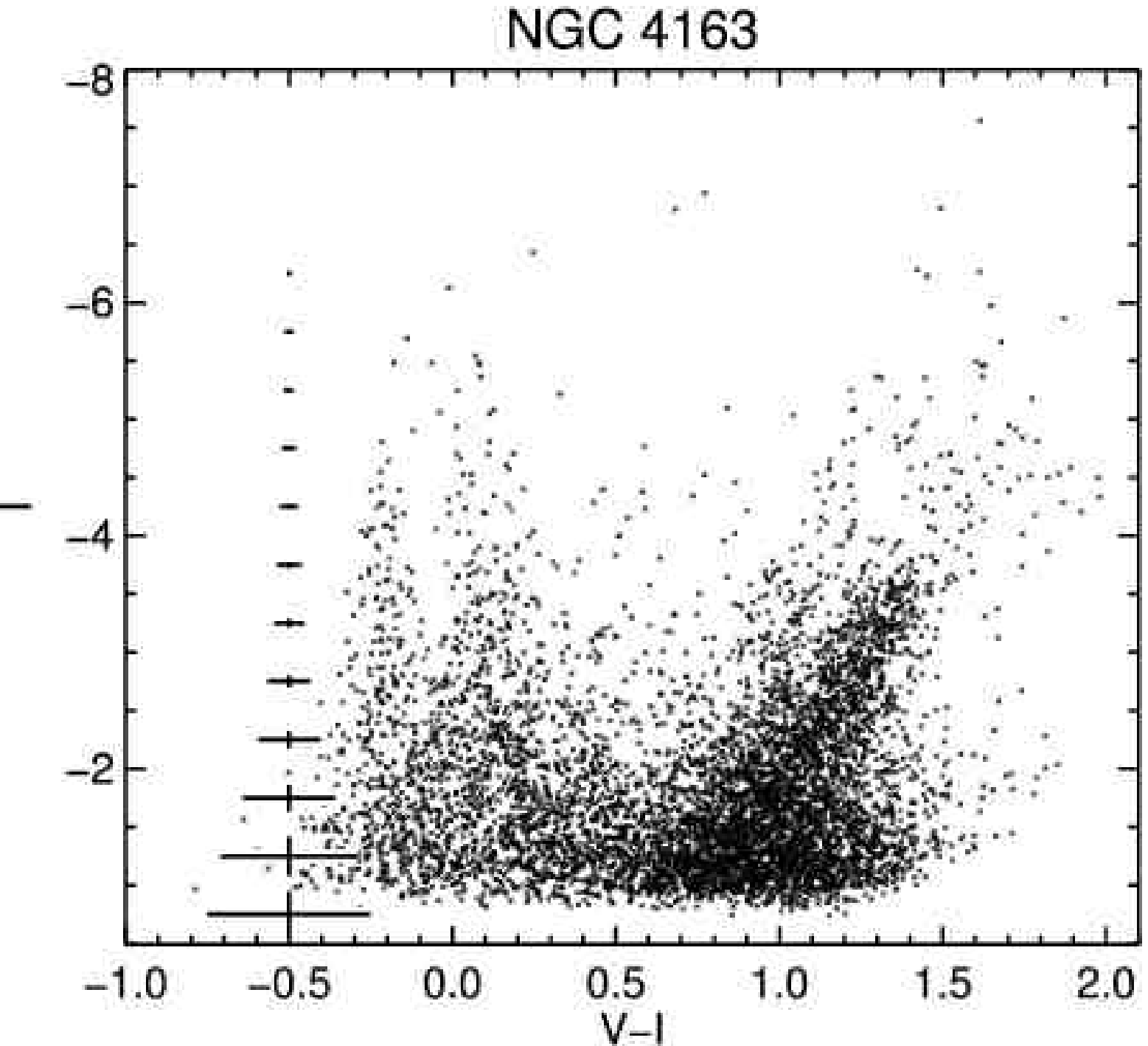,width=0.5\linewidth,clip=}
\end{tabular}
\caption{Color magnitude diagrams of Antlia, UGC~9128, UGC~4483, NGC~4163 with average photometric uncertainties per magnitude bin. The contours show the density of points where they would otherwise saturate the plot. Contour levels are 7.5, 15, 30, 60, 120, and 240 thousand points mag$^{-2}$. The distances used to calculate the absolute magnitudes are noted in Table~\ref{tab:galaxies}. The TRGB break occurs at $M_I = -4.0$. Note the differences in photometric depth and the presence and distribution of BHeB stars.}
\label{fig:cmds}
\end{figure}

CMDs created from our photometry are shown in Figure~\ref{fig:cmds}, with galaxies ordered from the faintest to the most luminous. Typical photometric errors per magnitude bin are shown. The different photometric depths across all the observations range from an absolute magnitude of $M_{I}\sim2.5$ for NGC~6822 to $M_{I}\sim-2$ for UGC~6456 and NGC~4214. The observations stop short of the depth of the red clump, with the exception of Antlia, NGC~1569, NGC~2366, and NGC~6822. With the exception of SBS~$1415+437$, all photometry reaches $\sim$2 mag below the tip of the red giant branch (TRGB) in the I band, as needed to provide constraints on the ancient SFRs (t$>8$ Gyr) \citep{Dolphin2002}, although our focus remains on the recent SFH.

\begin{figure}
\plotone{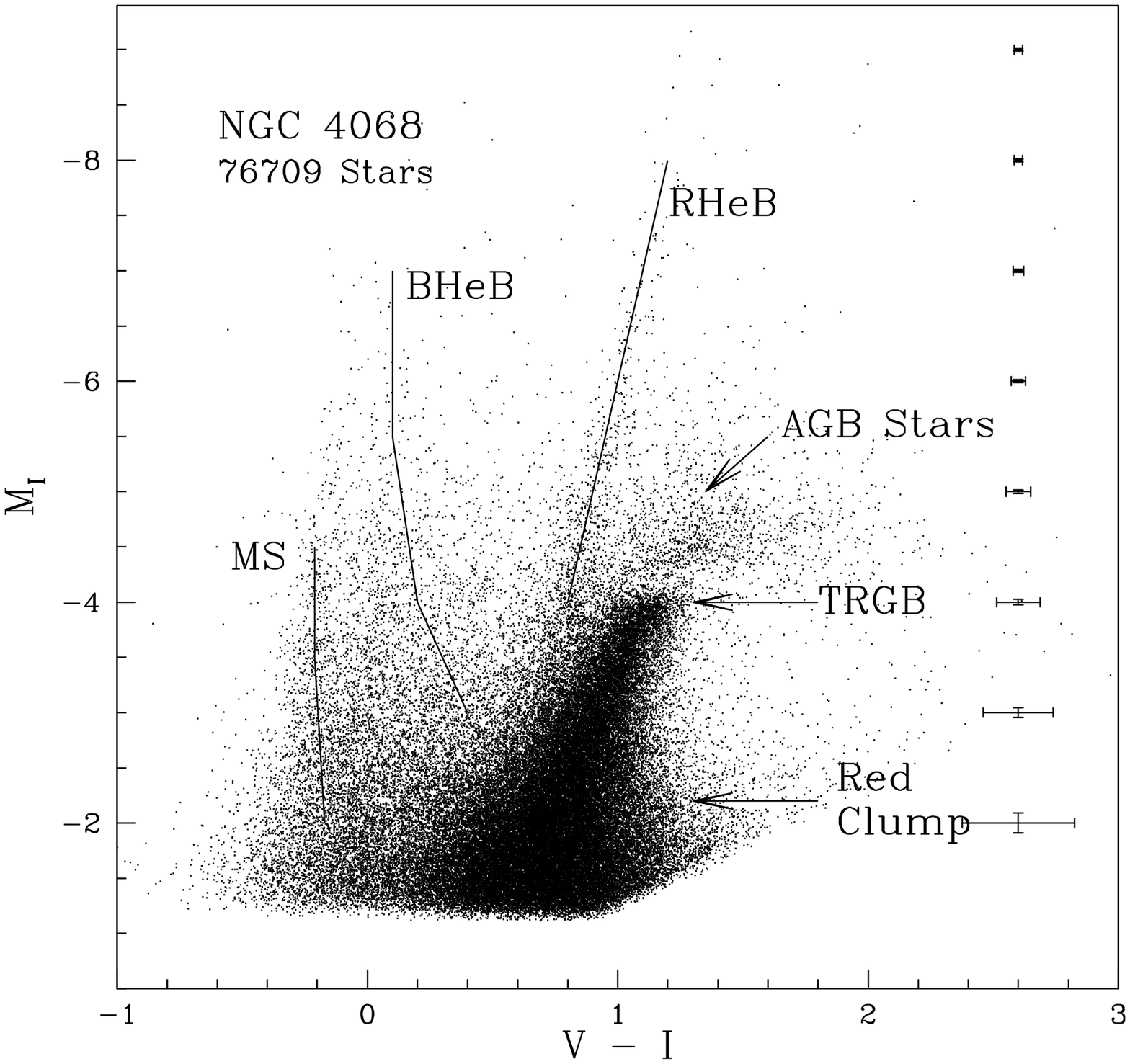}
\caption{CMD of NGC~4068 with the evolutionary stages of the stellar populations labeled. The MS, BHeB, RHeB, RGB, AGB, and red clump evolutionary stages are all easily identified in the stellar populations.}
\label{fig:evolution_stages}
\end{figure}

Multiple stellar evolutionary populations are identifiable by eye in the CMDs in Figure~\ref{fig:cmds}, including the main sequence (MS), the red giant branch (RGB), the red clump, the asymptotic giant branch (AGB), and the BHeB and red helium burning (RHeB) sequences. For didactic purposes, we present an enlargement of the CMD of NGC~4068 with the different stellar evolutionary populations overlaid in Figure~\ref{fig:evolution_stages}. While all of these evolutionary populations are used to derive the SFH, the helium burning (HeB) stars offer the clearest differentiation of starbursts, and are thus of the most interest in our study.

Stars on the HeB sequences have masses greater than $\sim$3~\msun\, and are burning helium in their cores \citep{Bertelli1994, Dohm-Palmer2002}. The ages of stars in this evolutionary sequence is between $5 - 1000$ Myr during which time the stars will turn off from the MS and migrate across the CMD from the RHeB sequence to the BHeB sequence and back again. HeB stars older than $\sim$600 Myr merge into the red clump; the actual age is somewhat metallicity dependent \citep{Bertelli1994}. The location of a HeB star on the CMD depends uniquely on age  \citep[i.e.,][]{Dohm-Palmer2002, Weisz2008, McQuinn2009} in contrast to the MS and RGB where multiple aged stars with different metallicities can occupy the same space on a CMD. Because the HeB lifetimes are relatively short, the presence of HeB stars indicates that a galaxy has experienced recent star formation. While this is true for both the RHeB and BHeB sequences, the distribution of the BHeB stars in the CMD is particularly informative due the larger number of BHeB stars \citep[e.g.,][]{Dohm-Palmer2002}. 

For example, a galaxy with a constant, low SFR over the past 500 Myr will have a sparse but smooth BHeB star distribution with the number of BHeB increasing at fainter magnitudes (i.e., lower masses). Similarly, a galaxy with a constant, but significantly higher SFR over the same time period, will have a similar distribution in the BHeB stars although the overall number of BHeBs will be higher at all magnitudes (e.g., IC~4662 in Figure~\ref{fig:cmds}). In contrast, for a galaxy with a variable SFR over the past 500 Myr, the BHeB distribution will be clumpy at different magnitudes such as in UGC~9128 (Figure~\ref{fig:cmds}) where an over-density of BHeB stars is identified at intermediate magnitudes. This effect will sometimes be accompanied by a truncation of the MS at bright magnitudes if SF has recently stopped. The properties of the BHeB population in both the distribution as well as the total number of stars can differentiate a galaxy with constant, low-levels of recent SF, from a galaxy with bursting levels of recent SF. 

\subsection{Methodology for Reconstructing SFHs \label{sfh_method}}

SFH recovery programs are sophisticated tools that use stellar evolutionary theories to reconstruct past rates of SF \citep[e.g.,][]{Tosi1989, Tolstoy1996, Gallart1996, Holtzman1999, Dolphin2002, Harris2001}. These techniques have been maturing over the past decade and are well-established in the literature \citep[see][and references therein]{Skillman2002, Skillman2003, Skillman2005, Tolstoy2009}. The primary inputs in reconstructing the past SFR are the photometry and artificial star recovery fractions (i.e., quantitative observational uncertainties and incompleteness) coupled with the stellar evolutionary models. We employ the numerical CMD fitting program, MATCH \citep{Dolphin2002} to construct a synthetic CMD based on the observed CMD varying the ages of the stellar populations to find the best fit to the observed CMD, using the stellar evolutionary models of \citet{Marigo07}. We assume the Salpeter single sloped power law initial mass function with a spectral index of $-1.35$ from $0.1-100$~\msun\ \citep{Salpeter1955} and a binary fraction of $35\%$ with a flat secondary mass distribution. A recent paper by \citet{Gogarten2009} includes a detailed description of the parameters and method employed here on similar data.

\begin{figure}
\plotone{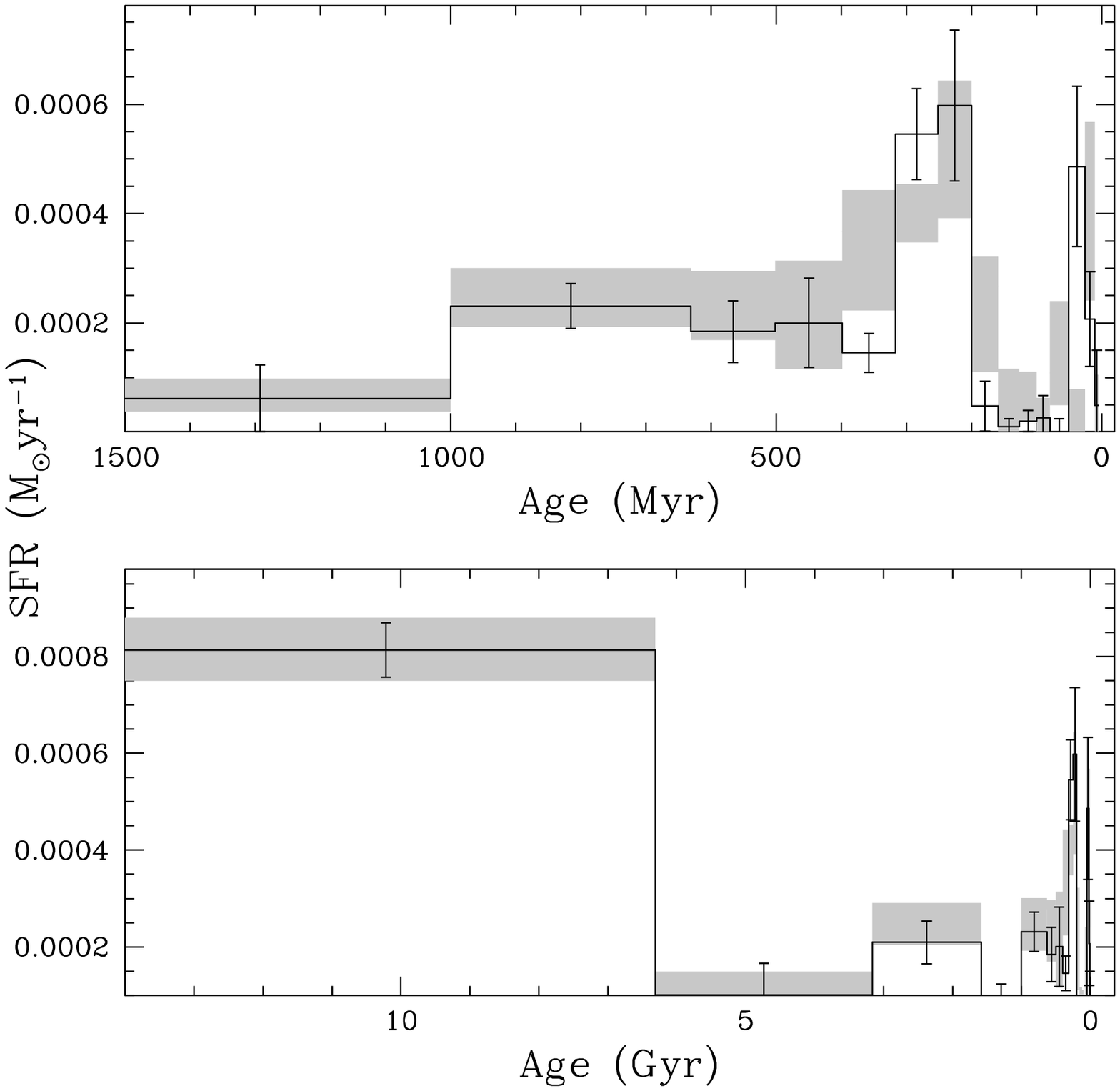}
\caption{The best SFH fit with metallicity constrained to increase in time for the Antlia dwarf galaxy is plotted as a solid line with the best SFH fit with unconstrained metallicity evolution overplotted in shaded gray. The solutions are in excellent agreement in both the ancient and recent time bins.}
\label{fig:antlia_metallicity}
\end{figure}

\begin{figure}
\plottwo{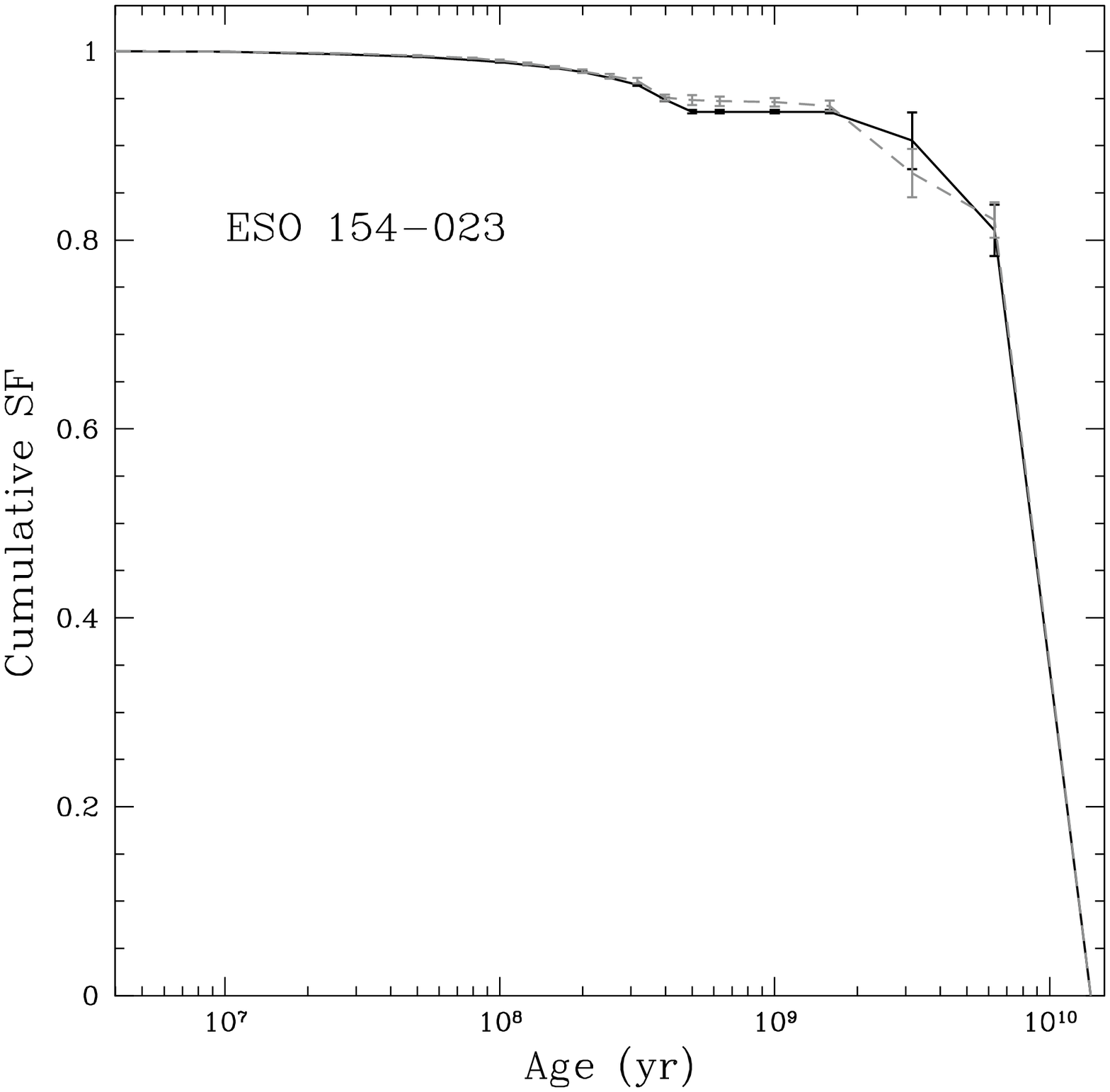}{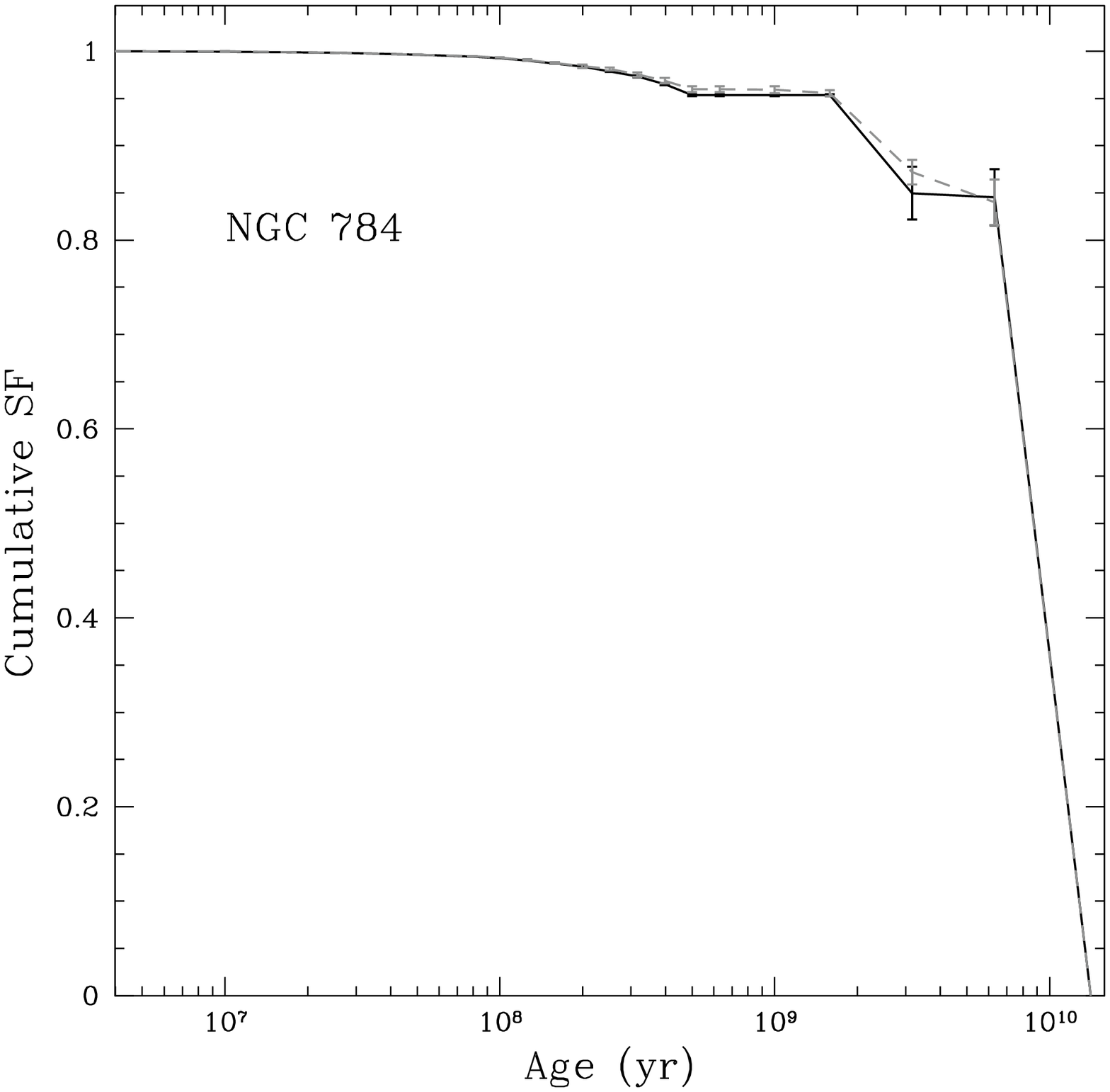}
\caption{The cumulative percentage of SF in ESO~154$-$023 and NGC~784 found both with the metallicity constrained to increase in time (black lines) and the metallicity unconstrained in the solutions (dashed gray lines). In both galaxies, the cumulative SF are in very good agreement.}
\label{fig:cumulative_sf}
\end{figure}

We constrained the metallicity, $Z(t)$, to increase as a galaxy evolves in time. This physically motivated constraint guides the metallicity evolution in the absence of observational constraints that would have been available with deeper photometric depths. We did not constrain the metallicity function for three galaxies (Antlia, NGC~2366, NGC~6822)\footnote{While the photometry of NGC~1569 is nearly as deep, the high levels of internal extinction dictated constraining the metallicity to increase with time.}, as the deeper photometry reaches the bottom of the red clump (NGC~2366) or two magnitudes below the red clump (Antlia, NGC~6822) providing enough information at recent times (t$<$1 Gyr) to constrain the metallicity. We present a comparison of the SFHs for both the unconstrained and the constrained metallicity solutions for the deepest photometric data, Antlia, in Figure~\ref{fig:antlia_metallicity}. The two solutions show excellent agreement in both the ancient and the recent time bins. For shallower photometry, the metallicity constrained and unconstrained solutions also showed good agreement with small deviations in SFRs seen in only a few of the time bins. We chose to use the constrained metallicity solutions in these datasets as this produces a more realistic metallicity evolution in the galaxies avoiding large jumps in metallicity over short time periods \citep{Williams2010}. The galaxies that showed the largest deviations were the galaxies with higher extinction or significant stellar crowding (i.e., NGC~1569). As an example, we present two examples of the cumulative SFH with the metallicity constrained and unconstrained for two galaxy with typical photometric depth of our sample, ESO154$-$023 and NGC~784, in Figure~\ref{fig:cumulative_sf}. As seen in the plots, the cumulative star formation is equivalent within the uncertainties.

Photometric errors and differential extinction can broaden features in a CMD. These effects are explicitly accounted for in the CMD fitting programs employed in this study. The photometric uncertainties are quantified both with the uncertainties in the photometric measurements and with a measure of the photometric incompleteness provided by artificial star tests. Extinction is a free parameter fit by the SFH recovery program. While these factors must be considered in any SFH derivation, they are of smaller consequence in the deriving recent SFHs for dwarf starburst galaxies. Photometric uncertainties are smaller on the relatively bright stars found in sites of recent SF. Foreground extinction is lower as the galaxies in our study are predominately found at high Galactic latitudes and internal extinction is lower in the low metallicity environments of the sample. Nevertheless, to ensure a robust treatment of these parameters, we studied different regions of the galaxies with a range in photometric errors, and allowed for varying foreground and differential reddening when fitting the CMD. We found that the SFRs derived were not significantly affected by varying photometric errors and low extinction levels ($A_{V}\ltsimeq0.5$). Higher levels of differential extinction is more problematic affecting the SFRs derived for galaxies such as NGC~1569 and NGC~6822 with known high Galactic and/or internal extinction. For these systems, our reported uncertainties in the derived SFRs are higher. A complete discussion of these three factors is given in \citet{McQuinn2009}. As an additional test of our results, we derived the SFH of one galaxy, NGC~4163, using a different SFH recovery program, IAC$-$pop \citep{Aparicio2009}. We found the SFHs to be equivalent within the uncertainties between the two methods. 

\begin{figure}
\plotone{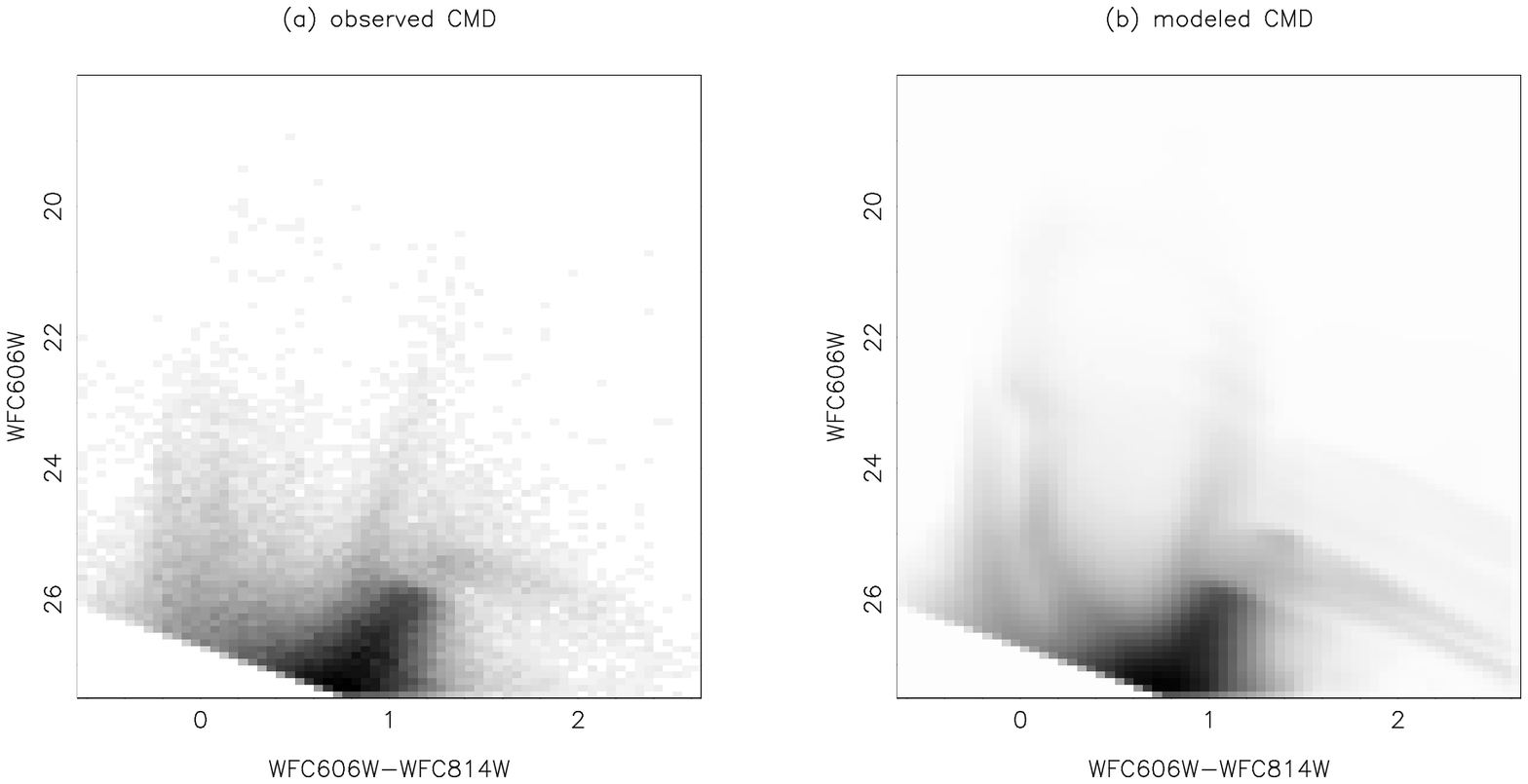}
\caption{The left panel shows the Hess diagrams for the ESO 154-023 observations. The right panel is the best-fit synthetic Hess diagram to the observations representative of a typical fit for the sample galaxies. While there are slight discrepancies, each of the evolutionary populations are well-described in the synthetic CMD.}
\label{fig:hess}
\end{figure}

To illustrate the recovery of the SFH, in Figure~\ref{fig:hess} we present a synthetic CMD created by MATCH for ESO~$154-023$, alongside the observed CMD on the same scale. The synthetic CMD represents a typical model fit to the data in our sample. The different evolutionary populations are well-reproduced in the synthetic CMD, in both the broad and fine features. The quality of the fit between the observed and modeled CMDs can be quantified with an effective $\chi^{2}$ parameter \citep{Dolphin2002}. The $\chi^{2}$ measures the likelihood of the SFH derived from the model CMD to be the true SFH of the observed galaxy given our inputs and models \citep[i.e.,][]{Dolphin2002}. The $\chi^{2}$s per degree of freedom for our analysis are presented in the last column of Table~\ref{tab:sfh_fits} and ranged from $1.12-2.67$. The three galaxies with a $\chi^{2}$ value above two suffer from high extinction (NGC~1569 and NGC~4449) and high crowding (NGC~1569, NGC~5253, and NGC~4449). To match these features, MATCH fits a distance modulus and total extinction value (i.e., foreground plus internal extinction) to each galaxy. In Table~\ref{tab:sfh_fits}, we compare these modeled values with distance moduli obtained from the literature and Galactic extinction values measured by \citet{Schlegel1998}. The distance moduli are well matched within a few tenths of a magnitude. The model extinction values show a wider spread when compared with the \citet{Schlegel1998} foreground extinction measurements. In many cases the modeled value is higher as is expected as the model estimates not only the foreground extinction but also the extinction internal to the galaxy. 

Crowding tends to be less in low surface brightness dwarf systems than in spiral and elliptical galaxies. However, regions of higher stellar density presented markedly lower completeness limits than the surrounding lower stellar density regions in four galaxies (IC~4662, NGC~1569, NGC~4449, NGC~5253). We accounted for this disparity by separating these galaxies into regions of higher and lower stellar density. In each region, we conducted the artificial star analysis thus more accurately describing the completeness across the changing condition of the galaxies. The SFHs were derived for each region separately and were summed creating a solution for the entire field of view. As a test, we performed the same analysis with two galaxies that did not exhibit the same extremes in stellar densities (NGC~4068 and NGC~4163) and found that the summed results were equivalent to the SFH derived from the entire field of view within the uncertainties. We therefore derived the SFHs from the global fields of view for the remainder of the sample which did not exhibit the same extremes in stellar density.

\subsection{Star Formation Histories \label{sfr_time}}

\begin{figure}
\plotone{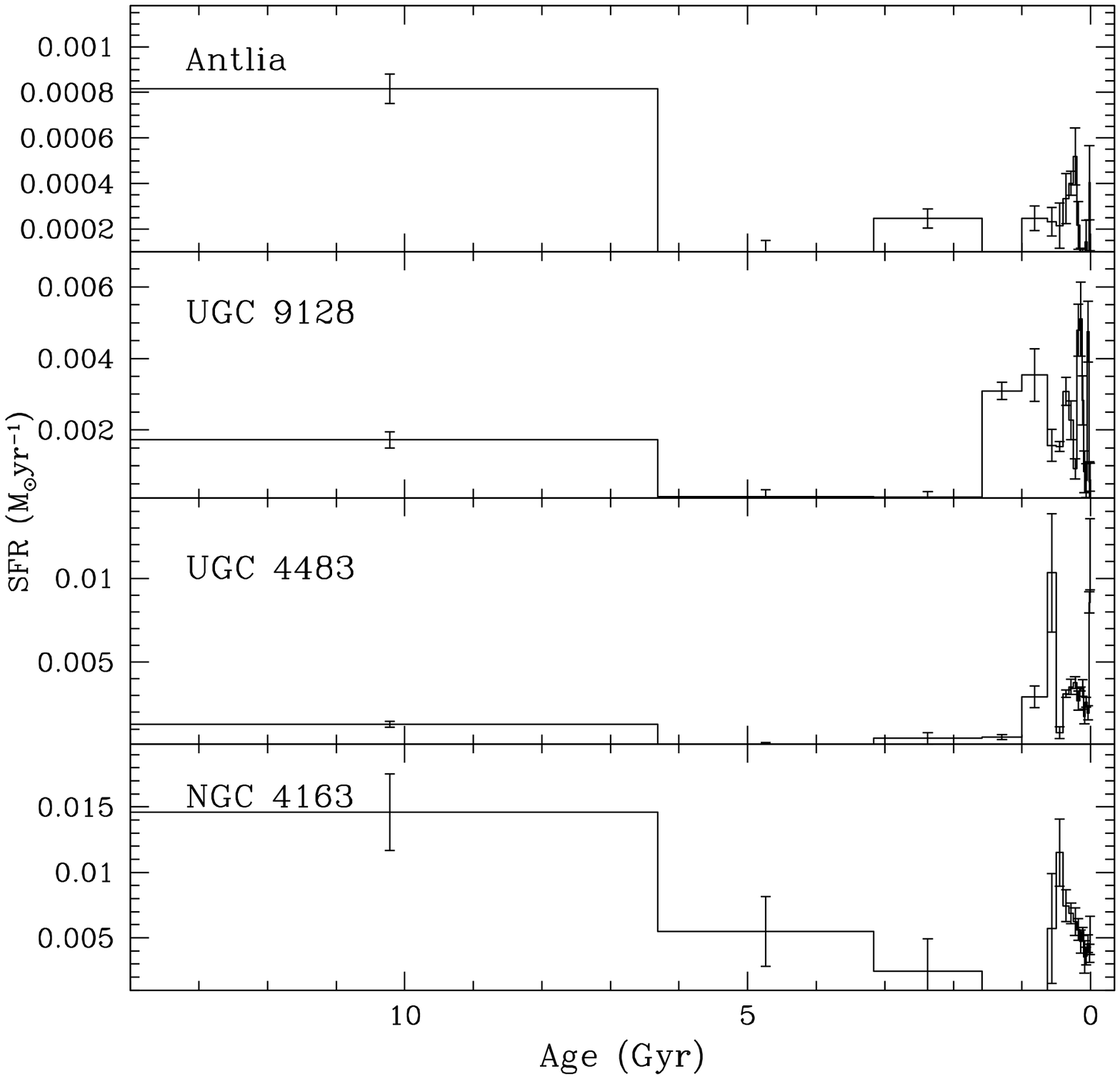}
\caption{The SFHs of Antlia, UGC~9128, UGC~4483, NGC~4163 derived from the HST optical observations of the resolved stellar populations for the eighteen galaxy sample.  The oldest time time ($6-14$ Gyr) is an average SFR that constrains the ancient SF giving context for the overall SFH yet these numbers are not used in our calculations. The uncertainties in this age bin are underestimated due to averaging over a longer time. The time resolution at more recent times is much finer and makes use of the unambiguous age dating of the HeB stars. Note the elevated levels of star formation in these most recent time bins. Expansions of the last 1 Gyr SFHs are presented in Figure~\ref{fig:sfh_1gyr}. The galaxies are ordered by $M_{B}$ luminosity from faintest to brightest.}
\label{fig:sfh_14gyr}
\end{figure}

The lifetime SFHs (i.e., SFR$(t)$) are presented in Figure~\ref{fig:sfh_14gyr}. No comparison with previously published SFHs are presented as none exist yet in the literature. The errors include both systematic and statistical uncertainties derived from Monte Carlo simulations. The average time resolution for deriving the SFH is $\delta$log$(t)\sim0.3$. However, the SFRs at large look-back times are not well constrained without photometry that reaches the oldest MS turn-off. Hence, we averaged the SFR results from the oldest time bins to create one coarse time bin of $6-14$ Gyr; loosely constraining the ancient SFH. The uncertainties in this older time bin are relatively small due to averaging over a longer time period. By averaging over a long time interval, we traded temporal resolution for more secure SFRs thus providing a baseline for comparison with current SFRs and a context for the intensity of the recent SF. A finer time resolution is used in deriving the SFH in the most recent $\sim500$ Myr leveraging the excellent temporal information from the HeB stars. The SFRs derived for NGC~6822 are lower limits due to the limited spatial coverage of the observations. As noted earlier in \S\ref{galaxies}, the ancient SFH for SBS~1415$+$437 is not well constrained due to the shallow photometry.

The uniform photometry and SFH recovery allow us to directly compare the results between galaxies. To first order, the SFHs share some common characteristics. Most show significant star formation in the oldest time bin indicative of galaxy assembly and the formation of stars in systems rich in gas. The galaxies also show elevated levels of recent star formation, as expected for systems previously identified as starbursts or fossil bursts.  

\begin{figure}
\plotone{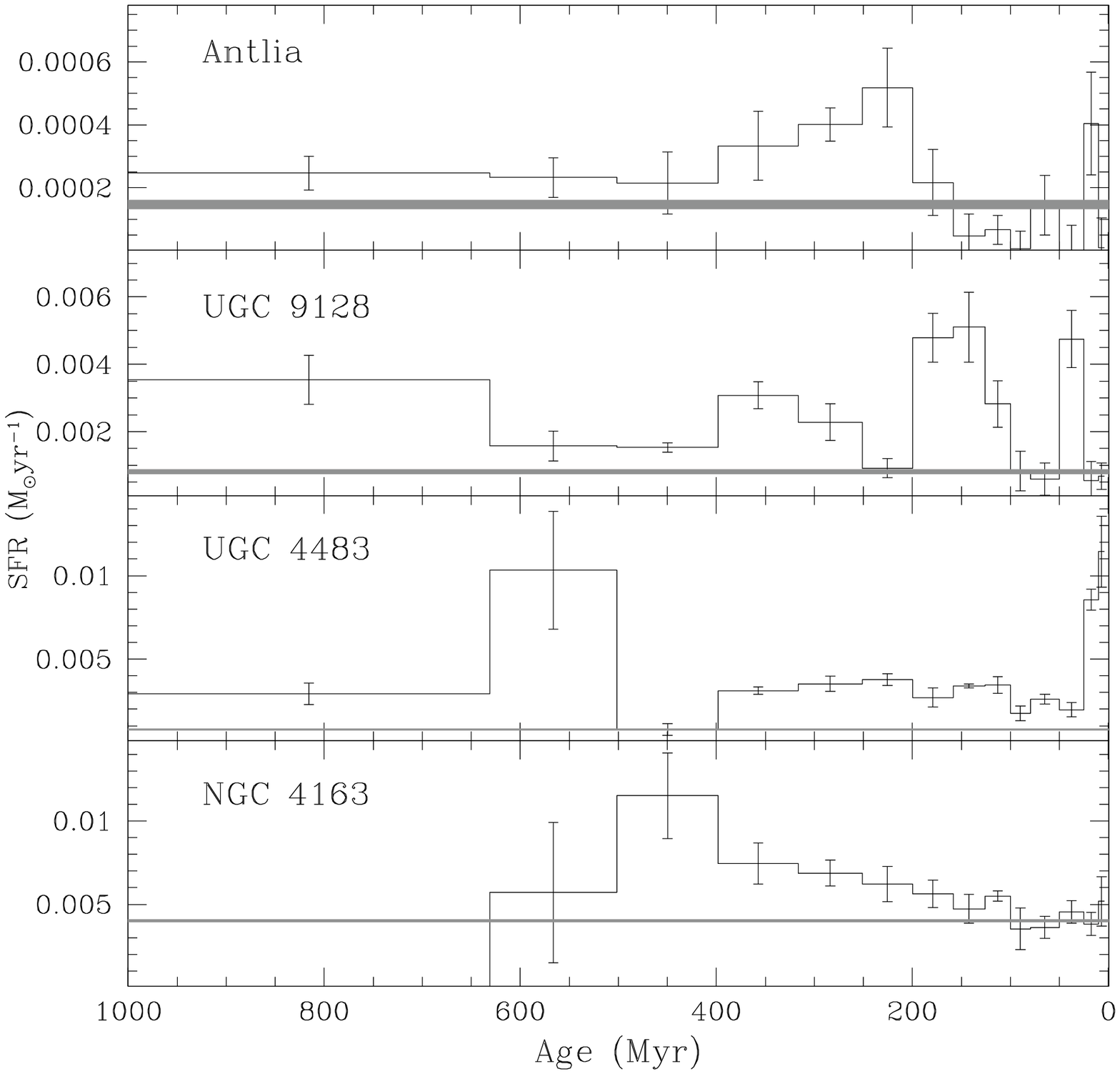}
\caption{The last 1 Gyr of the SFHs of Antlia, UGC~9128, UGC~4483, and NGC~4163 presented in Figure~\ref{fig:sfh_14gyr} are shown here in greater detail. The gray line represents the average SFR over the past 6 Gyr (${\mathrm{b_{recent}}}=1$). The thickness of the gray line represents the uncertainties in the average which are affected by the photometric depth, extinction, and photometric crowding. Each profile shows elevated SFRs at recent times yet the SFRs over the past few hundred Myr are diverse. All galaxies exhibit the common characteristic of having SFRs well above the last 6 Gyr average yet there is no specific pattern of SFR that characterizes a burst nor is there an absolute level of SF that sets a threshold for bursting SF. The galaxies are ordered by $M_{B}$ luminosity from faintest to brightest.}
\label{fig:sfh_1gyr}
\end{figure}

Deeper inspection reveals, however, that the details of the SFHs vary from galaxy to galaxy. While elevated levels of SF are sustained over a large interval of time ($\delta$t$>$few 100 Myr), there are variations and inhomogeneities in the SFR profiles on small temporal scales ($\delta$t$\sim10-20$ Myr). There does not appear to be a ``typical'' starburst SFR$(t)$ profile. The most notable differences in the recent SFRs are more readily apparent in Figure~\ref{fig:sfh_1gyr} where we present an expansion of the last gigayear of the SFHs for each of the galaxies. A number of systems show a sustained SFR over a period of a few hundred Myr (e.g., ESO~154$-$023, NGC~4068, and UGC~4483) while others exhibit fluctuations in their recent SFR (e.g., UGC~9128, NGC~1569, SBS~$1415+437$, NGC~2366, NGC~625, and NGC~4214). One galaxy, NGC~6822 shows the beginning of a burst with elevated levels of SF in the most recent 50 Myr. 

One of the advantages of using resolved stellar populations is the ability to identify and study galaxies whose bursts have ended. Five SFHs show a decline in SFR in the most recent times (i.e., Antlia, UGC~9128, NGC~625, NGC~4163, and NGC~6789) after experiencing elevated levels of SF for a period of a few hundred Myr. These fossil burst galaxies are excellent systems in which to measure the complete duration of a burst event (see Paper II), rather than a lower limit. Fossil burst galaxies also increase the number of nearby systems in which one can study the starburst phenomenon. The resolved stellar populations required for any detailed SFH history limits such studies to galaxies within $5-8$ Mpc. We encountered the limitations of studying farther galaxies even with observations of long integrations times (e.g., SBS~1415+437). Identifying fossil burst systems increases the sample of starburst galaxies in the nearby universe reducing our dependence on studying systems that push the limit of our observational capabilities. 

\section{When Is Star Formation a Starburst? \label{classify}}

SF activity within a burst is complex and consequently, there are many definitions in the literature to identify and quantify bursts. Here, we consider three metrics to identify starbursting systems and discuss their applicability/limitations.

\subsection{Gas Consumption Timescales}

A fundamental characteristic of a starburst is that the high rate of SF during the burst is short-lived compared to the lifetime of the host galaxy. A limit for how long this high SFR may be sustained can be determined by calculating how long it would take to turn all the raw material fueling the burst into stars. If  this ``gas consumption timescale'' \citep[$\tau_{gas}$,][]{Roberts1963} is significantly less than the Hubble time, then the burst must be unsustainable and thus must be short lived on a cosmic time scale, increasing the chance that the elevated SFR is temporary. 

One can calculate $\tau_{gas}$ by dividing the mass of the gas (i.e., the raw material that fuels the starburst) by the peak SFR (\msun\ yr$^{-1}$) during the starburst event. However, measuring an accurate and relevant gas mass is difficult for these low mass dwarf systems. The molecular phase of the ISM is rarely detectable in dwarfs \citep[e.g.,][]{Israel1995, Taylor1998, Barone2000}, and thus is unlikely to contribute significantly to the gas supply, unless the uncertain conversion factor from CO to H$_{2}$ is extremely high \citep[e.g.,][]{Wilson1995, Israel1997, Bolatto2008}. Atomic gas is thus likely to dominate the gaseous ISM. Indeed, the current best estimates suggest that the ratio of molecular gas mass to atomic gas mass is $0.3\pm0.05$ \citep[][and references therein]{Leroy2005} indicating that the atomic gas mass dominates. While molecular gas is needed for star formation to proceed, the atomic gas is what ultimately fuels the formation of molecular gas, and is the relevant quantity to consider for the gas consumption times of low mass systems. Yet, measurements of the atomic gas mass can also be problematic. The atomic gas is frequently far more extended than the star forming disk reaching up to several Holmberg radii at a density of $10^{19}$ atoms cm$^{-2}$ for dwarf irregular galaxies \citep[e.g.,][]{Huchtmeier1981}. Thus, the gas mass available for SF is lower than the measured \HI\ mass in a galaxy making $\tau_{gas}$ an upper limit. 

Using the extended \HI\ mass as the best available proxy for the gas fueling SF in the galaxies, we present the resulting gas consumption timescales in the absence of stellar recycling in Table~\ref{tab:truth_table}. We have calculated $\tau_{gas}$ from the peak SFR in the last Gyr for each galaxy (column 3) and atomic mass measurements taken from the literature (columns 4 and 5) after multiplying the \HI\ mass by a factor of 1.4 to account for the mass of helium and heavy elements \citep[e.g.,][]{Roberts1963, Kennicutt1994}. Of the sixteen galaxies for which we have found \HI\ measurements, fifteen show consumption timescales significantly less than a Hubble time ($0.44$ Gyr$<\tau_{gas}<5.2$ Gyr) indicating that the peak SFRs measured in these systems cannot be sustained for the entire history of the galaxy. Thus, while the gas consumption timescales are only upper limits, these fifteen galaxies still fit this simplistic starburst metric. One galaxy, NGC~6822, has a gas consumption timescale longer than a Hubble time ($\tau_{gas}=26$ Gyr); however, not only is the atomic gas known to be unusually extended ($>600$~arcmin$^2$) relative to the optical disk \citep{deBlok2000}, but the optical observations cover only a fraction ($\sim17$~arcmin$^2$) of the star forming disk (see observational footprint in Figure~\ref{fig:images}).

Our simple calculation supports the gas consumption timescale as a useful first order metric for confirming that fifteen of the sixteen galaxies fit this starburst rubric. However, the $\tau_{gas}$ metric provides only a coarse evaluation of these systems as starbursts, due to the dwarfs' spatially extended \HI\ halo that may not be available for SF. An additional drawback of this simplistic gas consumption timescale metric is that is does not provide any additional insight into the nature of the burst.

\subsection{Starburst Thresholds \label{thresholds}}

Starbursts are defined by high levels of SF activity, yet the exact level of SF that constitutes ``bursting'' is subjective. Setting an absolute SFR as a threshold to identify bursts would bias any results toward larger galaxies with intrinsically higher SFRs. For an unbiased metric, one must consider the SFR relative to what an individual galaxy has experienced in the past. Previously, \citet{Scalo1986} and \citet{Kennicutt1998} formulated the birthrate parameter (b) which compares the current SFR with the the average SFR over the lifetime of a galaxy (i.e., b$\equiv$~SFR~/~$<$SFR$>$). McQuinn \etal\ (2009) modified this parameter using the average SFR during the last 6 Gyr,instead of the average SFR over the lifetime of the galaxy (i.e., ${\mathrm{b_{recent}}}\equiv $~SFR~/~$<$SFR${\mathrm{_{0-6~Gyr}}}>$). This calculation decouples any SF activity during the initial assembly period of galaxies from the SF during later epochs providing a more relevant baseline in the current epoch. Note that 6 Gyr is still a substantial fraction of the lifetime of a galaxy, and thus is a robust measure of the average SFR. In addition, the normalized SFRs are more securely anchored with SFRs from the more recent time bins (see \S\ref{sfr_time}). 

Using a ${\mathrm{b_{recent}}}>2$ to identify a burst and defining the beginning and end points of a burst at the time when ${\mathrm{b_{recent}}}=1$ \citep{McQuinn2009}, the duration of the starburst events in our sample last over a few hundred Myr, with the shortest duration being 450 Myr and the longest being 1.3 Gyr. The accuracy of these measurements depends in part on our inherent time resolution of the affected SFH. The temporal resolution over the last few 100 Myr is fine enough that we can discern and measure fluctuations in the SFR over short timescales (from $\delta$t$25-50$ Myr in the last 100 Myr to $\delta$t$\sim125$ Myr 500 Myr ago). In contrast, older bins have coarser inherent time resolution (from $\delta$t$\sim350$ Myr in the $650-1000$ Myr time bin), giving less accuracy on the interval burst duration. Coarse time bins not only reduce the accuracy of the burst duration, but also reduce our sensitivity to bursts. With larger time bins, short duration increases are not detectable if their durations are much shorter than the width of the time bin in our derived SFH. We are therefore most sensitive to starbursts that have occurred in the last 500 Myr where are temporal resolution is the finest.  

The average SFR for recent times (${\mathrm{b_{recent}}}=1$) are drawn in gray across the SFHs in Figure~\ref{fig:sfh_1gyr}; the width of the gray line represents the uncertainties in ${\mathrm{b_{recent}}}=1$. We identify starbursts if the birthrate parameter is greater than two \citep{Kennicutt2005}. All galaxies show significant levels of recent star formation that are well above ${\mathrm{b_{recent}}} = 2$ in multiple time bins. In the case of NGC~6822, the high levels of SF occur in the last 50 Myr, indicating that this burst is likely just beginning. The four additional galaxies we included in our sample based on the HeB distributions in their CMDs (Antlia, UGC~9128, ESO~$154-023$, and NGC~784) are confirmed as starbursts. All show ${\mathrm{b_{recent}}}>2$ over multiple time bins. 

Because of the long temporal baseline available with CMD studies, our analysis also identifies fossil bursts whose \textit{current} SFR is below our starburst threshold (i.e., Antlia, UGC~9128, NGC~625, NGC~4163, and NGC~6789). Four of the five fossil bursts (Antlia, UGC~9128, NGC~4163, and NGC~6789) do not show the typical burst characteristics of intense H$\alpha$, UV, or optical emission. For example, looking at a current optical image of the lowest mass galaxy in our sample, the Antlia dwarf galaxy, it is not at all obvious that it has undergone a period of significant star formation in its recent past. The absolute values of the SFR in Antlia are the lowest in our sample, but taken in the context of its own SFH, the recent burst of SF is significant and measurable. 

The fifth fossil burst, NGC~625, is unique among our sample. Although NGC~625 produces strong radio continuum and H$\alpha$ emission often associated with starbursts \citep{Cannon2003, Cannon2004}, the SFRs over the past 500 Myr fall below our starburst threshold of ${\mathrm{b_{recent}}}=2$. Yet, NGC~625 shows strong evidence at other wavelengths of a starburst. This discrepancy can be understood if we look at the lifetime SFH in Figure~\ref{fig:sfh_14gyr}. The SFH shows a low level of SF for the first 11 Gyr without the high levels of SF often associated with initial galaxy assembly, followed by multiple bursts in SF over the past 3 Gyr. This ``suppression'' of SF at early times is similar to the SFH reported for the dwarf galaxy Leo A \citep{Cole2007}. Our ${\mathrm{b_{recent}}}$ metric assumes that the high levels of SF associated with early stages of SF activity in a galaxy occurs at t$>$6 Gyr; NGC~625 does not fit this assumption but we include it for comparison to the others in our sample. Note that suppressed initial SF over a significant period of time is also seen in two other systems in our sample, NGC~2366 and NGC~4214.

\begin{figure}
\plotone{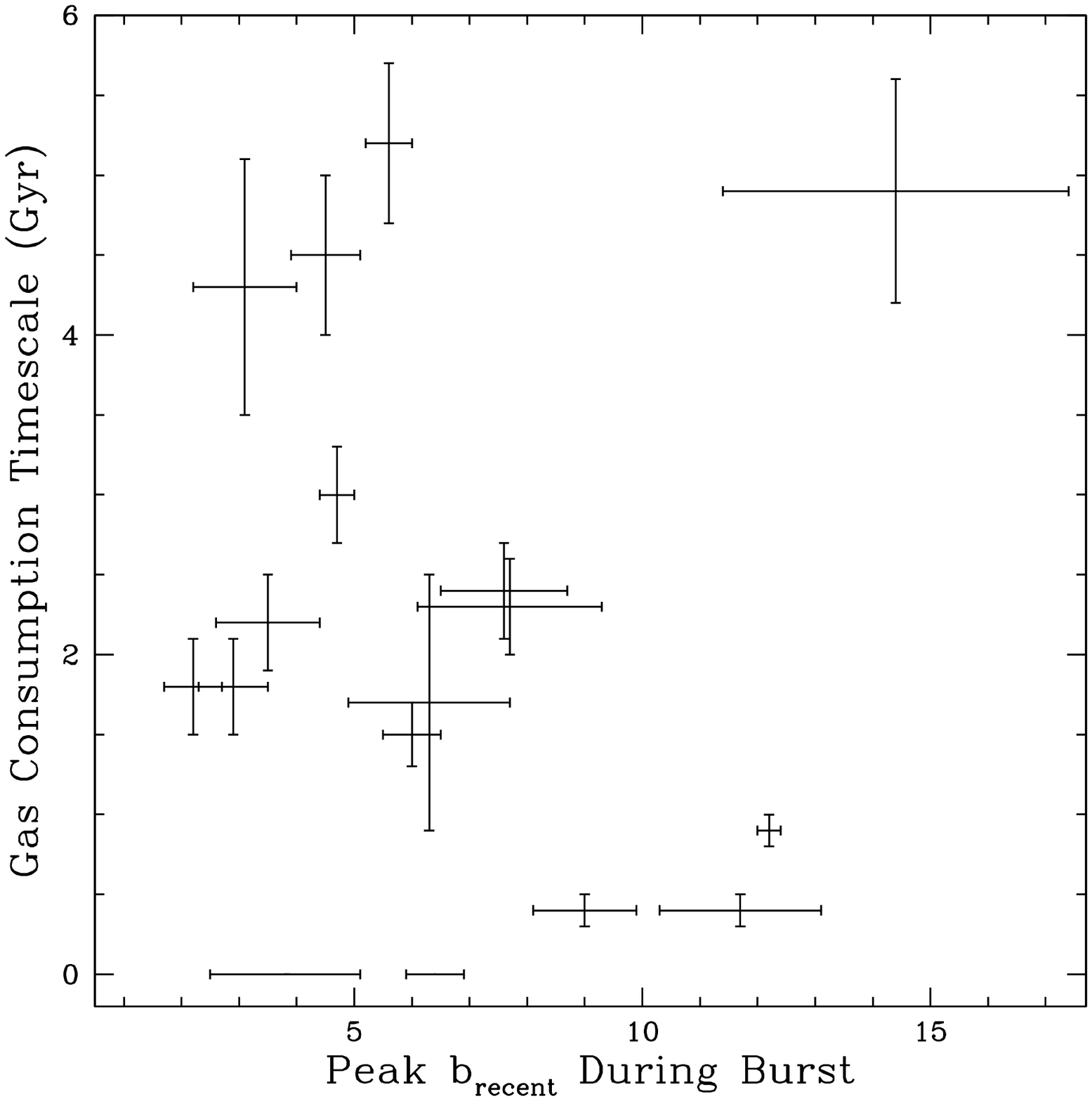}
\caption{The gas consumption timescale is plotted as a function of the maximum value of ${\mathrm{b_{recent}}}$ during the most recent burst.  Gas consumption timescales are based on \HI\ mass measurements, which are available for all but two of the galaxies in our sample. We consider the gas consumption times to be upper limits, as it is unlikely that the significant fraction of gas at large radii \citep{Huchtmeier1981} will be available for star formation. The uncertainties in $\tau_{gas}$ were calculated assuming a 10\% uncertainty in the atomic gas masses. The plot indicates that the gas consumption timescale is not a reliable indicator of the strength of a starburst, given the lack of correlation with ${\mathrm{b_{recent}}}$.  On the other hand, all of the consumption times are less than a Hubble time, indicating that the recent SF cannot be sustained indefinitely.}
\label{fig:bburst_gas}
\end{figure}

Figure~\ref{fig:bburst_gas} presents the peak ${\mathrm{b_{recent}}}$ values from each burst against the gas consumption timescale for fifteen  of the systems. The observational field of view for one galaxy (NGC~6822) did not cover the majority of the optical disk so the gas consumption timescale for this galaxy yields an upper limit of $\sim26$ Gyr; it not plotted as it is off the scale of this figure. There is little correlation between the gas consumption timescales and peak SFRs in the sample illustrating the limitations in using $\tau_{gas}$ for more than a simple diagnostic for classifying a galaxy as a starburst. The extended gas disks of galaxies may not be available for SF within the optical disk rendering a gas consumption timescale a limited criterion with which to evaluate the properties of starbursts.

\subsection{H${\alpha}$ Emission from Starbursts}
The SF activity in a burst creates a short$-$lived population of high mass stars that emits copious amounts of radiation. This strong emission can be observed and used as a starburst indicator. Specifically, the high mass stars formed in a starburst create high levels of UV emission (M$>$ 3~\msun) and H$\alpha$ emission (M$>$ 20~\msun) over the lifetime of these stars. Our time-resolved SFHs afford an opportunity to compare to the SFHs derived from their resolved stellar populations with the different timescales of these starburst indicators. We will explore the emission and timescales from the UV emission in a later paper; several of these galaxies are targets for a GALEX legacy program (P.I. Skillman, Proposal number 60026). Here, we compare the H$\alpha$ emission from \citet{Lee2009}  with our SFHs. Since the H$\alpha$ emission is expected to be emitted over a timescale $\ltsimeq$ 5 Myr, we compare the H$\alpha$ fluxes to the SFRs in our most recent time bin of $4-10$ Myr. We do not include younger time bins because the evolution of the most massive stars with lifetimes $\ltsimeq 4$ Myr are not calculated in the stellar evolutionary models employed in the SFH program.

\begin{figure}
\plotone{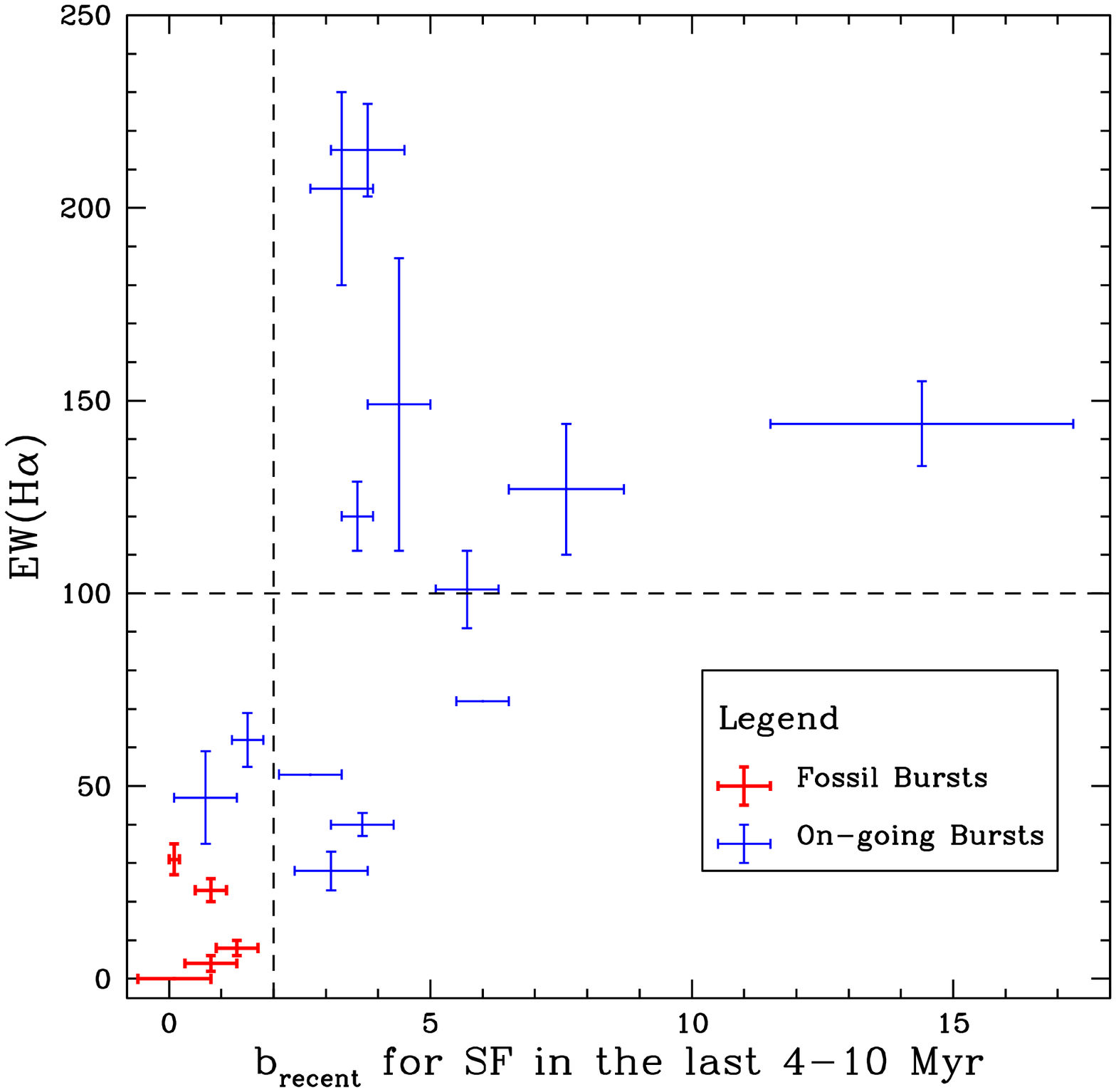}
\caption{The EW(H$\alpha$) (\AA) \citep{Lee2009} is plotted against the ${\mathrm{b_{recent}}}$ values from our time bin of $4-10$ Myr; the most closely matched with the H$\alpha$ emission timescale. The horizontal dotted line represents the 100 \AA~threshold set by \citet{Lee2009} delineating a starburst galaxy and the vertical dotted line represents our limit of ${\mathrm{b_{recent}}}=2$ for the same purpose based on resolved stellar population analysis. Seven galaxies classified as H$\alpha$ starbursts also show bursting levels of SF in our most recent time bin. Seven galaxies that do not meet the H$\alpha$ starburst threshold, also show lower levels of SF in the past $4-10$ Myr. Four galaxies show discrepancies between the two metrics indicating that the SF is changing on short timescales of a few Myr. These galaxies clearly show bursting SF from a stellar populations perspective but do not show starburst levels of H$\alpha$ emission.}
\label{fig:b4_halpha}
\end{figure}

\citet{Lee2009} identify a galaxy as a starburst if it has an H$\alpha$ equivalent width (EW(H$\alpha$)) of greater than 100 \AA. They suggest that this corresponds to a birthrate value of 2 or 3 \citep{Kennicutt1998}. This EW(H$\alpha$) starburst metric can be directly compared with our starburst indicator ${\mathrm{b_{recent}}}$, as shown in Figure~\ref{fig:b4_halpha}. The horizontal dotted line represents the delineation of a starburst in EW(H$\alpha$) and the vertical dotted line represent our delineation of a starburst (${\mathrm{b_{recent}}}\geq2$). Galaxies presently bursting are plotted as blue points while galaxies with fossil bursts are plotted in thicker, red points. The measured EW(H$\alpha$) and ${\mathrm{b_{recent}}}$ values are listed in Table~\ref{tab:truth_table}.

The comparison of these two starburst indicators in Figure~\ref{fig:b4_halpha} highlights both the usefulness and the limitations of the techniques. Seven galaxies have EW(H$\alpha$) $>$ 100 \AA~classifying these systems as H$\alpha$ starbursts. These systems also meet our starburst threshold of a ${\mathrm{b_{recent}}}$ $>2$ in our most closely matched time bin of $4-10$ Myr. Eleven galaxies have a measured EW(H$\alpha$) below the 100 \AA~threshold and were not classified as H$\alpha$ starbursts. These eleven galaxies can be broken into two groups when considering our SFH results. In the first group, seven of these systems had a measured ${\mathrm{b_{recent}}}<$ 2 in the last $4-10$ Myr suggesting that the starbursts in these galaxies are currently weak or have ended (red points). This is congruous with lower H$\alpha$ emission. The second group, however, contains four galaxies (NGC~4068, ESO~154$+$023, NGC~784, NGC~4449) that show ${\mathrm{b_{recent}}}>2$ in the last $4-10$ Myr. In other words, these four galaxies show recent starburst characteristics from analysis of their stellar populations but do not meet the H$\alpha$ emission starburst threshold. Note that our most recent time bin overlaps with, but is not coincident with, the timescale of H$\alpha$. The SFR in this final group of four galaxies must therefore be changing on timescales shorter than a few Myr to explain the discrepancy. Fluctuations of SFR on short timescales are also seen in the SFHs in Figure~\ref{fig:sfh_1gyr} at recent times where our temporal resolution is the finest.

A broader observation can be drawn from Figure~\ref{fig:b4_halpha}. Analysis of stellar populations gives a longer timescale to consider SF activity within a galaxy. From this perspective, the starbursts are events lasting $at~least$ a few 100 Myr. If the starburst durations we measure are typical of dwarf galaxies, then short term fluctuations of SF activity are as fundamental as the longer term characteristics of the SFR. The inherently short timescale of the H$\alpha$ emission is limited to identifying starbursts based only on the most current SF activity which, as we see in Figure~\ref{fig:b4_halpha}, may or may not be representative of the recent SFH of a galaxy. H$\alpha$ emission will miss identifying an important number of starburst systems. 

\section{Are Starbursts ``Self-Quenching''?\label{quench}}

Feedback from massive stars in the form of UV ionizing radiation and SNe impact the surrounding gas cloud from which stars are being formed. The idea that this feedback can regulate or even quench in situ SF via the heating and mechanical disruption of the gas has been put forth by previous authors \citep[e.g.,][]{Nishi2000}. Higher metallicity clouds are more effective at cooling so the quenching effect would theoretically be more important in low-metallicity systems \citep{Omukai1999}. 

The ``self-quenching'' process has been extrapolated to SF on galactic scales increasing with galaxy mass \citep{Kaviraj2007}. \citet{Ferguson1998} posit that feedback from a burst limits the burst duration to short timescales of $\sim 10$ Myr. Numerous other studies have reported that starbursts last $\ltsimeq 10$ Myr based on: $(1)$ fitting starburst models to UV spectra \citep{Tremonti2001}; $(2)$ studying stellar clusters within starburst galaxies \citep{Harris2004}; and $(3)$ emission from Wolf-Rayet stars \citep{Schaerer1999, Mas-Hesse1999}, to name a few. These short durations would seem to corroborate the idea of ``self-quenching'' derived by calculations balancing the kinetic energy output of SNe with the binding energy of the gas \citep{Thornley2000}. If starburst durations are comparable to or shorter than the crossing time, they are self-extinguishing explosions destroyed by their energy output in a non-equilibrium fashion \citep{Meurer2000}.

In Paper II, we measure the durations of starbursts in the sample galaxies from our time-resolved SFHs presented in Figures~\ref{fig:sfh_14gyr}$-$\ref{fig:sfh_1gyr}. The durations range from $450\pm50$ to $640\pm190$ Myr in fourteen galaxies and up to $1300\pm290$ Myr in four galaxies. These longer durations are comparable to or longer than the rotational period of the host galaxies and are in agreement with other studies measuring durations on the order of 100 Myr \citep{Calzetti1997, Meurer2000, Lee2009}. The SFH show that starbursts events last a few hundred Myr, much longer than the lifetime of O stars. Rapid ``self-quenching'' does not apply to starbursts in our sample. 

The shorter timescales cited in the literature above likely result from measuring only part of a burst event, corresponding to the lifetime of individual star clusters formed. This ``flickering'' is seen in the dips and variations in the SFHs on timescales of 10 Myr and suggests the small scale SF may be disrupted or quenched via stellar feedback. SF activity on these short timescales is associated with small spatial scales, and can be significantly affected by local conditions, making it a stochastic process. The non-equilibrium energy and mass transfer creating the small scale self-quenching may trigger SF in adjacent areas, creating self-propagating bursts of longer durations. We will investigate SF propagation within a burst in Paper III of this series using the spatial information of the stellar populations, regional SFHs, and stellar evolutionary time sequences.

\section{Conclusions \label{conclusions} }

We have used optical data obtained from the HST data archive to reconstruct the SFHs of eighteen nearby starburst dwarf galaxies. The SFHs show large variations from galaxy to galaxy, and, while the galaxies represent a class of objects undergoing significant recent star formation, there is no ``idealized'' SFR profile that characterizes the starburst phenomenon.  Not only are the changes in the SFR during a burst varied but the peak SFRs during the bursts span nearly three orders of magnitude in value indicative of the range in physical conditions present in each of the galaxies studied. Thirteen of the galaxies show ongoing starbursts and five present fossil bursts. The SFRs fall on a continuum between bursting and ``regular'' with the distinctions between the two dependent on the individualized history of each host system and the complicated feedback mechanisms created during the bursts. With this in mind, we examined three metrics for identifying starbursts with the following results and conclusions:

$\bullet$ The gas consumption timescale provides a coarse evaluation of whether a galaxy's SFR is sustainable over a cosmic timescale. For the sixteen galaxies with \HI\ mass measurements, fifteen systems can be classified as starbursts based on their gas consumption timescales being significantly shorter than the Hubble time. The gas consumption timescales are upper limits as the atomic gas mass used in the calculation measure the mass in the extended \HI\ halo; much of this gas may not be available for SF in the optical disk thus limiting the usefulness of such a metric. The remaining galaxy (NGC~6822) has an upper limit of the gas consumption timescales of 26 Gyr calculated from SFRs derived from a limited field of view of the optical disk.

$\bullet$ In fourteen galaxies, the application of the EW(H$\alpha$) criterion for identifying bursting SF over the past $\sim$5 Myr \citep{Lee2009} agrees with our measured ${\mathrm{b_{recent}}}$ values in our closest matching time bin of $4-10$ Myr. In four galaxies, the ${\mathrm{b_{recent}}}$ values from $4-10$ Myr ago indicate bursting SF while the EW(H$\alpha$) measurements are below the starburst threshold indicating that while the bursts are long-lasting events, the SFR can change on timescales of only a few Myr. If the longer durations of a few hundred Myr seen in this galaxy sample are typical of starbursts, this longer lived phenomenon is best studied at wavelengths correlating with comparable emission timescales; the short timescale of H$\alpha$ emission will miss identifying bursting galaxies experiencing short-lived fluctuations / dips in the SFR.

$\bullet$ We compared how the SFR changes over time within each galaxy normalized with an average SFR from the past 6 Gyr (${\mathrm{b_{recent}}}$=SFR~/~$<$SFR$>{\mathrm{_{0-6~Gyr}}}$); avoiding a comparison of absolute values of SFR biased towards larger systems. The recent SFRs are higher than twice the average SFR over the last 6 Gyr (${\mathrm{b_{recent}}}$ $>$ 2) in seventeen galaxies confirming that these galaxies are starbursts by this stellar population metric. The exception is NGC~625 which presents a unique SFH. The high SFRs seen in most galaxies in our sample in the ancient time bins are not seen in NGC~625. The SFH shows quiescent levels of SF until $\sim3$ Gyr ago, after which time SF proceeded at elevated levels until the current epoch. This galaxy shows evidence of a starburst at other wavelengths although it does fit our burst metric because of the high SFR baseline over the last 3 Gyr.

$\bullet$ The starburst events seen in the SFHs last at least $450$ Myr on timescales comparable to or longer than the rotational period of the host galaxies ruling out ``self-quenching'' of these bursts from stellar feedback mechanisms.

\section{Acknowledgments}
Support for this work was provided by NASA through grants AR-10945 and AR-11281 from the Space Telescope Science Institute, which is operated by Aura, Inc., under NASA contract NAS5-26555. E.~D.~S. is grateful for partial support from the University of Minnesota. J.~J.~D. was partially supported as a Wyckoff fellow. K.~B.~W.~M. gratefully acknowledges Matthew, Cole, and Carling for their support. This research made use of NASA's Astrophysical Data System and the NASA/IPAC Extragalactic Database (NED) which is operated by the Jet Propulsion Laboratory, California Institute of Technology, under contract with the National Aeronautics and Space Administration. Finally, we would like to acknowledge the usage of the HyperLeda database (http://leda.univ-lyon1.fr).

We would like to thank the anonymous referee for a prompt and very helpful report.

{\it Facilities:} \facility{Hubble Space Telescope}

\clearpage

\begin{deluxetable}{llllllrrrccccr}
\tabletypesize{\tiny}
\rotate
\tablewidth{0pt}
\tablecaption{Observation Summary and Global Parameters \label{tab:galaxies}}
\tablecolumns{14}
\setlength{\tabcolsep}{0.02in}
\tablehead{
\colhead{}	 		&
\colhead{Starburst}		&
\colhead{Example} 		&
\colhead{} 			&
\colhead{}			&
\colhead{}			& 
\colhead{$\lambda_{606}$}	&
\colhead{$\lambda_{814}$}	&
\colhead{D}		 	&
\colhead{m$-$M}			&
\colhead{}			&
\colhead{d$_{25}$}		&
\colhead{A$_{R}$}	 	&
\colhead{M$_{B}$}		\\
\colhead{Galaxy}		&
\colhead{ID Method}		&
\colhead{Ref.}			&
\colhead{RA}			&
\colhead{Decl.}			&
\colhead{HST ID.}		&
\colhead{(sec)}			&
\colhead{(sec)}			&
\colhead{(Mpc)}			&
\colhead{(mag)}			&
\colhead{Ref}			&
\colhead{(arcmin)}		&
\colhead{(mag)}			&
\colhead{(mag)}			\\
\colhead{(1)}			&
\colhead{(2)}			&
\colhead{(3)}			&
\colhead{(4)}			&
\colhead{(5)}			&
\colhead{(6)}			&
\colhead{(7)}			&
\colhead{(8)}			&
\colhead{(9)}			&
\colhead{(10)}			&
\colhead{(11)}			&
\colhead{(12)}			&
\colhead{(13)}			&
\colhead{(14)}			
}

\startdata
ANTLIA	 	& CMD 	& present work	& 10:04:04.10s& $-$27:19:52s	& GO-10210	&  985	&  1170	& $1.25\pm0.1$	& 25.49	& 13	& $1.86\pm0.13$	& 0.212	& $-9.8$\tablenotemark{a}  \\ 
UGC~9128 	& CMD 	& present work	& 14:15:56.5s & $+$23:03:19s	& GO-10210	&  985	&  1170	&  2.24		& 26.75	& 13	& $1.02\pm0.11$	& 0.065	& $-12.34$\tablenotemark{b} \\ 
UGC~4483 	& $\mu$, morphology & \citet{Loose1986} & 08:37:03.0s & $+$69:46:31s & GO-8769 	& 9500 (555W) &  6900	& $3.2\pm0.2$ 	& 27.53 & 3	& $1.12\pm0.12$	& 0.091 & $-12.53\pm0.23$  	\\
NGC~4163 	& UBV colors & \citet{Gallagher1986} & 12:12:09.1s & $+$36:10:09s	& GO-9771 & 1200&   900	& 3.0 	& 27.36	& 10	& $1.62\pm0.11$	& 0.052 & $-13.66$ 	  	\\
UGC~6456 	& K band, $\mu$, color & \citet{GildePaz2003}	& 11:27:59.9s & $+$78:59:39s	& GTO-6276& 4200 (555W)	&  4200	& $4.3\pm0.1$	& 28.19 & 11	& $0.79\pm0.11$	& 0.096 & $-13.69\pm0.19$ 	\\ 
NGC~6789 	& K band, $\mu$, color & \citet{GildePaz2003}	& 19:16:41.1s & $+$63:58:24s 	& GO-8122 & 8200 (555W) &  8200	& $3.6\pm0.2$ 	& 27.78	& 4	& $0.98\pm0.11$	& 0.187 & $-14.3$\tablenotemark{a} \\
\\
NGC~1569 	& multi-wavelength obs. & \citet{deVaucouleurs1974} & 04:30:49.0s & $+$64:50:53s& GO-10885& 9587 	&  4892	& $3.36\pm0.2$	& 27.63	& 7	& $3.98\pm0.11$ & 1.871 & $-14.74$ 	   	\\
NGC~4068 	& UBV colors & \citet{Gallagher1986} & 12:04:00.8s & $+$52:35:18s	& GO-9771 & 1200&   900	& 4.3 	& 28.17 & 10	& $1.86\pm0.11$	& 0.058 & $-14.87$	   	\\
SBS1415+437 	& K band, $\mu$, color & \citet{GildePaz2003}	& 14:17:01.4s & $+$43:30:05s 	& GO-9361 & 35280 	& 35280 & $13.6\pm1.4$ 	& 30.7	& 1	& $0.35\pm0.12$	& 0.024 & $-15.07\pm0.46$ 	\\
IC 4662 	& multi-wavelength obs. & \citet{Heydari-Malayeri1990}  & 17:47:08.8s & $-$64:38:30s 	& GO-9771 & 1200&   900 & 2.4 		& 26.94	& 10	& $2.09\pm0.11$	& 0.188 & $-15.09$ 	  	\\
ESO154-023 	& CMD 	& present work	& 02:56:50.38s& $-$54:34:17s	& GO-10210& 1000	&  1300 & 5.76		& 28.80	& 13	& $4.79\pm0.11$	& 0.045	& $-16.14$\tablenotemark{b}\\ 
NGC~2366 	& M$_{B}$, emission lines & \citet{Thuan1981} & 07:28:54.6s & $+$69:12:57s	& GO-10605 & 4780 (555W) &  4780	& $3.2\pm0.4$ 	& 27.52	& 8	& $4.37\pm0.11$	& 0.097 & $-16.17\pm0.36$ 	\\
\\
NGC~625 	& H$_{\alpha}$, UBVI images & \citet{Marlowe1997} & 01:35:04.6s & $-$41:26:10s 	& GO-8708 & 5200 (555W) & 10400 & $3.9\pm0.4$	& 27.95	& 2	& $6.61\pm0.10$	& 0.044 & $-16.19\pm0.18$ 	\\
NGC~784  	& CMD 	& present work	& 02:01:17.0s & $+$28:50:15s	& GO-10210&  930	&  1230	&  5.19		& 28.58	& 13	& $4.17\pm0.11$ & 0.158	& $-16.5$\tablenotemark{a} \\
NGC~5253 	& H$_{\alpha}$, UBVI images & \citet{Marlowe1997} & 13:39:55.9s & $-$31:38:24s 	& GO-10765& 2400 (555W)	&  2360	& $3.5\pm0.4$  	& 27.88	& 12	& $5.01\pm0.10$	& 0.186 & $-16.74\pm0.18$  \\
NGC~6822	& CMDs of Stellar Clusters	& \citet{Hodge1980} & 19:44:56.6s & $-$14:47:21s& GO-6813 & 3900 (555W)	&  3900	& $0.5\pm0.04$	& 23.3	& 6	& $17.38\pm0.11$& 0.632	& $-16.84$	   	\\
NGC~4214 	& multi-wavelength obs. & \citet{Huchra1983} & 12:15:39.2s & $+$36:19:37s	& GO-6569 & 1300 (555W) &  1300	& $2.7\pm0.2$ 	& 27.13	& 5	& $6.76\pm0.11$	& 0.058 & $-16.93\pm0.22$  	\\
NGC~4449 	& optical and radio data & \citet{Hunter1982} & 12:28:11.9s & $+$44:05:40s	& GO-10585& 4920 (555W)	&  4120	& $4.2\pm0.5$ 	& 28.12 & 9	& $4.68\pm0.11$	& 0.051 & $-17.94\pm0.33$  	\\
\enddata

\tablecomments{Cols. (3) Note that there are multiple citations for many of these galaxies identifying them by different methods as starbursts. We show but one example citation per galaxy. (4) and (5): R.A. and Decl. in J2000 coordinates. Col. (12) Diameter of the projected major axis of a galaxy at the isophotal level 25 mag arcsec$^{-2}$ in the B-band \citet{Paturel2003} HyperLeda database (http://leda.univ-lyon1.fr). Col. (13) \citet{Schlegel1998}. Col. (14) Absolute magnitudes uncorrected for Galactic extinction.}

\tablenotetext{a}{\citet{deVaucouleurs1991}}
\tablenotetext{b}{~Local Volume Legacy (LVL) project, \citet{Lee2008}}

\tablerefs{
(1) \citet{Aloisi2005}; (2) \citet{Cannon2003}; (3) \citet{Dolphin2001}; (4) \citet{Drozdovsky2001}; (5) \citet{Drozdovsky2002}; (6) \citet{Gieren2006}; (7) \citet{Grocholski2008}; (8) \citet{Karachentsev2002}; (9) \citet{Karachentsev2003}; (10) \citet{Karachentsev2006}; (11) \citet{Mendez2002}; (12) \citet{Sakai2004}; (13) \citet{Tully2006};
}
\end{deluxetable}

\clearpage
\begin{deluxetable}{lccccc}
\tabletypesize{\scriptsize}
\tablewidth{0pt}
\tablecaption{Comparison of Distance, Extinction, and Fit Values \label{tab:sfh_fits}}
\tablecolumns{6}
\tablehead{
\colhead{}	 		&
\colhead{Best Fit}		&
\colhead{Lit. Value}		&
\colhead{Total A$_{V}$}		&
\colhead{Galactic A$_{R}$}	&
\colhead{}			\\
\colhead{Galaxy}		&
\colhead{(m $-$ M)}		&
\colhead{(m $-$ M)}		&
\colhead{(mag)}			&
\colhead{(mag)}			&
\colhead{$\chi^{2}$}
}
\startdata
ANTLIA				& 25.42$\pm.06$	& 25.49	& 0.20$\pm.04$	& 0.212	& 1.18 \\
UGC 9128			& 26.70$\pm.04$	& 26.75	& 0.20$\pm.04$	& 0.065 & 1.36 \\
UGC 4483			& 27.86$\pm.07$	& 27.53	& 0.10$\pm.04$	& 0.091	& 1.12 \\
NGC 4163	 		& 27.40$\pm.04$	& 27.36 & 0.05$\pm.04$	& 0.050 & 1.31 \\
UGC 6456			& 28.30$\pm.04$	& 28.19	& 0.05$\pm.04$	& 0.096	& 1.08 \\
NGC 6789			& 27.85$\pm.04$	& 27.78	& 0.15$\pm.04$	& 0.187	& 1.29 \\
NGC 1569: HSB\tablenotemark{a}	& 27.53$\pm.06$	& 27.63	& 1.8$\pm0.1$	& 1.87	& 1.60 \\
NGC 1569: LSB\tablenotemark{b}	& 27.72$\pm.06$	& 27.63	& 1.5$\pm.1$	& 1.87	& 2.14 \\
NGC 4068		 	& 28.25$\pm.04$	& 28.17 & 0.00$\pm.04$ 	& 0.060 & 1.59 \\
SBS 1415+437			& 30.70$\pm.04$	& 30.70	& $-.05\pm.04$	& 0.024	& 1.46 \\
IC 4662: HSB\tablenotemark{a} 	& 26.80$\pm.04$ & 26.94 & 0.45$\pm.04$ 	& 0.190 & 1.22 \\
IC 4662: LSB\tablenotemark{b} 	& 26.75$\pm.04$ & 26.94 & 0.50$\pm.04$ 	& 0.190 & 1.74 \\
ESO154-023			& 28.80$\pm.04$	& 28.80	& 0.10$\pm.04$	& 0.045	& 1.63 \\
NGC 2366: Field 1		& 27.48$\pm.06$	& 27.52	& 0.20$\pm.04$	& 0.097	& 1.33 \\
NGC 2366: Field 2		& 27.44$\pm.07$	& 27.52	& 0.15$\pm.04$	& 0.097	& 1.30 \\
NGC 625				& 28.05$\pm.04$	& 27.95	& $-.05\pm.04$	& 0.044	& 1.40 \\
NGC 784				& 28.70$\pm.04$	& 28.58	& 0.10$\pm.04$	& 0.158	& 1.84 \\
NGC 5253: HSB\tablenotemark{a}	& 27.75$\pm.04$	& 27.70	& 0.15$\pm.04$	& 0.186	& 1.26 \\
NGC 5253: LSB\tablenotemark{b}	& 27.65$\pm.04$	& 27.70	& 0.20$\pm.04$	& 0.186	& 2.35 \\
NGC 6822: Field 1		& 23.38$\pm.05$	& 23.30	& 1.1$\pm.1$	& 0.632	& 1.31 \\
NGC 6822: Field 2		& 23.39$\pm.04$	& 23.30	& 1.1$\pm.1$	& 0.632	& 1.42 \\
NGC 6822: Field 3		& 23.45$\pm.06$	& 23.30	& 0.70$\pm.04$	& 0.632	& 1.28 \\
NGC 4214			& 27.20$\pm.04$	& 27.13	& 0.15$\pm.04$	& 0.058	& 1.26 \\
NGC 4449: HSB\tablenotemark{a}	& 27.95$\pm.04$	& 28.12	& 0.25$\pm.04$	& 0.051	& 3.00 \\
NGC 4449: LSB\tablenotemark{b}	& 28.05$\pm.04$	& 28.12	& 0.15$\pm.04$	& 0.051	& 2.41 \\
\enddata

\tablecomments{Cols. (2) Distance Modulus best fit by the CMD fitting program. The uncertainties are lower bounds as they include only statistical uncertainties and (3) Distance Modulus reported by various authors in literature. See Table~\ref{tab:galaxies} for references. (4) Foreground and internal extinction best fit by the CMD fitting program. The uncertainties are lower bounds as they include only statistical uncertainties.  (5) Galactic foreground extinction at $\lambda_{R} = 650$ nm reported by \citet{Schlegel1998}. (6) Best $\chi^{2}$ value of CMD fitting program.}

\tablenotetext{a}{`HSB' refers to the high surface brightness region.}
\tablenotetext{b}{`LSB' refers to the low surface brightness region.}
\end{deluxetable}



\begin{deluxetable}{lrrcccrr}
\tabletypesize{\scriptsize}
\tablewidth{0pt}
\tablecaption{Classification of Starbursts \label{tab:truth_table}}
\tablecolumns{8}
\tablehead{
\colhead{}				&
\colhead{Peak ${\mathrm{b_{recent}}}$}	&
\colhead{Peak SFR of Burst}		&
\colhead{Atomic Mass}			&
\colhead{}				&
\colhead{$\tau_{gas}$		}	&
\colhead{${\mathrm{b_{recent}}}$}	&
\colhead{EW(H$\alpha$)}			\\
\colhead{Galaxy}			&
\colhead{of Burst}			&
\colhead{($10^{-3}$ \msun yr$^{-1}$)}	&
\colhead{($\times10^{7}$ \msun)}	&
\colhead{Ref.}				&
\colhead{(Gyr)}				&
\colhead{last $4-10$ Myr}		&
\colhead{\AA}				
}     

\startdata
Antlia Dwarf	&  $3.5  \pm0.9$ &       $0.52   \pm0.01$&       0.08    &       8       &       2.2     &   $0.06   \pm0.65$ &	   No Detect.	   \\
UGC 9128	&  $6.3  \pm1.4$ &       $5.1    \pm1.0$ &       2       &       7       &       1.7     &   $0.83   \pm0.49$ &	   $4	   \pm2$   \\
UGC 4483	&  $14	 \pm3$   &       $11     \pm2$ 	 &       4       &       9       &       4.9     &   $14     \pm3$    &	   $144    \pm11$  \\
NGC 4163	&  $2.9  \pm0.6$ &       $12     \pm3$   &       1.5     &       13      &       1.8     &   $1.3    \pm0.4$ &	   $8	   \pm2$   \\
UGC 6456	&  $7.6  \pm1.1$ &       $23     \pm3$   &       4       &       1       &       2.4     &   $7.6    \pm1.1$ &	   $127    \pm17$  \\
NGC 6789	&  $3.8  \pm1.3$ &       $15     \pm5$   &       ...     &       ...     &       ...     &   $0.78    \pm0.34$&	   $23     \pm3$   \\
\\
NGC 1569	&  $21 	 \pm1$   &       $240    \pm10$  &       7.5     &       4       &	 0.44    &   $3.8    \pm0.7$ &	   $215    \pm12$  \\
NGC 4068	&  $4.7  \pm0.3$ &       $46     \pm3$   &       10      &       13      &       3.0     &   $3.1    \pm0.7$ &	   $28     \pm5$   \\
SBS 1415+437	&  $12	 \pm2$   &       $150    \pm10$  &       10      &       12      &       0.93    &   $3.3    \pm0.6$ &	   $205    \pm25$  \\
IC 4662		&  $7.7  \pm1.6$ &       $76     \pm15$  &       13      &       11      &       2.3     &   $5.7    \pm0.6$ &	   $101    \pm10$  \\
ESO 154-023	&  $6.4  \pm0.5$ &       $120    \pm10$  &       ...     &       ...     &       ...     &   $3.7    \pm0.6$ &	   $40     \pm3$   \\
NGC 2366	&  $5.6	 \pm0.4$ &       $160    \pm10$  &       60	 &   	 4       &       5.2	 &   $1.5    \pm0.3$ &	   $149    \pm38$  \\
\\
NGC 625	  	&  $1.4  \pm0.1$ &	 $40.    \pm2$   &       11      &       2       &       1.8     &   $0.15   \pm0.07$ &	   $31     \pm4$   \\
NGC 784		&  $4.5  \pm0.6$ &       $120	 \pm20$  &       38      &       7       &       4.5     &   $2.7    \pm0.6$ &	   $53     $	   \\
NGC 5253	&  $9.0  \pm0.9$ &       $400    \pm40$  &       12      &       10      &       0.42    &   $3.6    \pm0.3$ &	   $120    \pm9$   \\
NGC 6822	&  $3.1  \pm1.1$ &       $7.3    \pm2.5$ &       13      &       3       &       26\tablenotemark{a}&$0.69 \pm0.55$ &	   $47     \pm12$  \\
NGC 4214	&  $3.1  \pm0.9$ &       $130    \pm40$  &       41      &       4       &       4.3     &   $1.5    \pm0.3$ &	   $62     \pm7$   \\
NGC 4449	&  $6.0  \pm0.5$ &       $970    \pm70$  &       110     &       4       &       1.5     &   $7.1    \pm0.6$ &	   $72     $	   \\
\enddata
\tablerefs{
(1) Begum et al. 2008 (FIGGS); (2) Cannon et al. 2004; (3) de Blok 2000;
(4) Walter et al. 2008 (THINGS SURVEY); (6) van Eymeren et al. 2009;
(7) van Zee 2001; (8) Whiting et al. 1997.; (9) Thuan \& Seitzer 1979; (10) L{\'o}pez-S{\'a}nchez et al. 2008;
(11) L{\'o}pez-S{\'a}nchez et al.  2009; (12) Huchtmeier et al. 2005; (13) Bottinelli et al. 1990; (14) Swaters \& Balcells (2002) }

\tablecomments{The gas consumption timescales ($\tau_{gas}$) were calculated using peak SFR and the \HI\ gas mass adjusted by a factor of 1.4 to account for helium and heavy elements; they are upper limits as the atomic gas extends up to several Holmberg radii farther than the optical disk for dwarf irregular galaxies \citep{Huchtmeier1981}. The uncertainties in $\tau_{gas}$ were calculated by adding in quadrature the SFR uncertainties with an assumed uncertainty of 10\% in the atomic gas mass measurements. The EW(H$\alpha$) measurements were taken from \citet{Lee2009}.}

\tablenotetext{a}{The gas consumption timescale for NGC~6822 is an extreme upper limit. Not only is the atomic gas unusually extended relative to the optical disk, but the field of view of the optical observations used to derive the SFRs does not cover the main optical disk.}

\end{deluxetable}

\clearpage

\clearpage
\begin{figure}
\figurenum{\ref{fig:images}}
\centering
\begin{tabular}{cc}
\epsfig{file=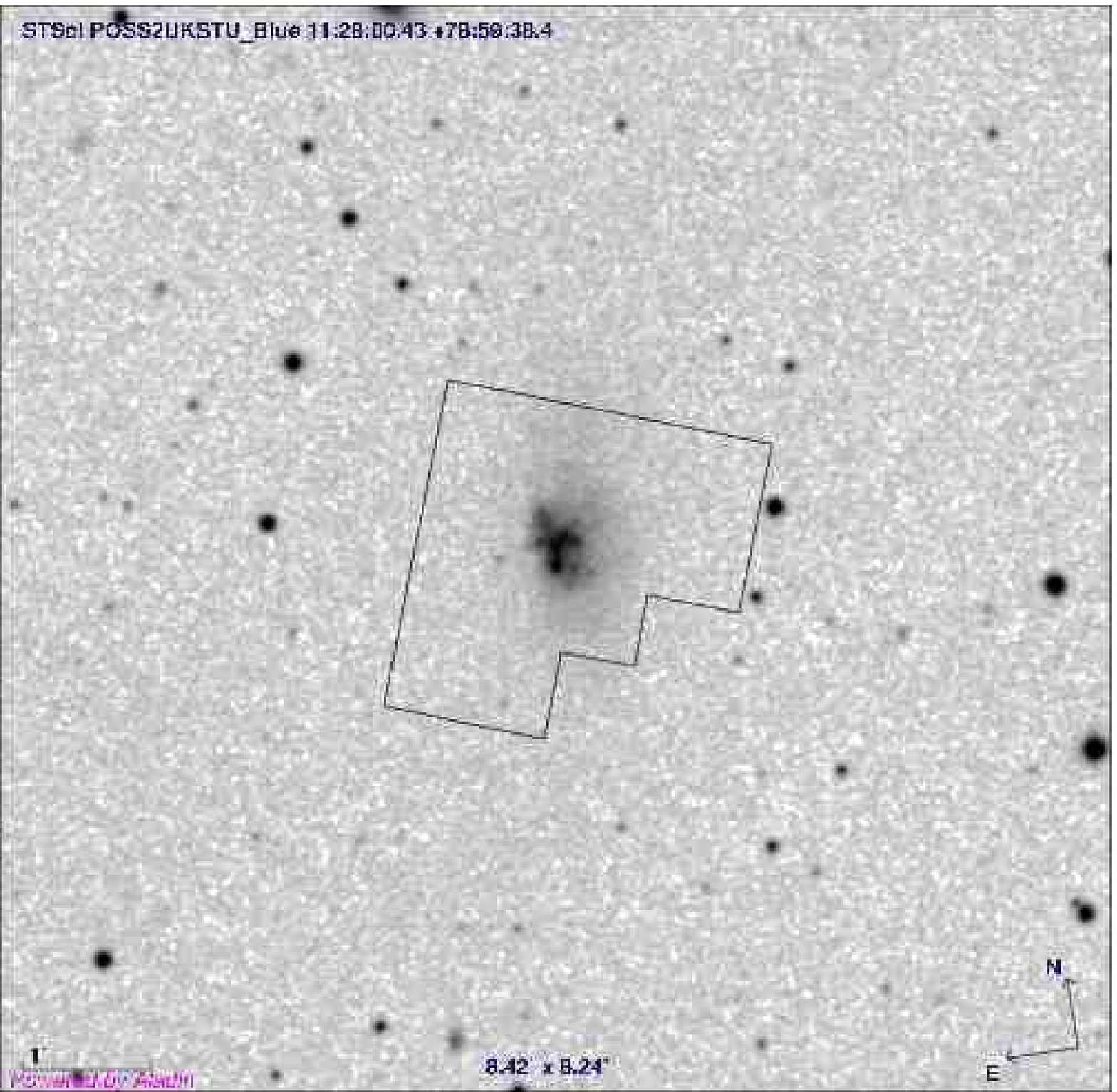,width=0.5\linewidth,clip=} & 
\epsfig{file=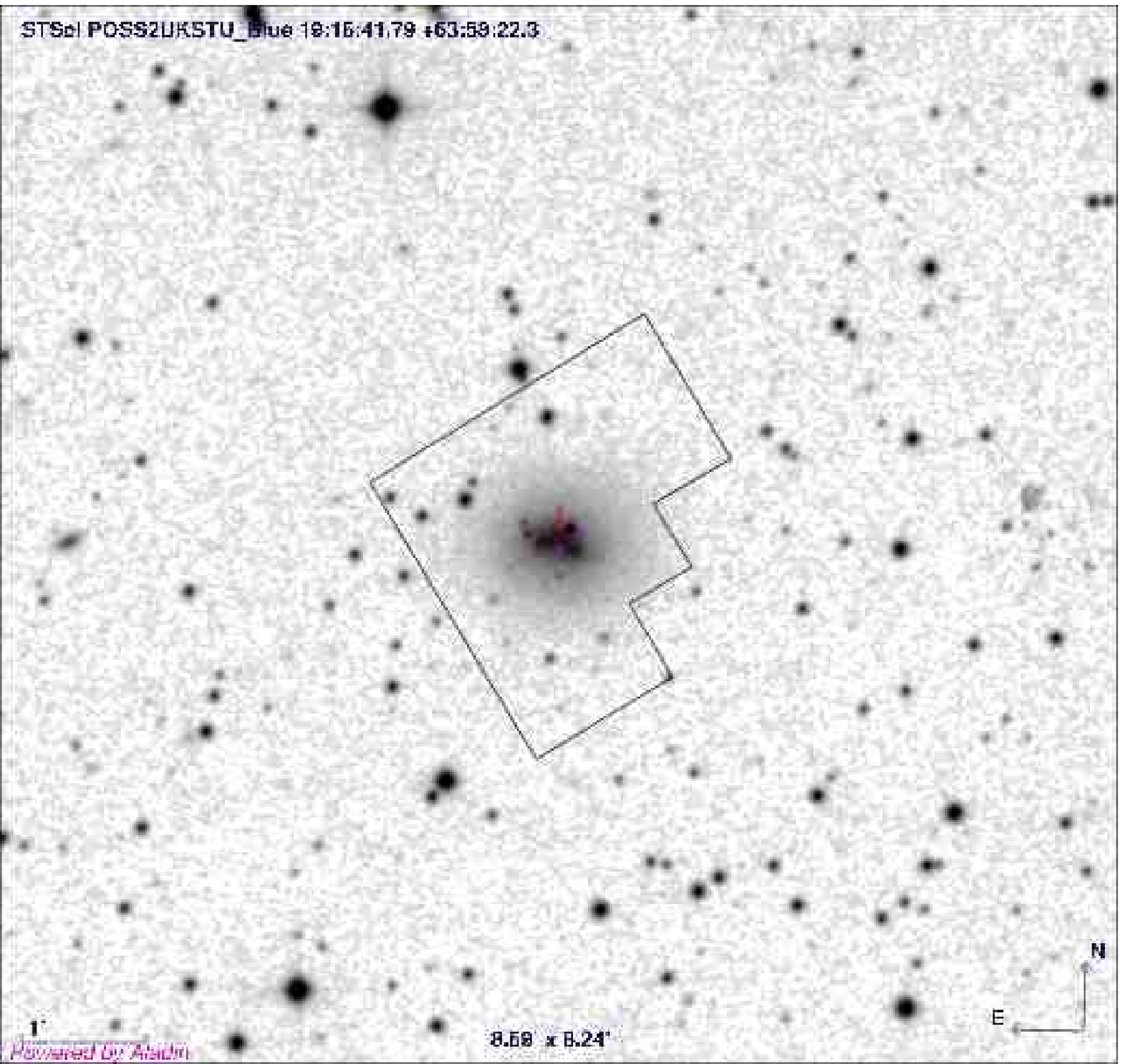,width=0.5\linewidth,clip=} \\
\epsfig{file=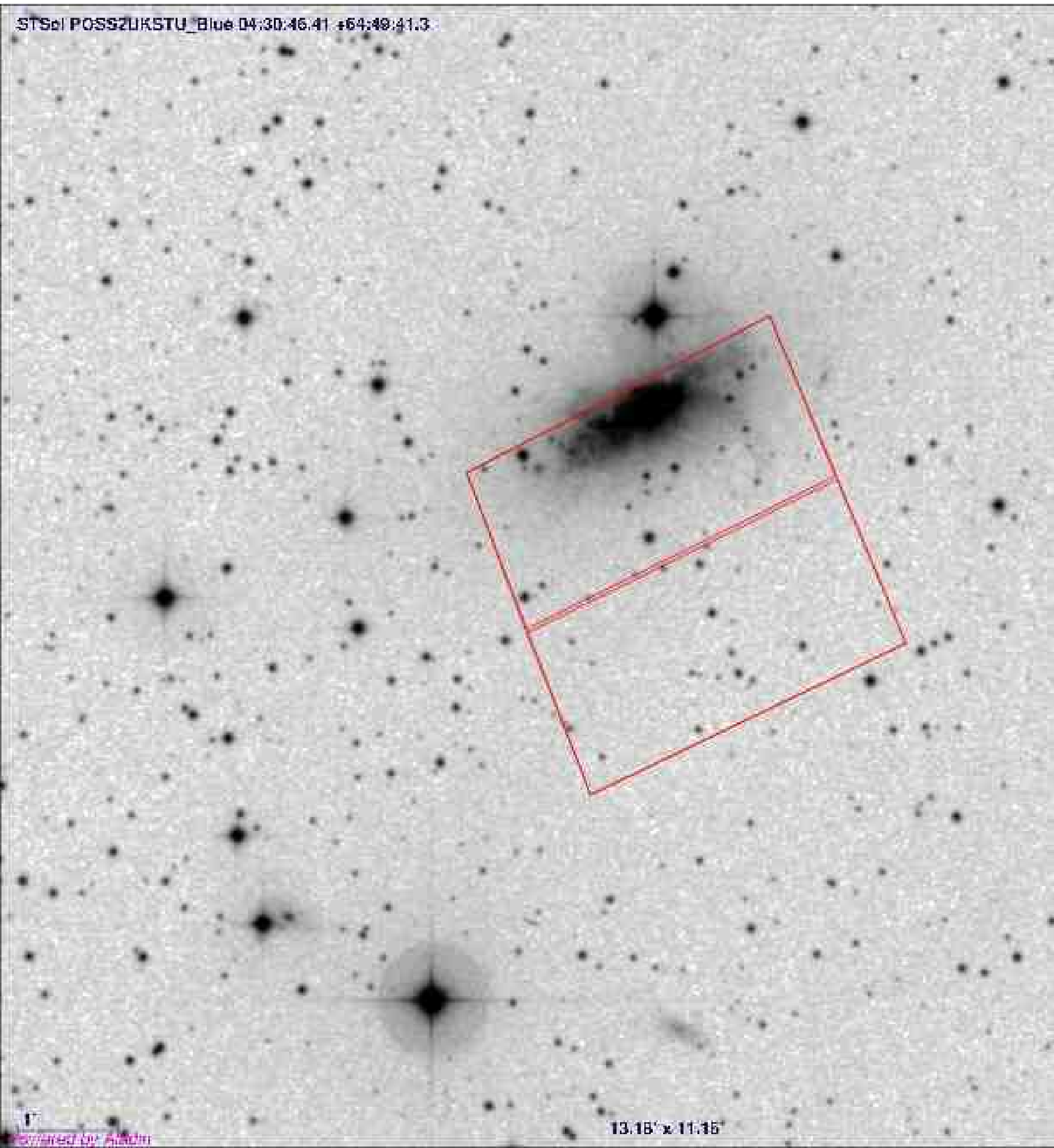,width=0.5\linewidth,clip=} &
\epsfig{file=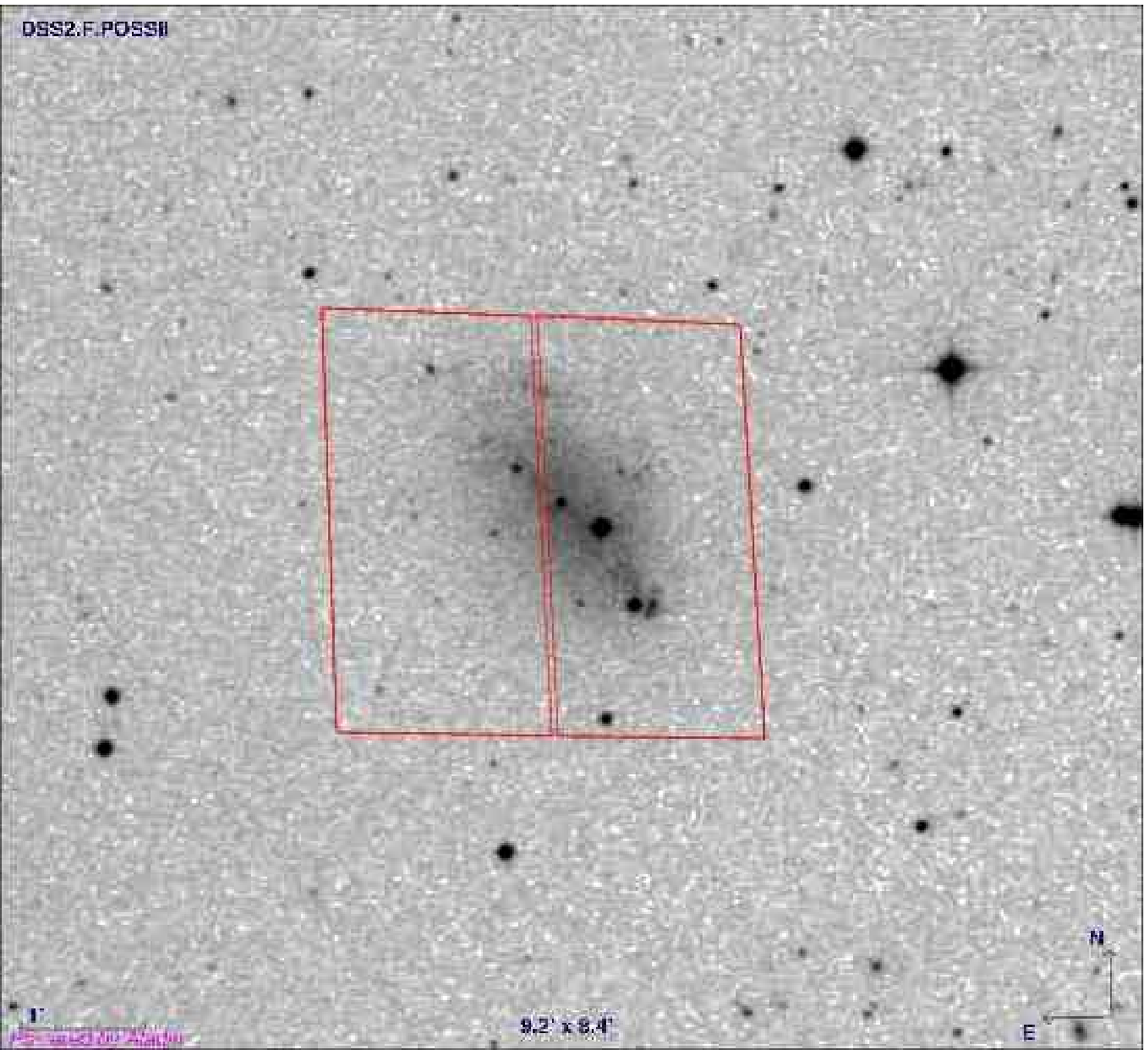,width=0.5\linewidth,clip=}
\end{tabular}
\caption{\textit{HST footprints continued: UGC~6456, NGC~6789, NGC~1569, NGC~4068}}
\end{figure}

\clearpage
\begin{figure}
\figurenum{\ref{fig:images}}
\centering
\begin{tabular}{cc}
\epsfig{file=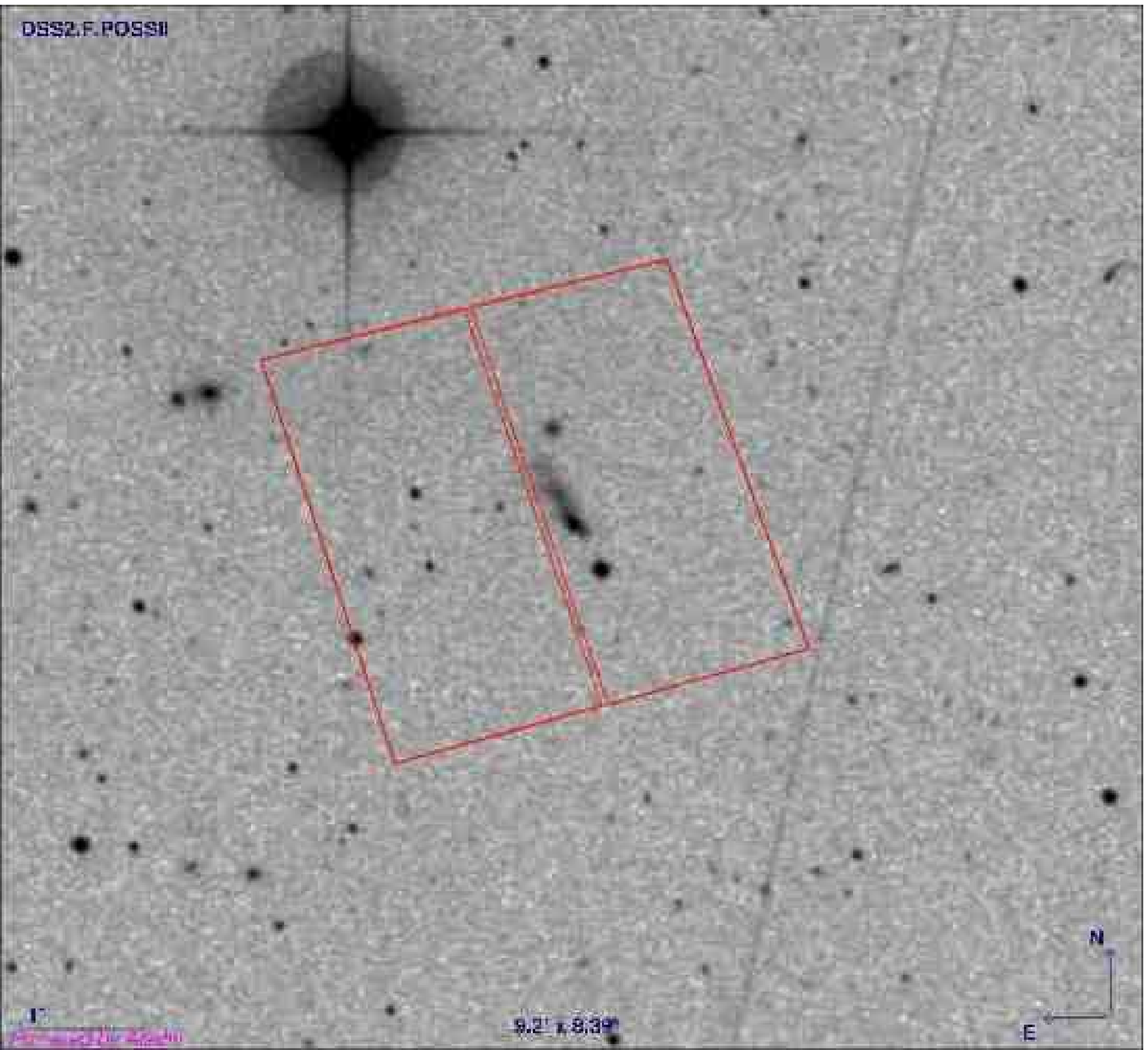,width=0.5\linewidth,clip=} & 
\epsfig{file=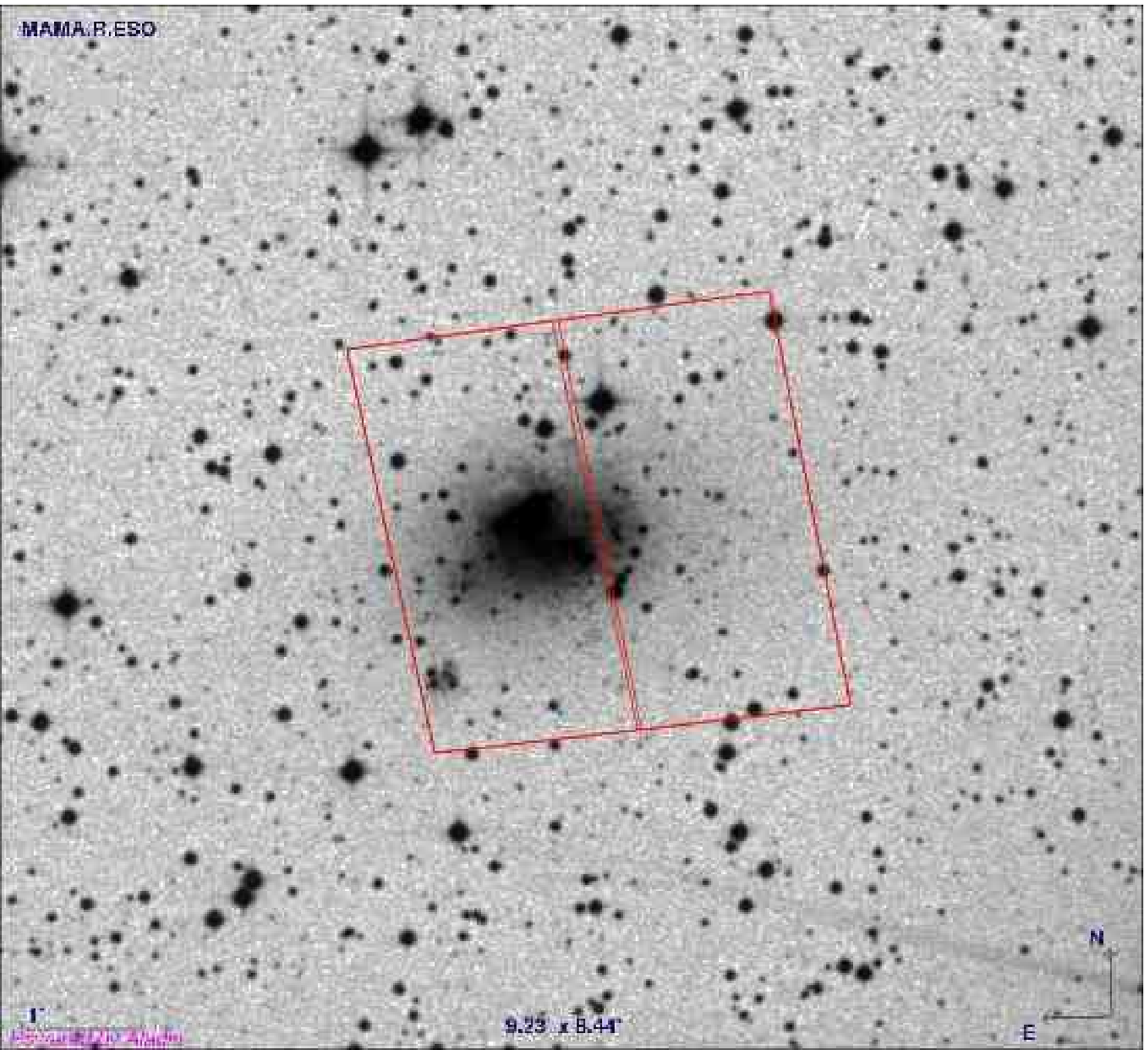,width=0.5\linewidth,clip=} \\
\epsfig{file=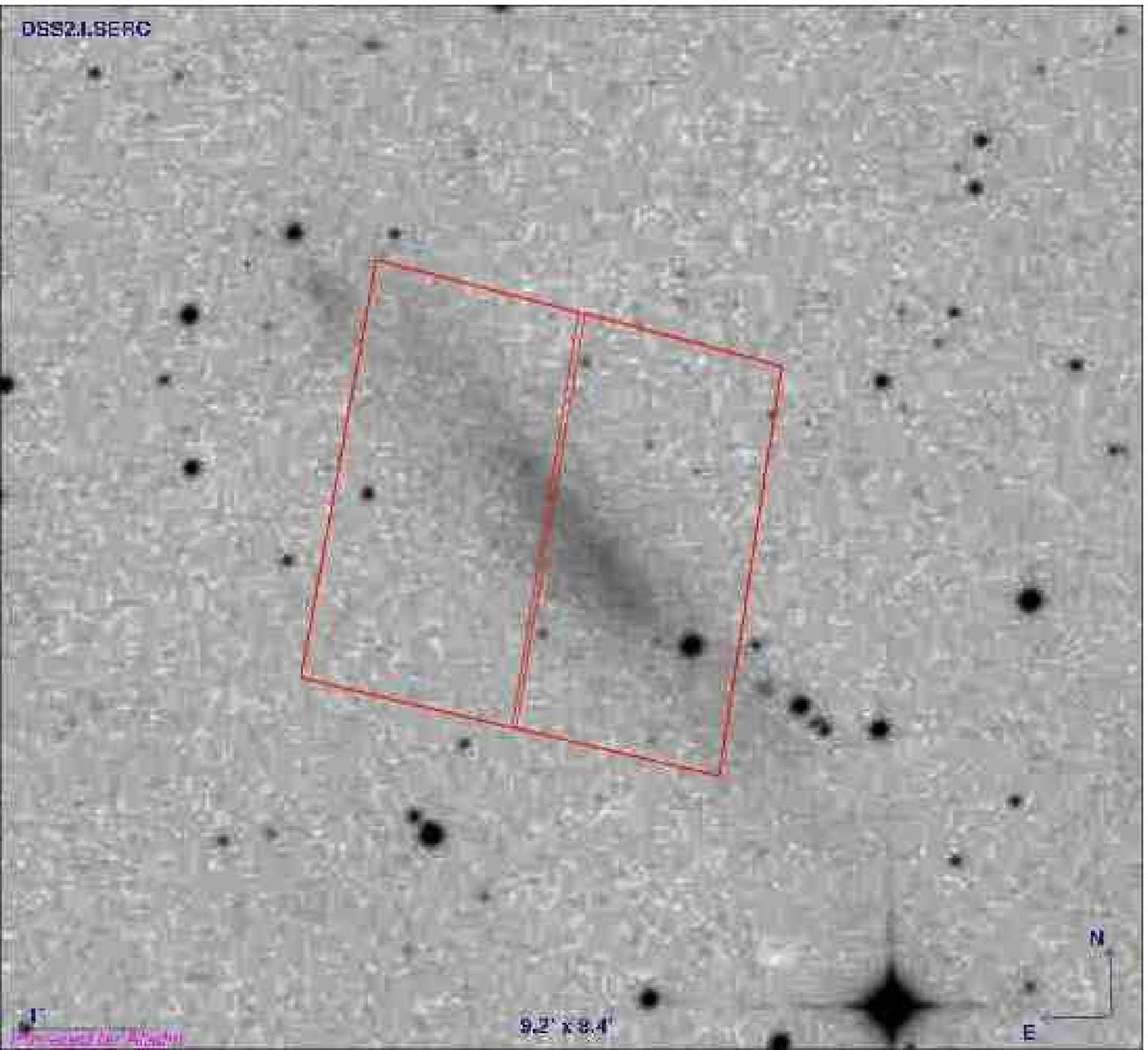,width=0.5\linewidth,clip=} &
\epsfig{file=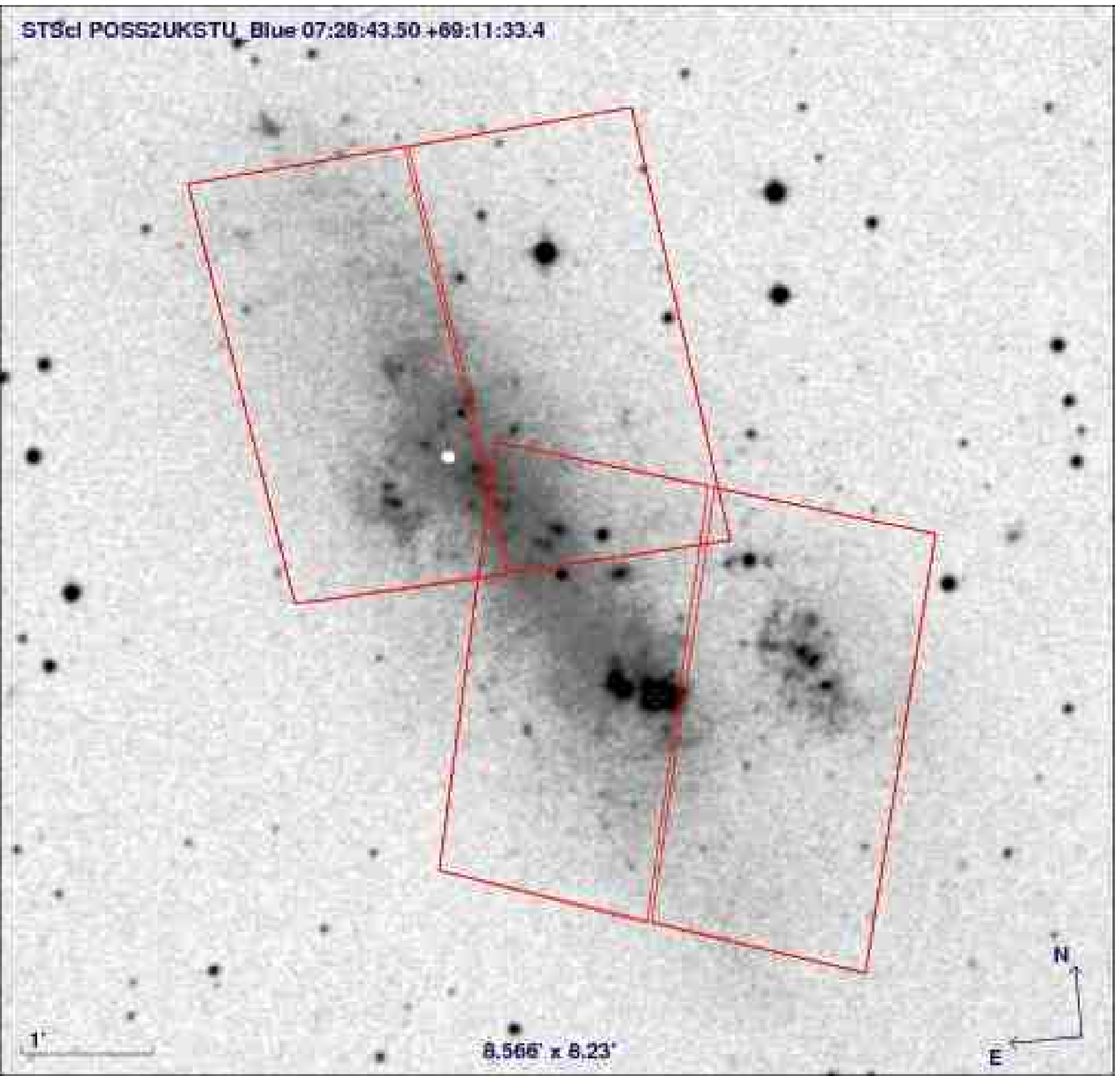,width=0.5\linewidth,clip=}
\end{tabular}
\caption{\textit{HST footprints continued: SBS1415$+$437, IC~4662, ESO154$-$023, NGC~2366}}
\end{figure}

\clearpage
\begin{figure}
\figurenum{\ref{fig:images}}
\centering
\begin{tabular}{cc}
\epsfig{file=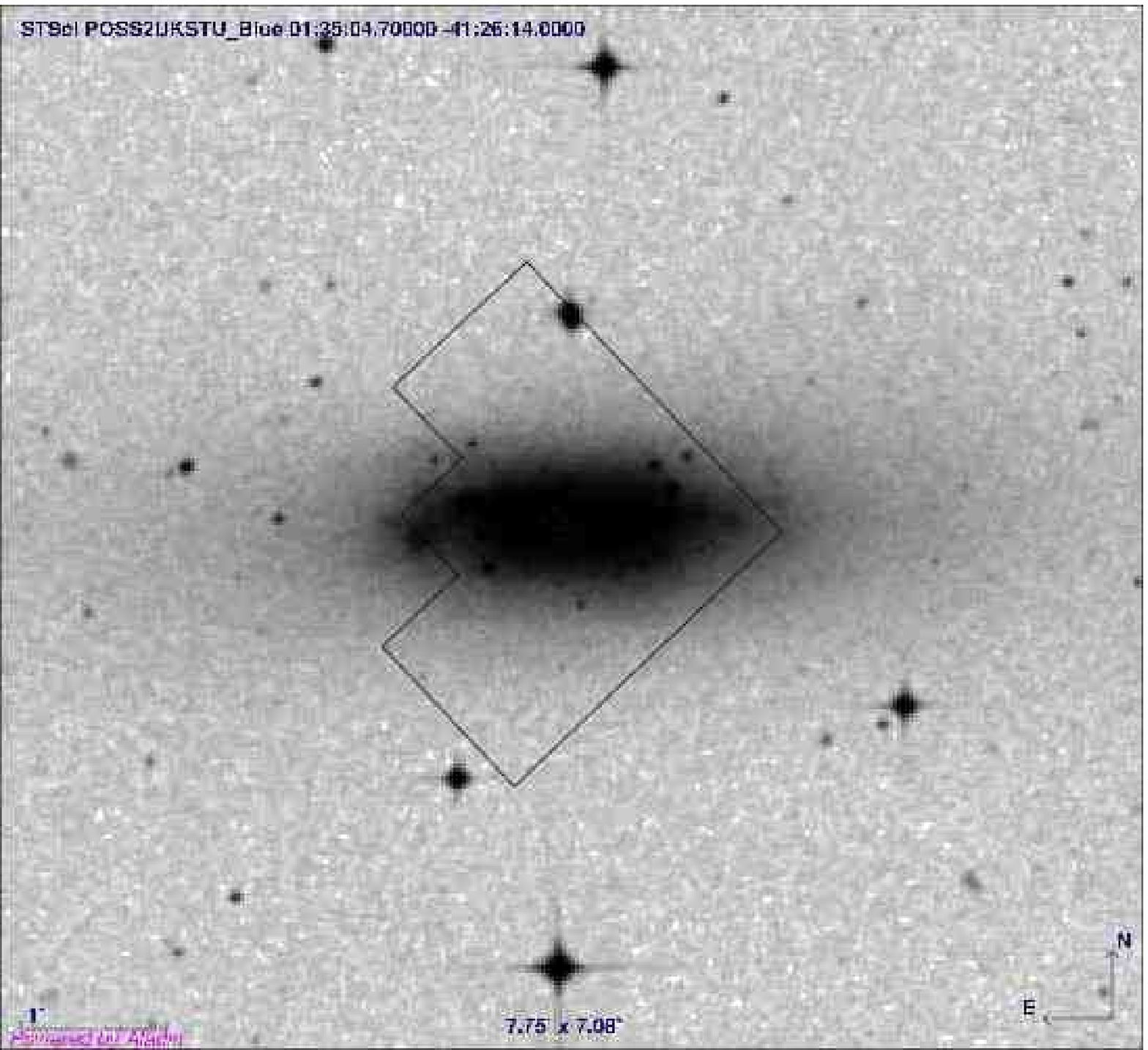,width=0.5\linewidth,clip=} & 
\epsfig{file=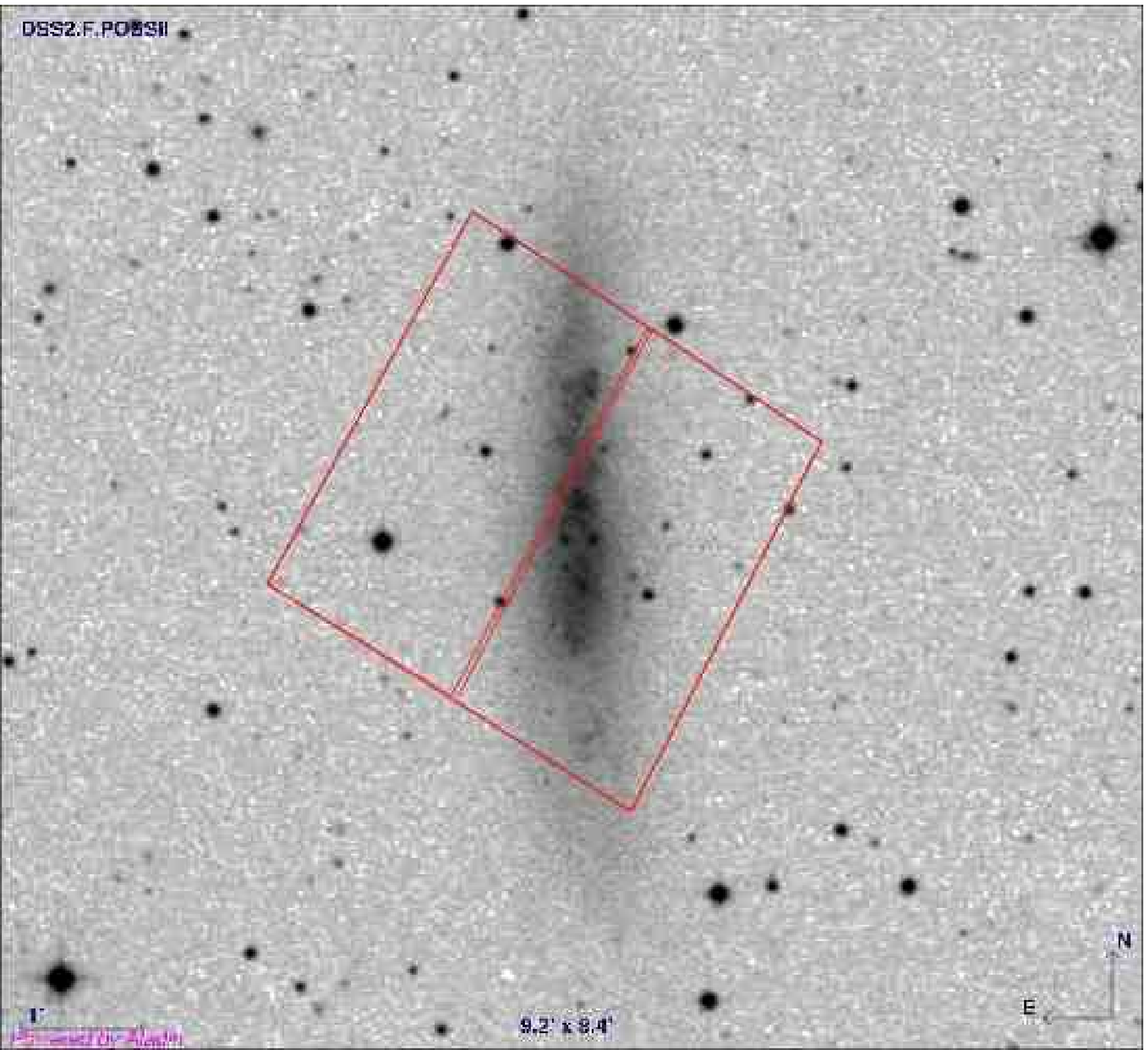,width=0.5\linewidth,clip=} \\
\epsfig{file=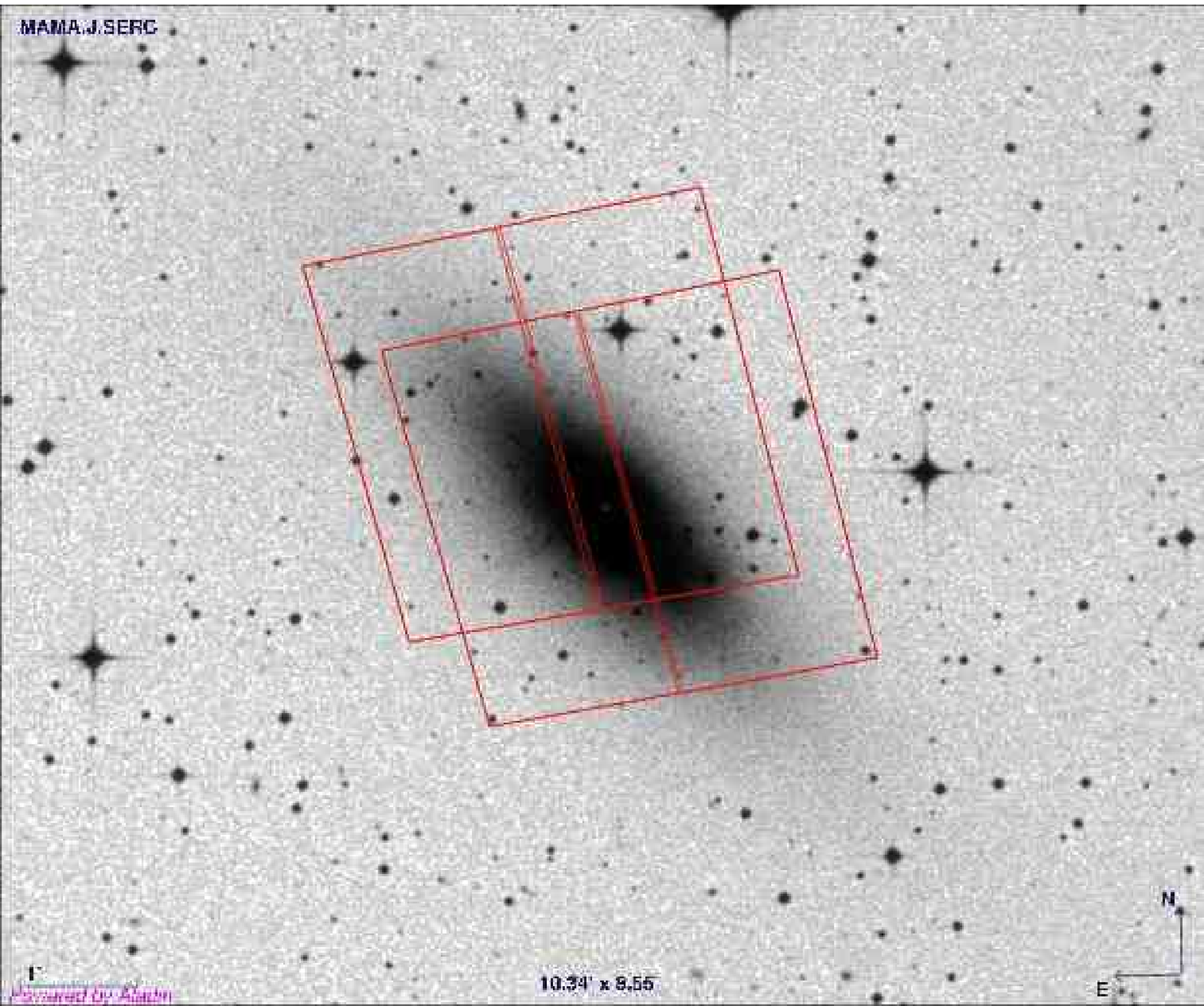,width=0.5\linewidth,clip=} &
\epsfig{file=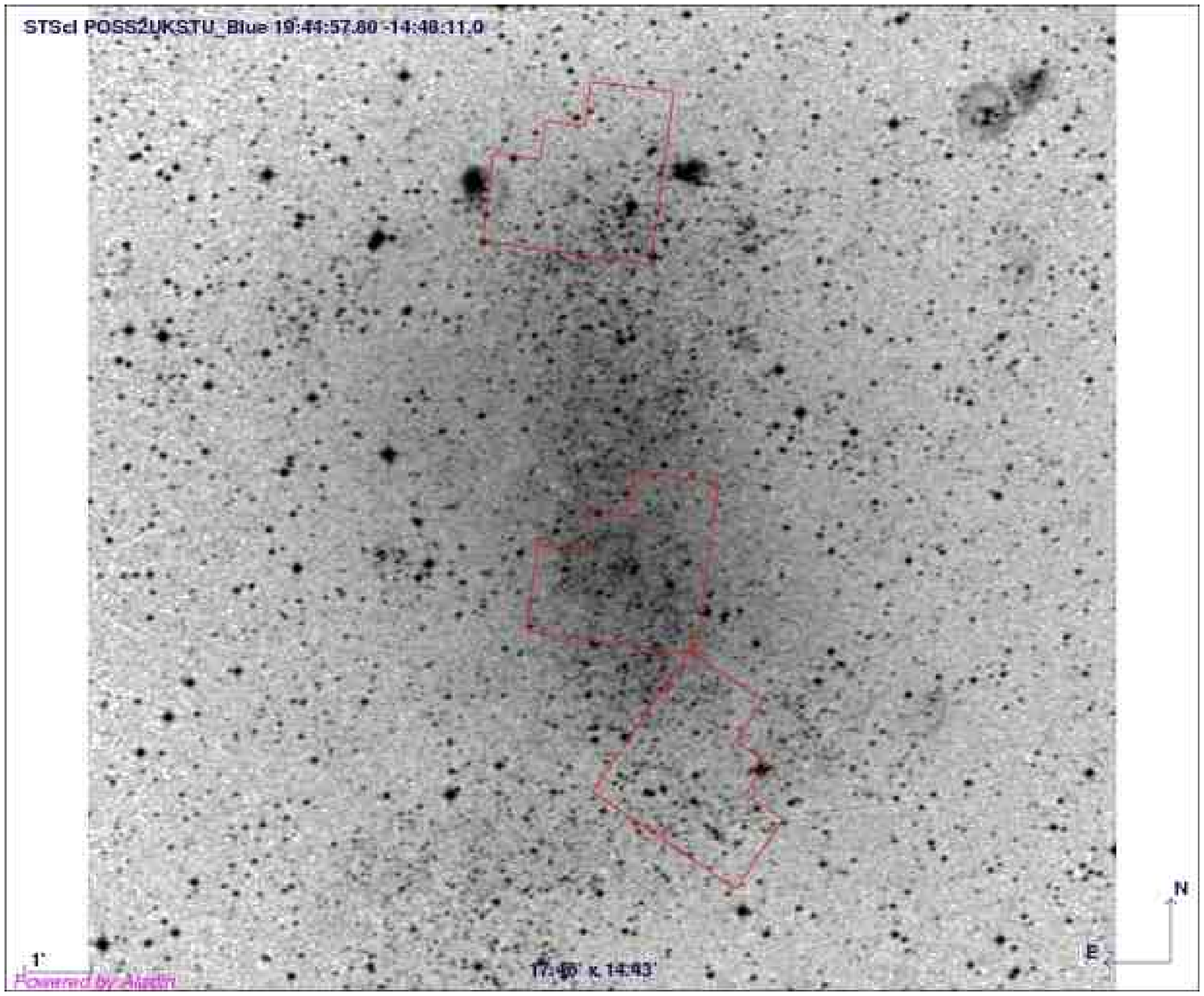,width=0.5\linewidth,clip=}
\end{tabular}
\caption{\textit{HST footprints continued: NGC~625, NGC~784, NGC~5253, NGC~6822}}
\end{figure}

\clearpage
\begin{figure}
\figurenum{\ref{fig:images}}
\centering
\begin{tabular}{cc}
\epsfig{file=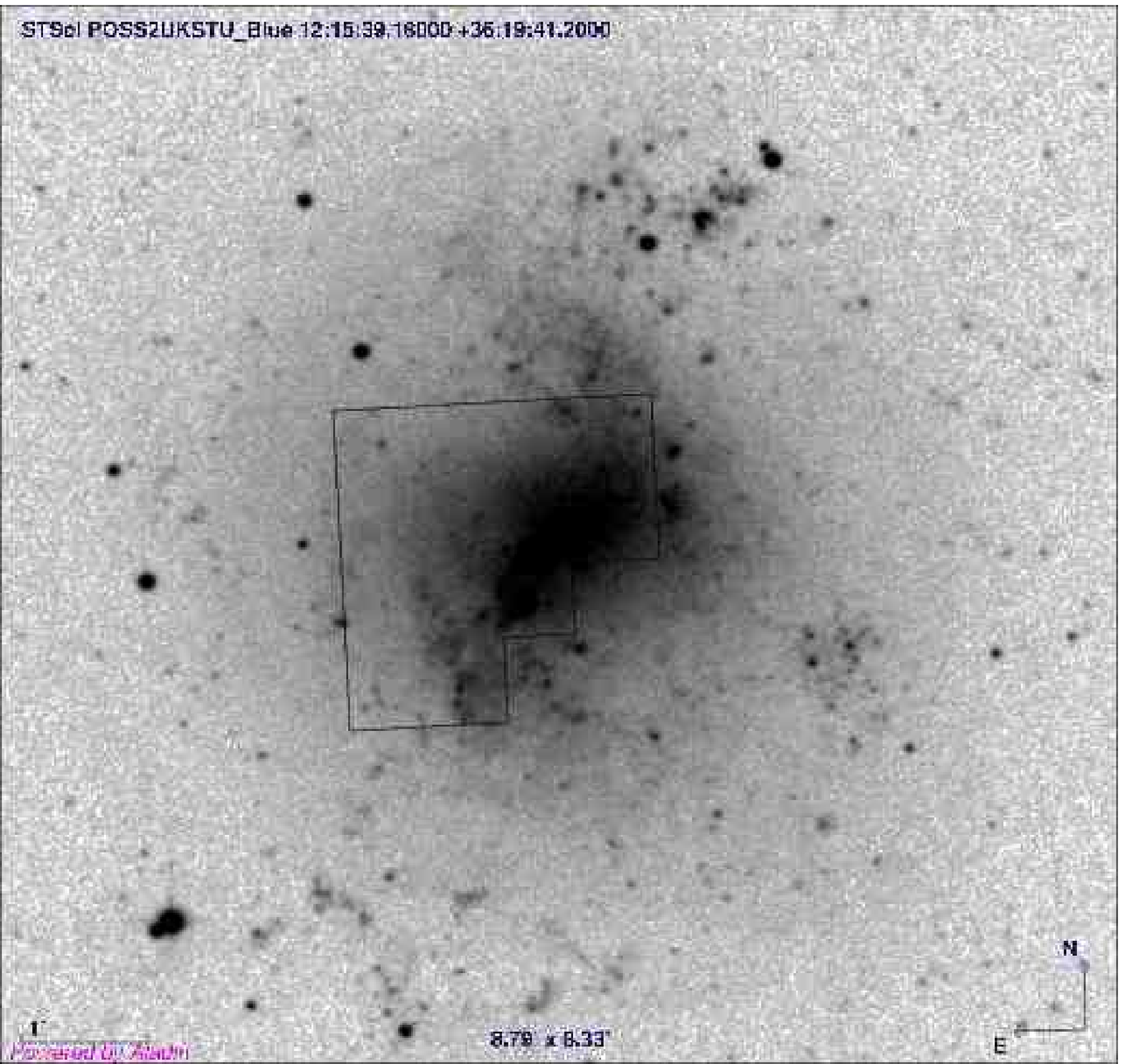,width=0.5\linewidth,clip=} & 
\epsfig{file=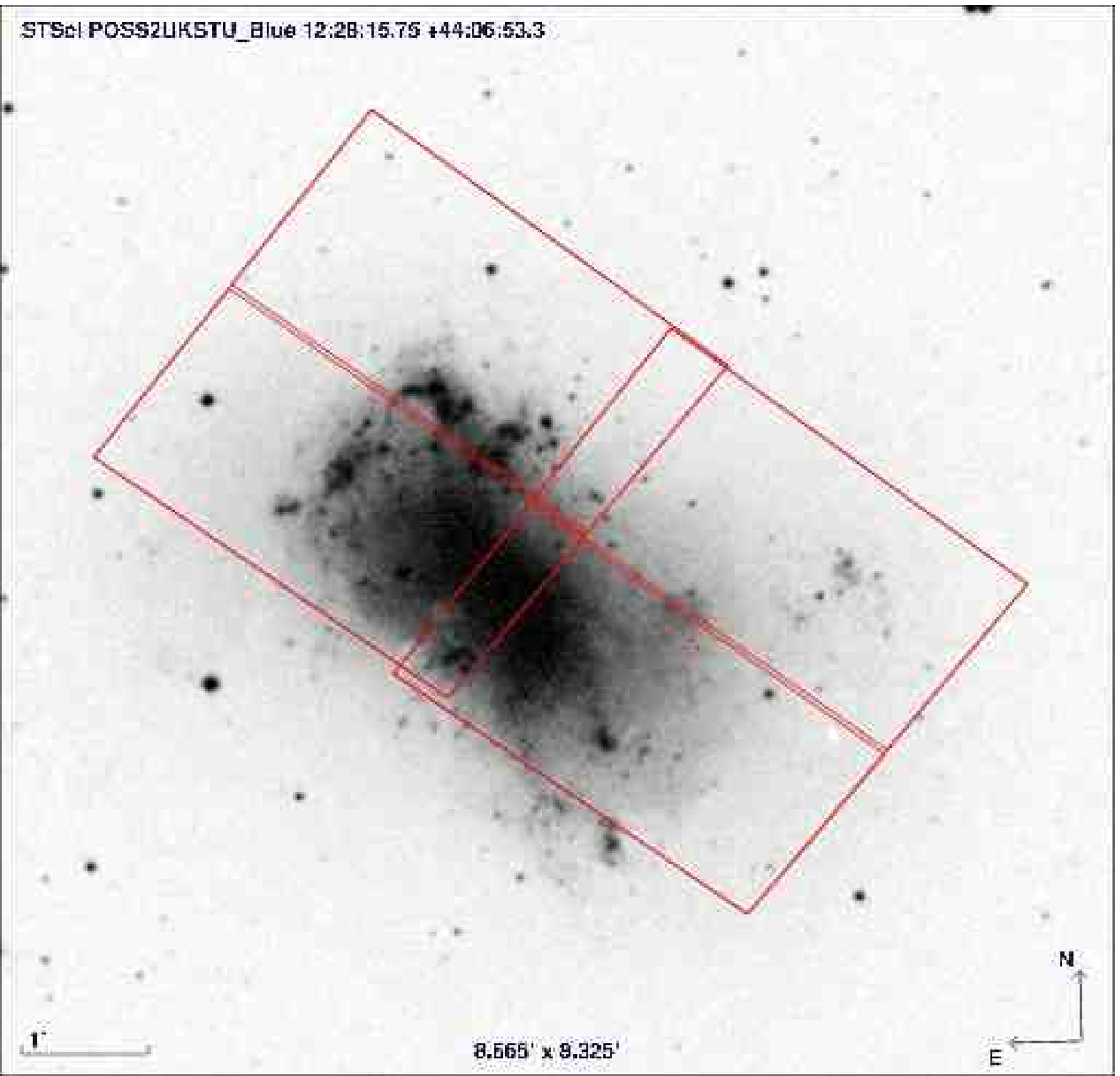,width=0.5\linewidth,clip=} 
\end{tabular}
\caption{\textit{HST footprints continued: NGC~4214, NGC~4449}}
\end{figure}

\clearpage

\clearpage

\clearpage
\begin{figure}
\figurenum{\ref{fig:cmds}}
\centering
\begin{tabular}{cc}
\epsfig{file=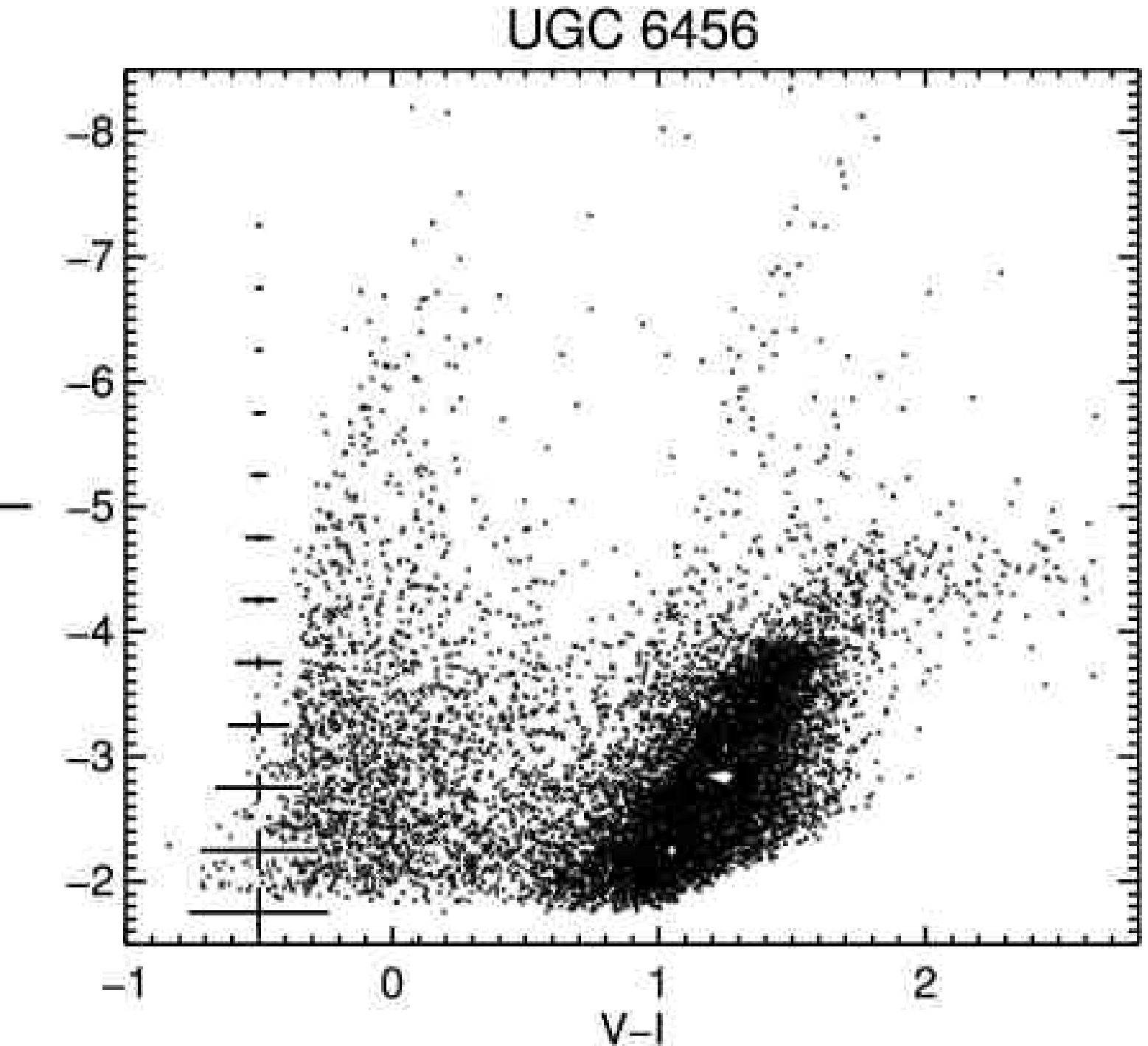,width=0.5\linewidth,clip=} & 
\epsfig{file=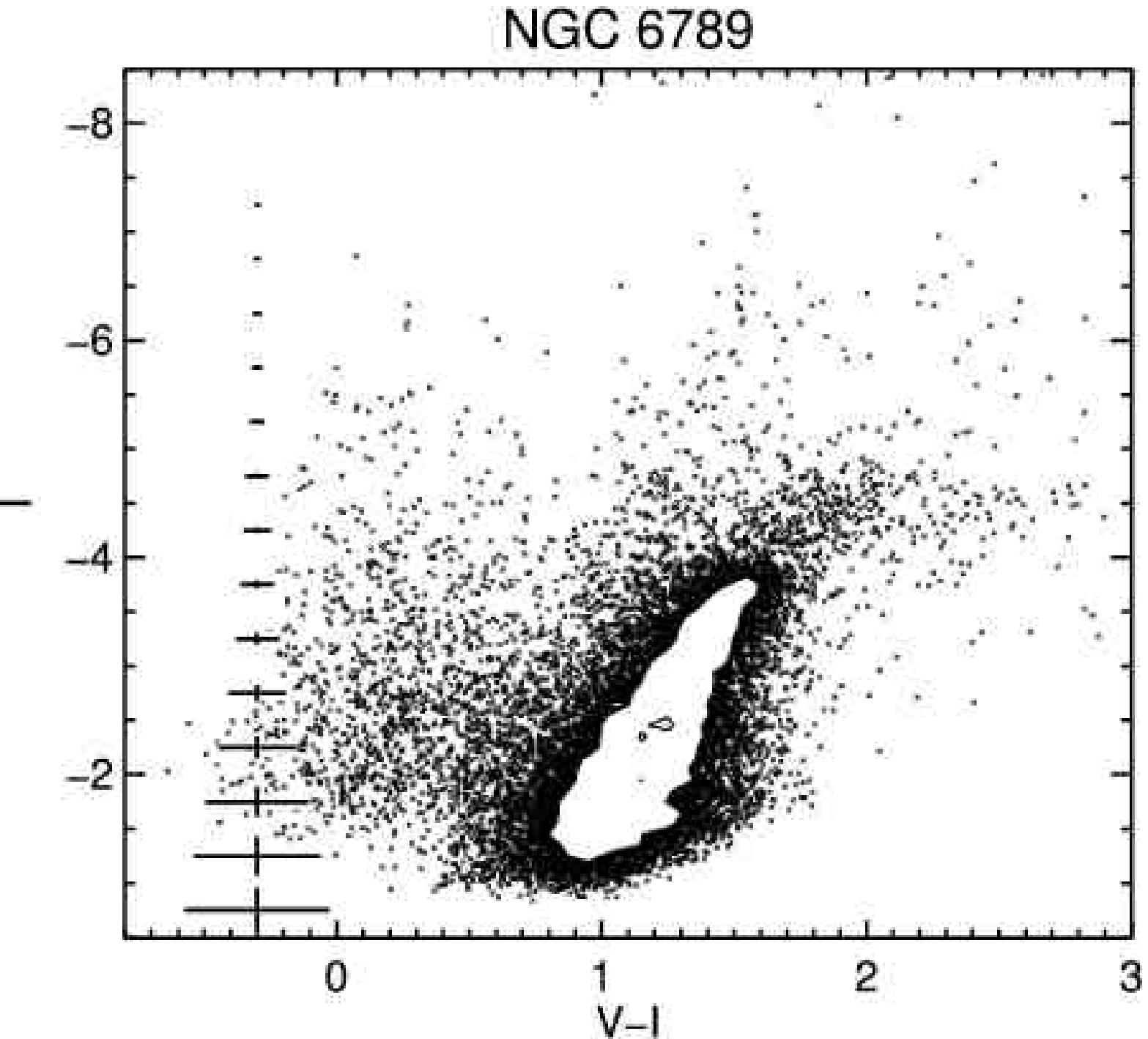,width=0.5\linewidth,clip=} \\
\epsfig{file=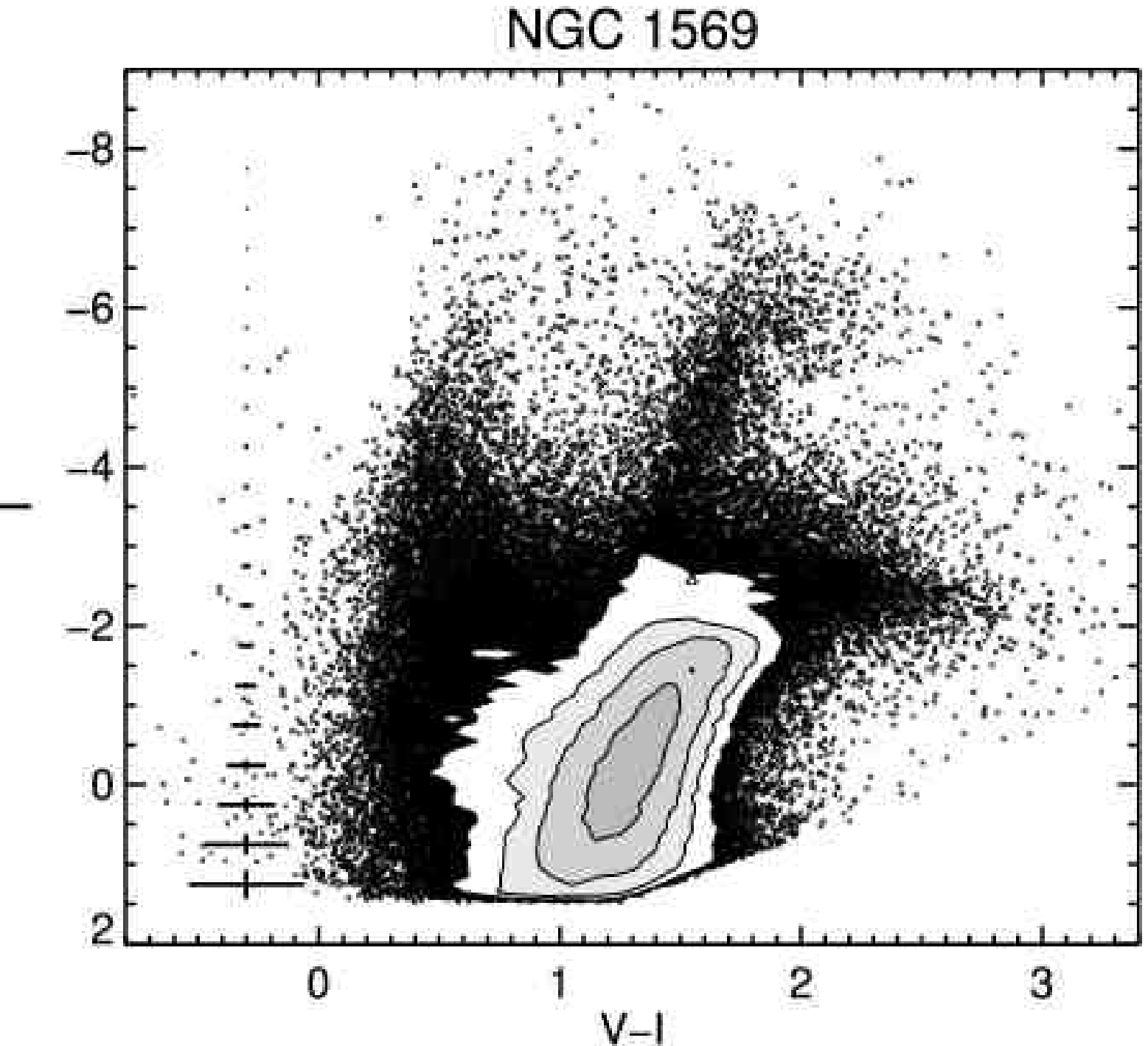,width=0.5\linewidth,clip=} &
\epsfig{file=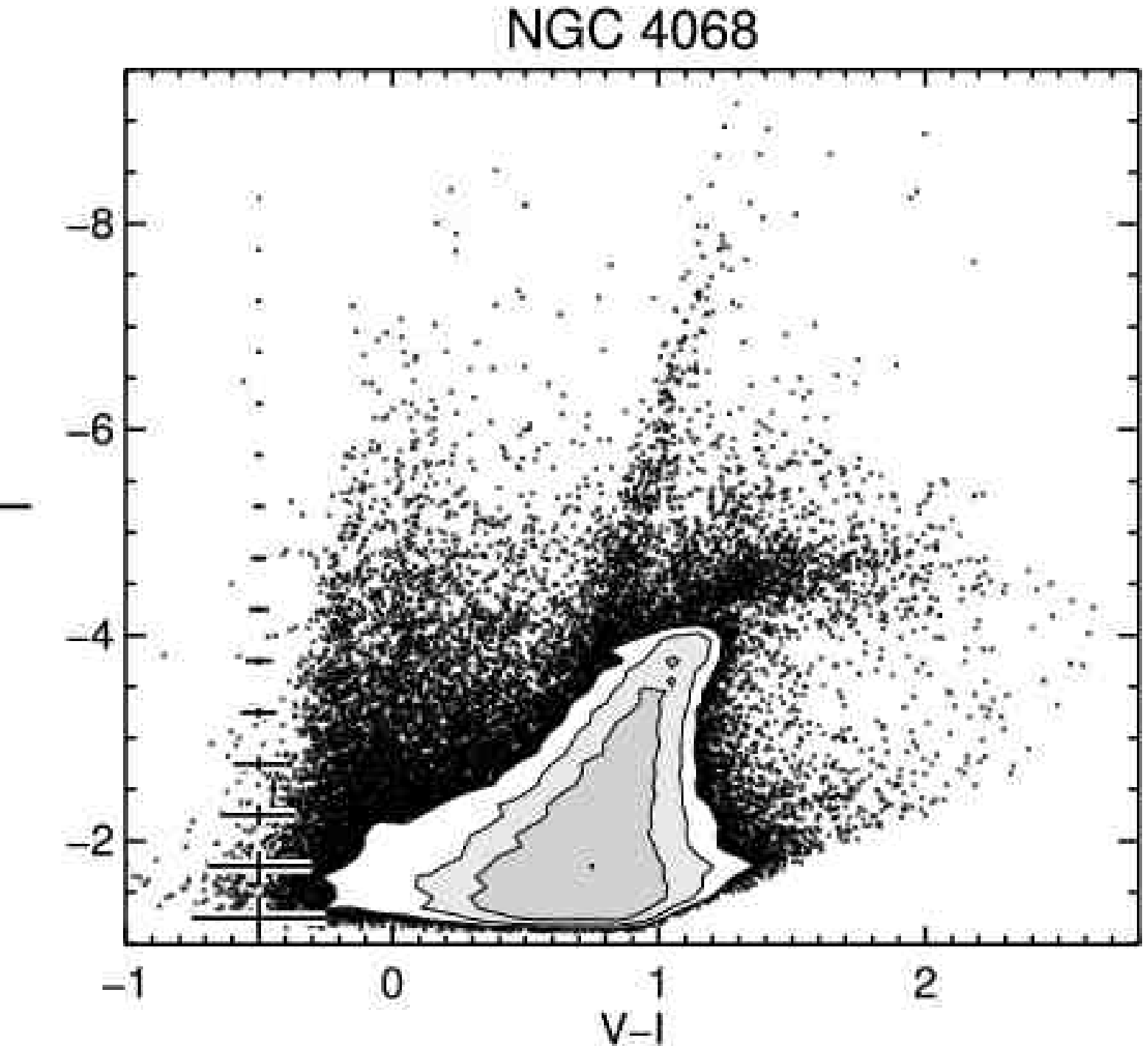,width=0.5\linewidth,clip=}
\end{tabular}
\caption{\textit{CMDs continued: UGC~6456, NGC~6789, NGC~1569, NGC~4068}}
\end{figure}

\clearpage
\begin{figure}
\figurenum{\ref{fig:cmds}}
\centering
\begin{tabular}{cc}
\epsfig{file=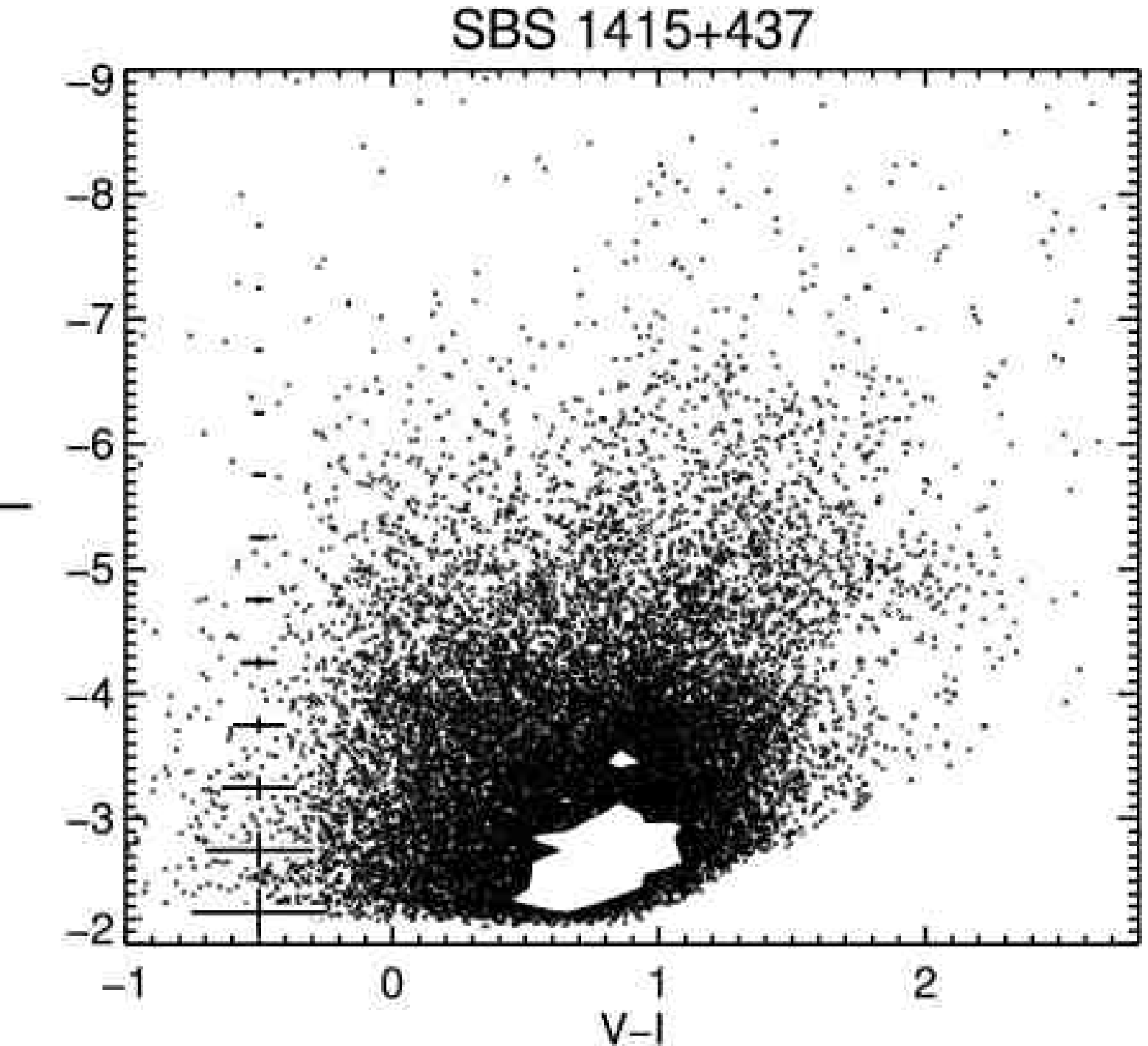,width=0.5\linewidth,clip=} & 
\epsfig{file=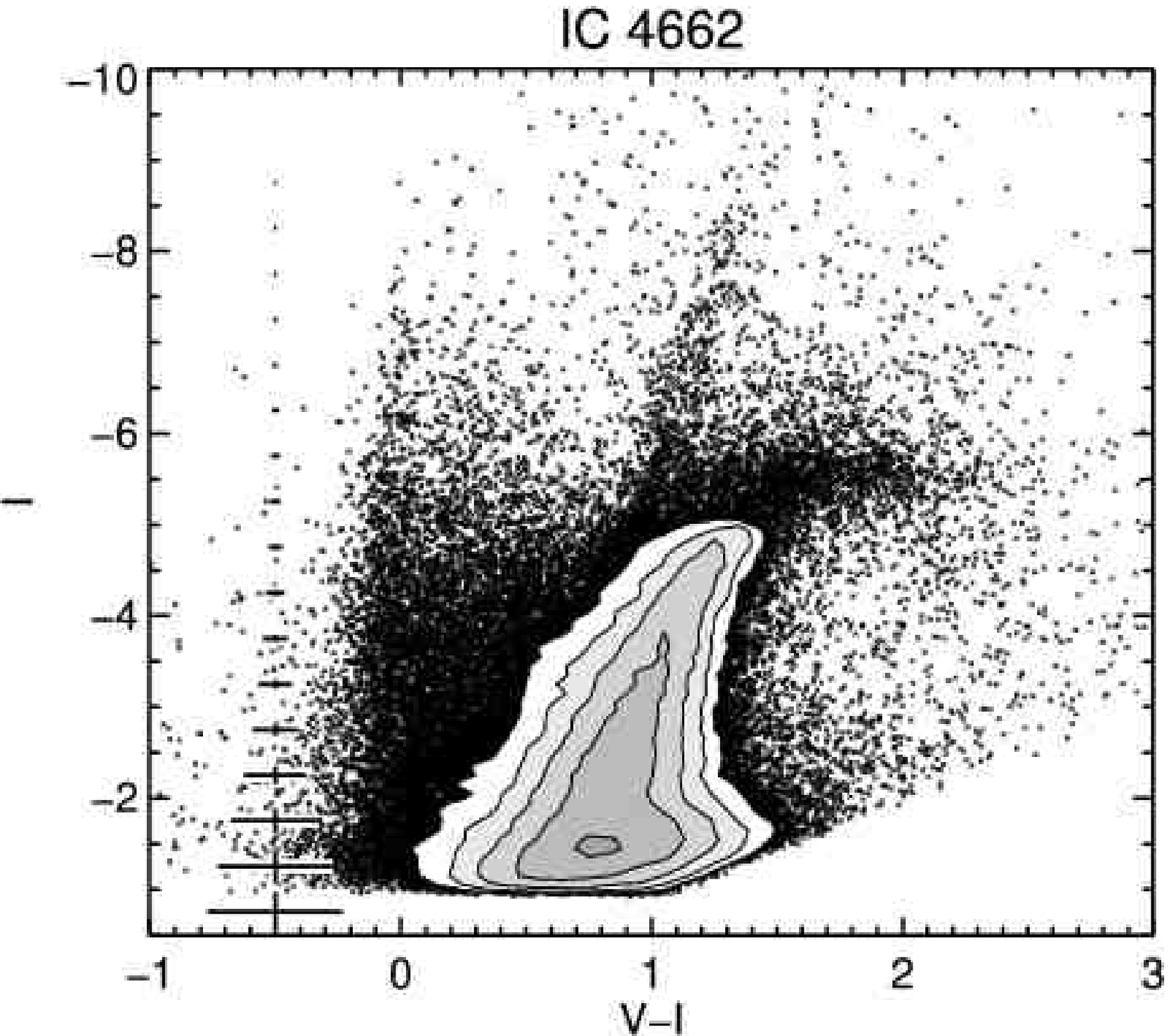,width=0.5\linewidth,clip=} \\
\epsfig{file=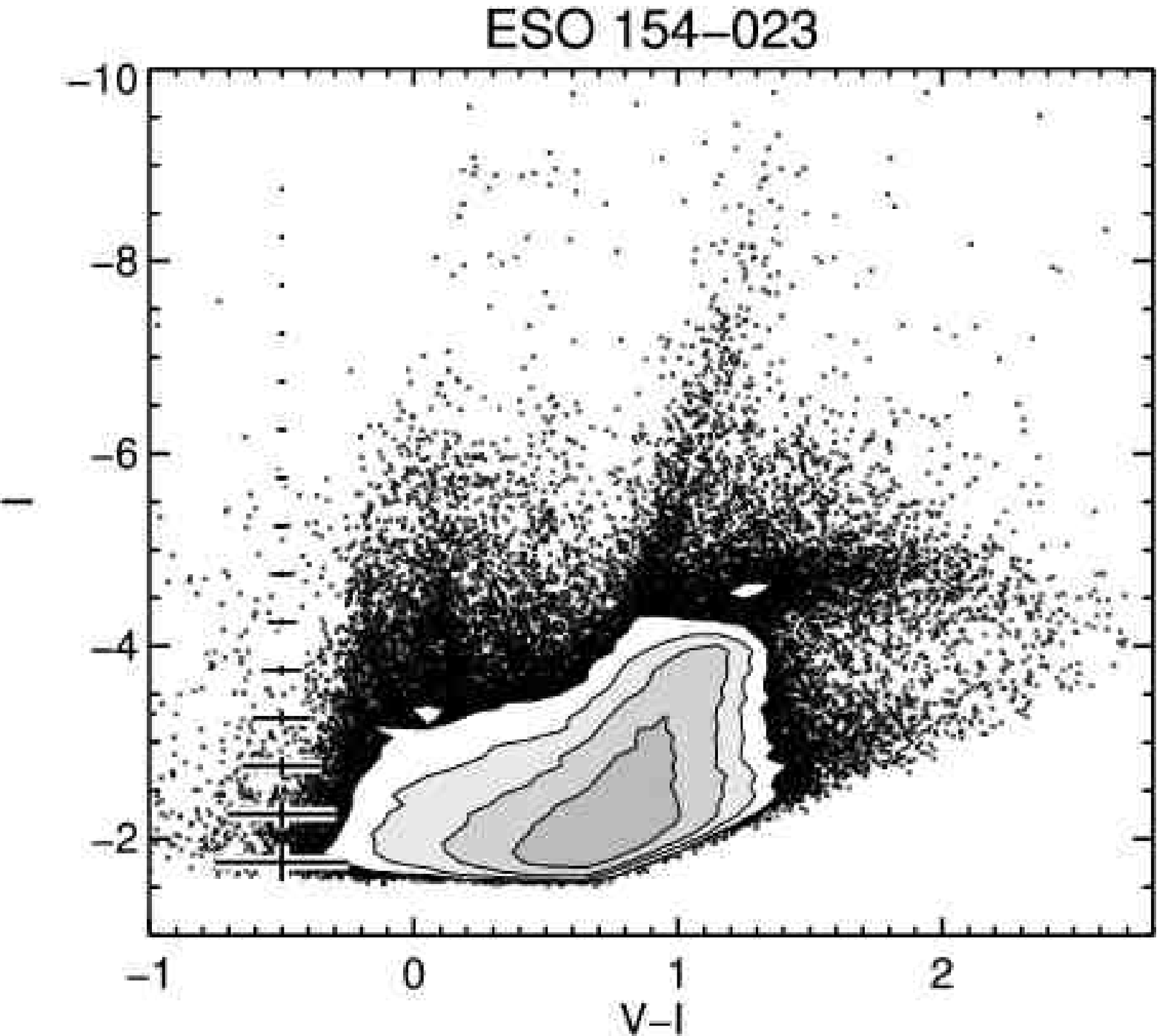,width=0.5\linewidth,clip=} &
\epsfig{file=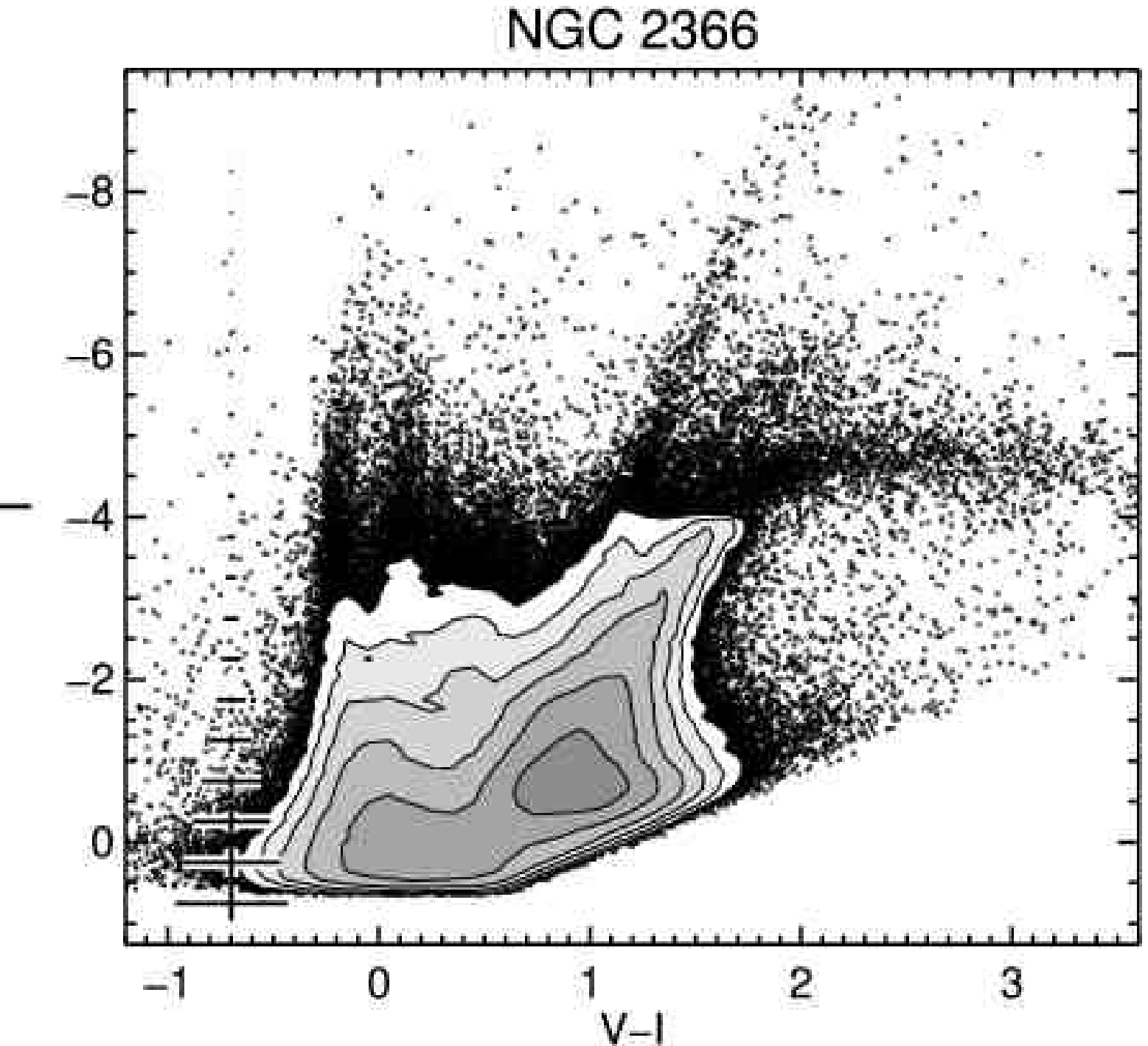,width=0.5\linewidth,clip=}
\end{tabular}
\caption{\textit{CMDs continued: SBS1415$+$437, IC~4662, ESO154$-$023, NGC~2366}}
\end{figure}

\clearpage
\begin{figure}
\figurenum{\ref{fig:cmds}}
\centering
\begin{tabular}{cc}
\epsfig{file=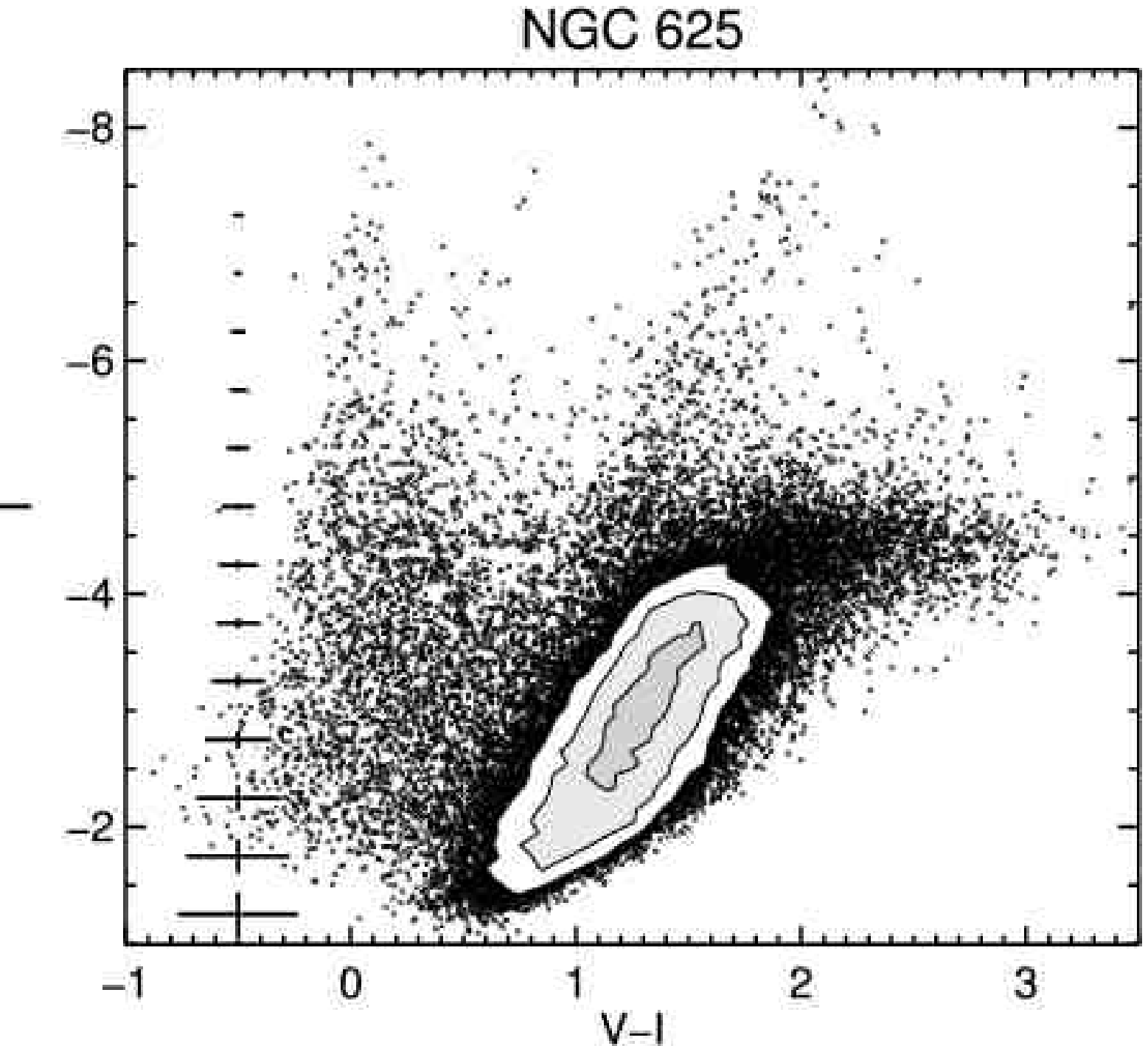,width=0.5\linewidth,clip=} & 
\epsfig{file=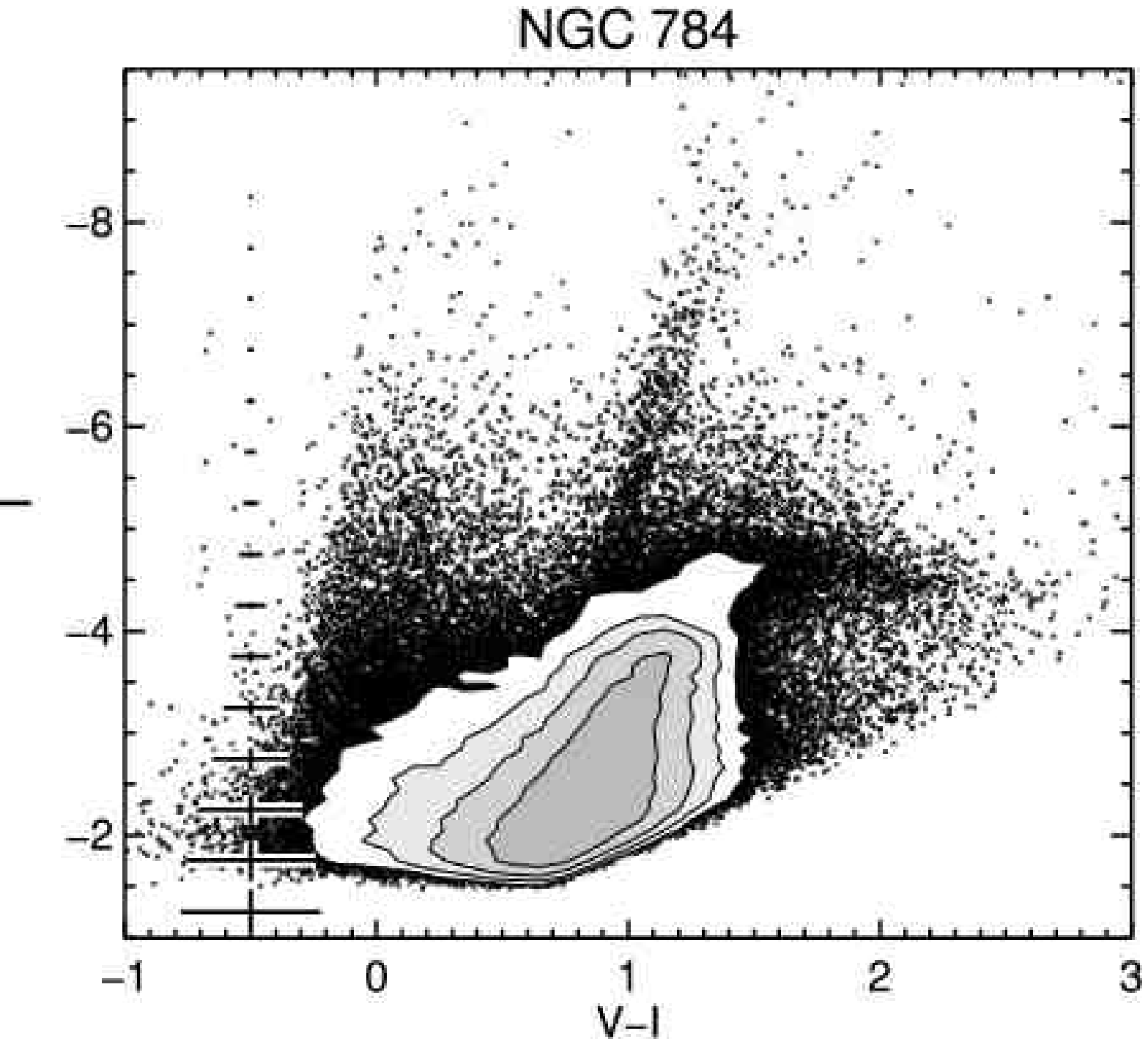,width=0.5\linewidth,clip=} \\
\epsfig{file=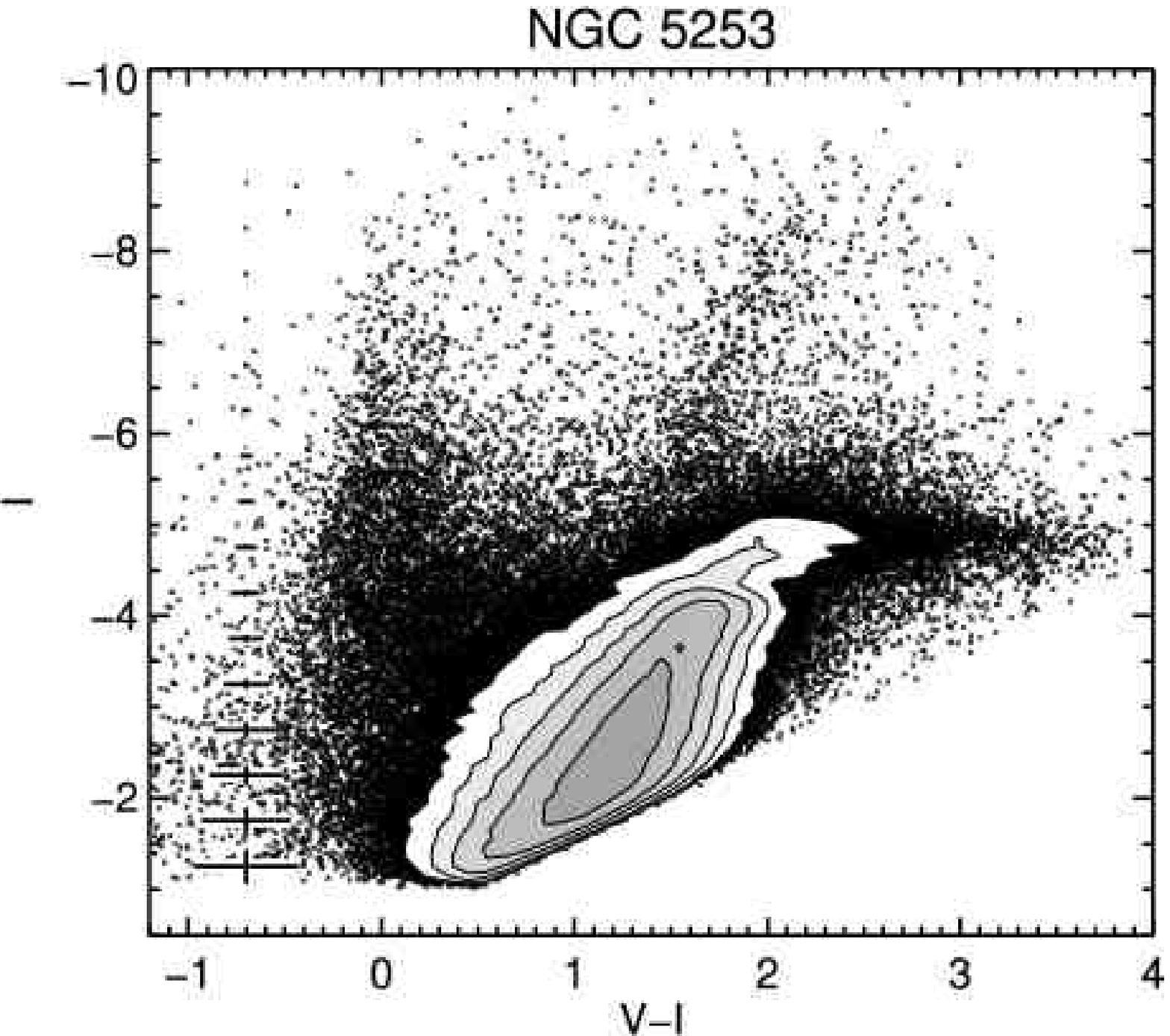,width=0.5\linewidth,clip=} &
\epsfig{file=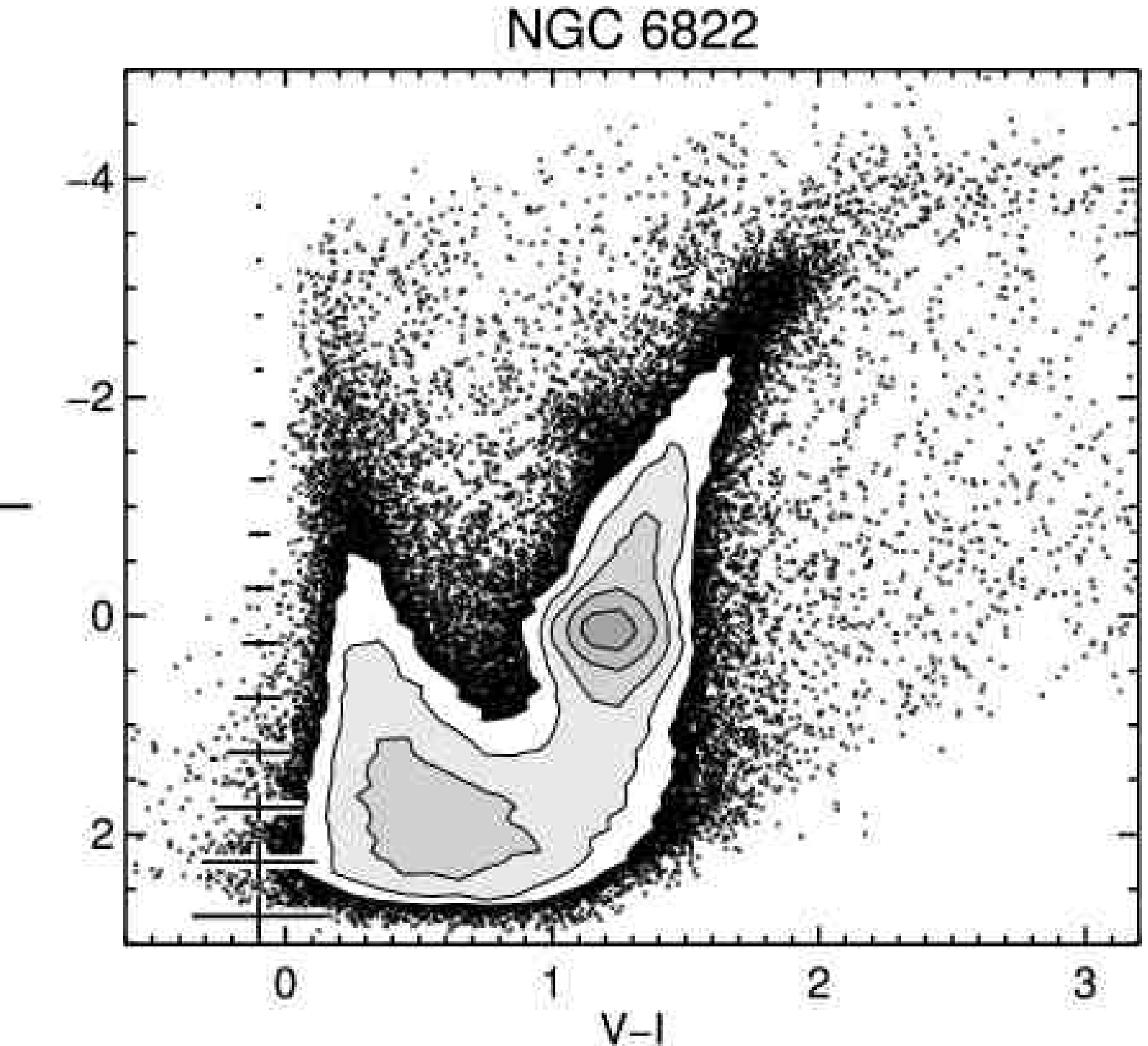,width=0.5\linewidth,clip=}
\end{tabular}
\caption{\textit{CMDs continued: NGC~625, NGC~784, NGC~5253, NGC~6822}}
\end{figure}

\clearpage
\begin{figure}
\figurenum{\ref{fig:cmds}}
\centering
\begin{tabular}{cc}
\epsfig{file=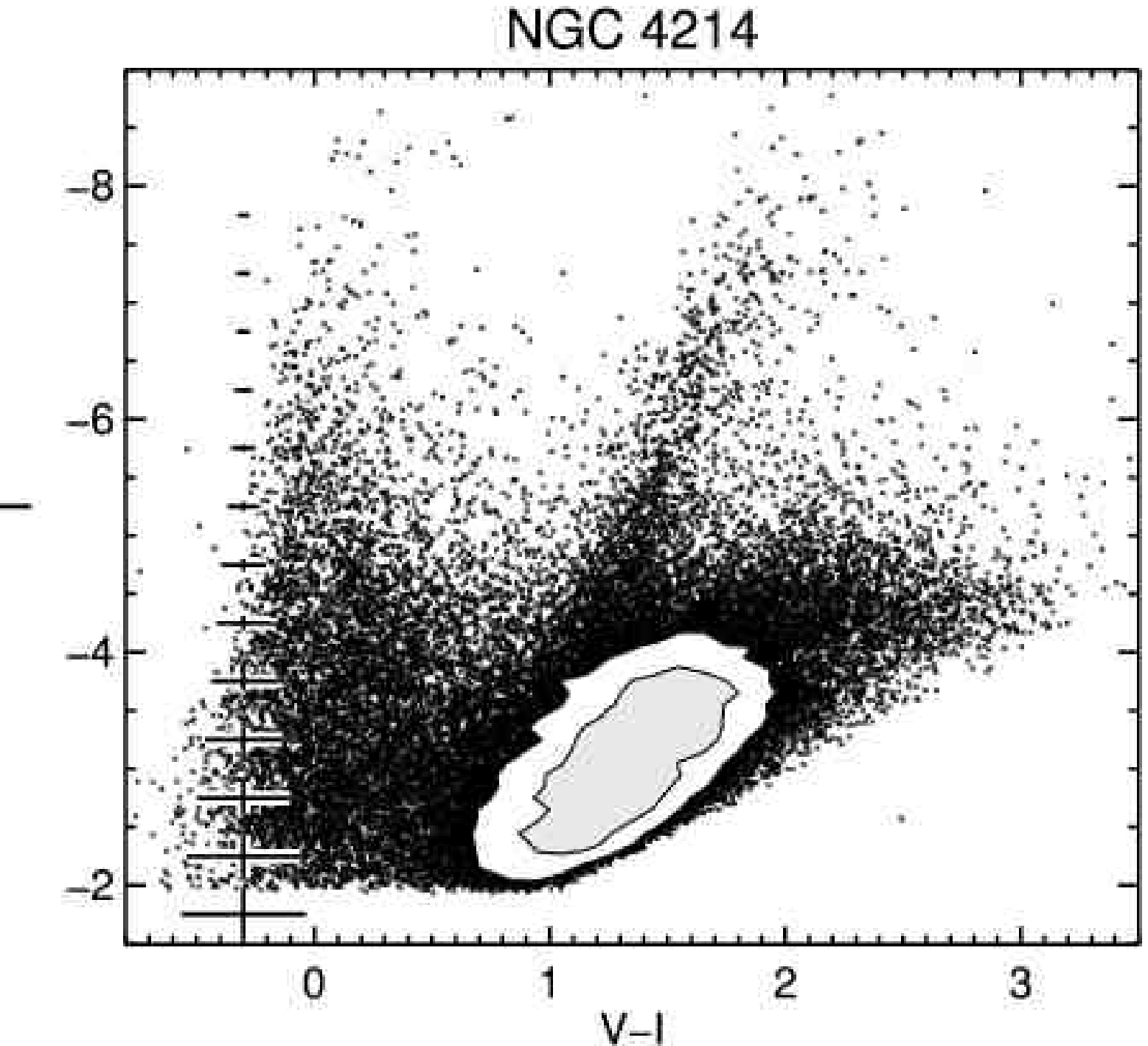,width=0.5\linewidth,clip=} & 
\epsfig{file=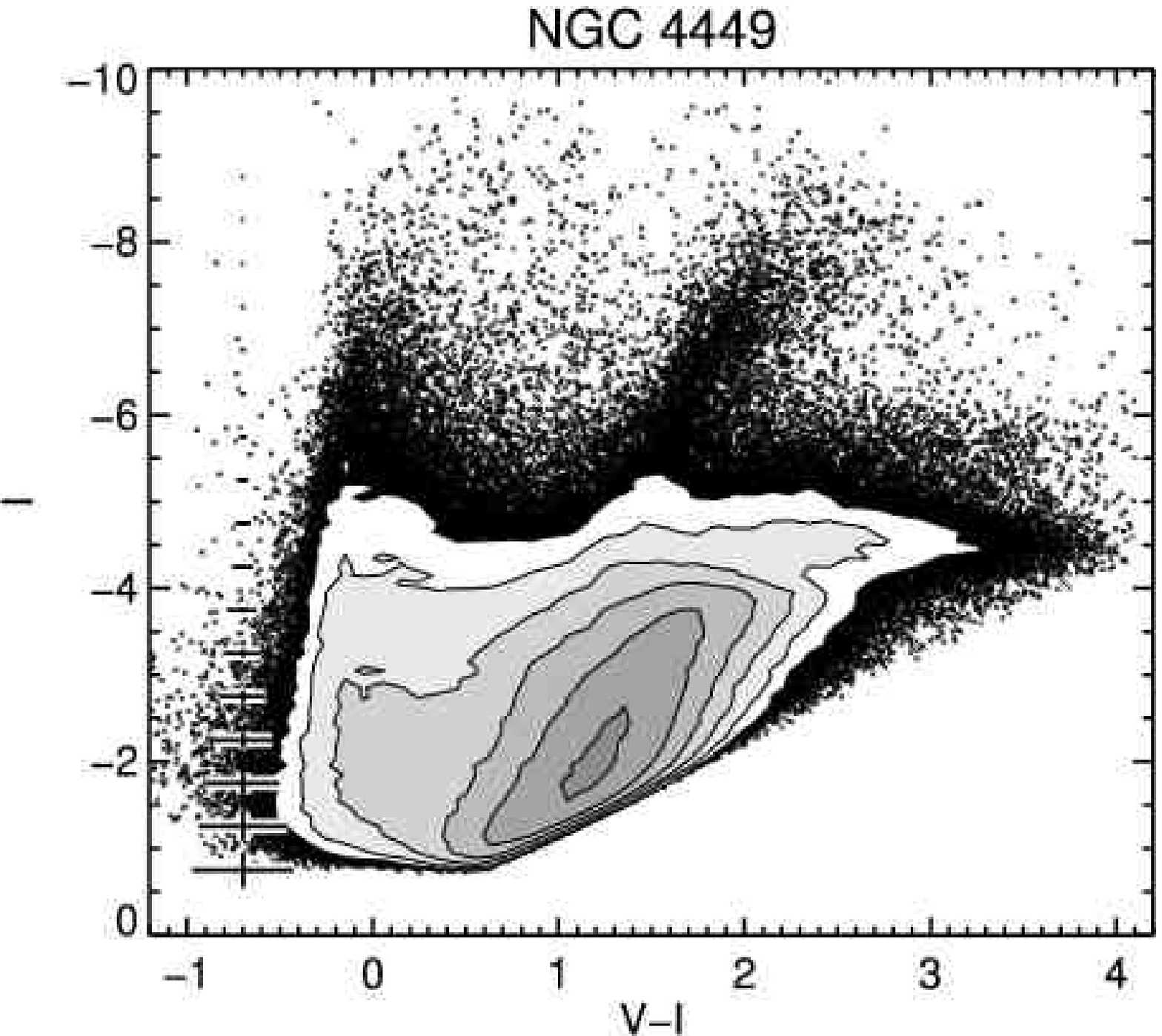,width=0.5\linewidth,clip=} 
\end{tabular}
\caption{\textit{CMDs continued: NGC~4214, NGC~4449}}
\end{figure}

\clearpage

\clearpage

\clearpage

\clearpage

\clearpage

\clearpage
\begin{figure}
\figurenum{\ref{fig:sfh_14gyr}}
\plotone{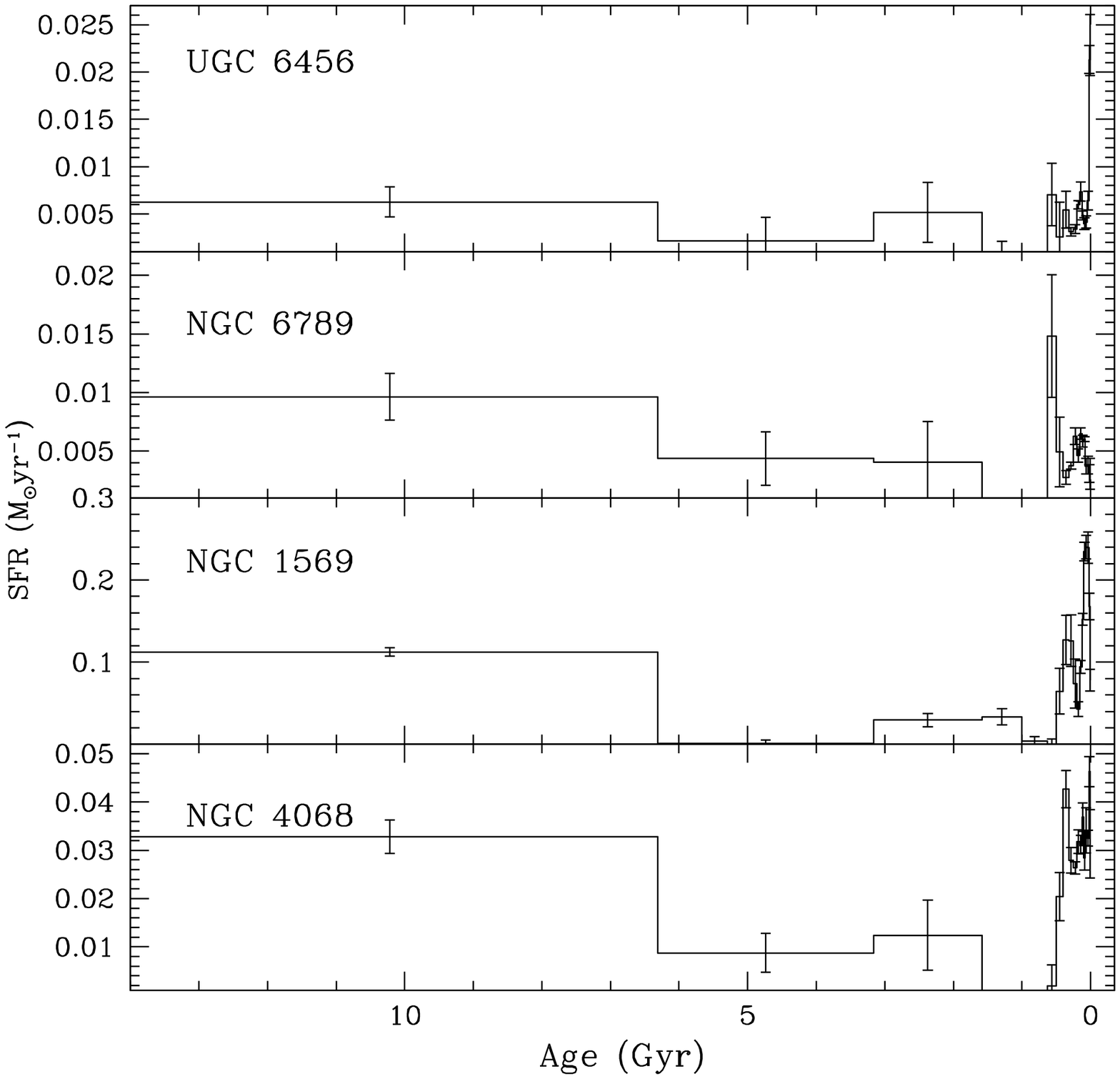}
\caption{\textit{Lifetime SFHs continued: UGC~6456, NGC~6789, NGC~1569, NGC~4068}}
\end{figure}

\clearpage
\begin{figure}
\figurenum{\ref{fig:sfh_14gyr}}
\plotone{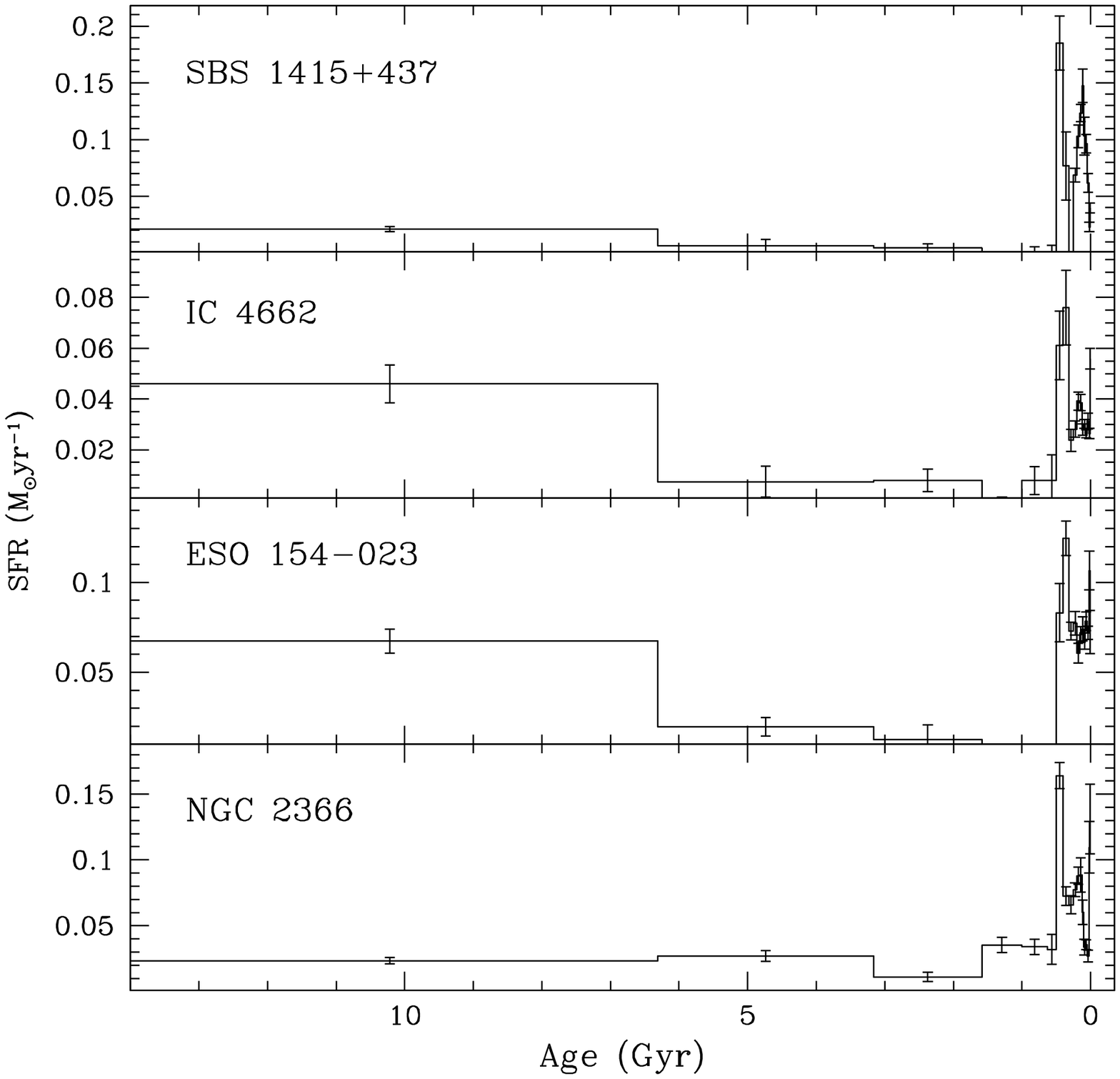}
\caption{\textit{Lifetime SFHs continued: SBS1415$+$437, IC~4662, ESO154$-$023, NGC~2366.} Note the ancient SFH for SBS~1415$+$437 is not well constrained due to the shallow photometry.}
\end{figure}

\clearpage
\begin{figure}
\figurenum{\ref{fig:sfh_14gyr}}
\plotone{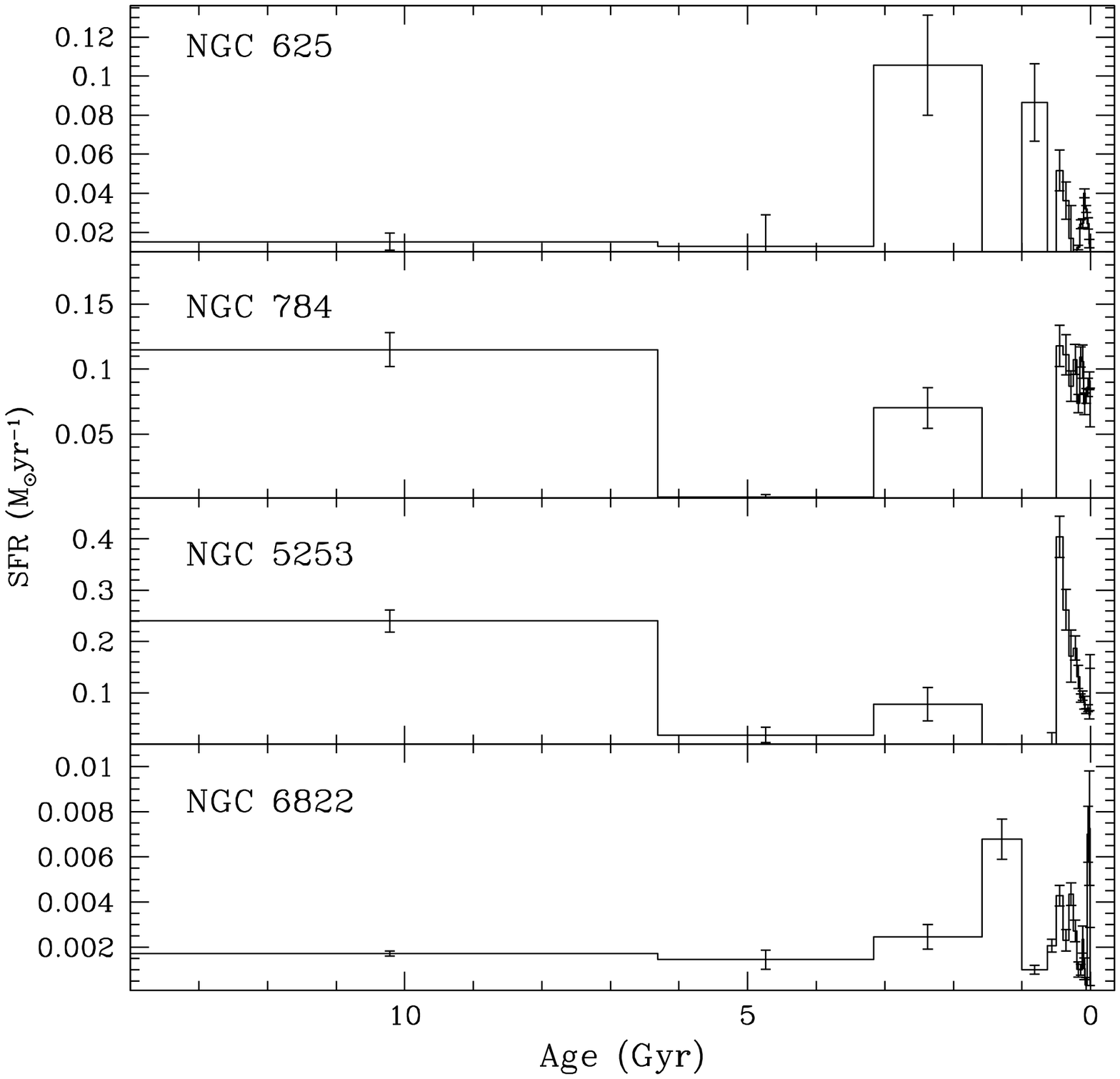}
\caption{\textit{Lifetime SFHs continued: NGC~625, NGC~784, NGC~5253, NGC~6822.} Note the SFRs derived for NGC~6822 are lower limits due to the limited spatial coverage of the observations.}
\end{figure}

\clearpage
\begin{figure}
\figurenum{\ref{fig:sfh_14gyr}}
\plotone{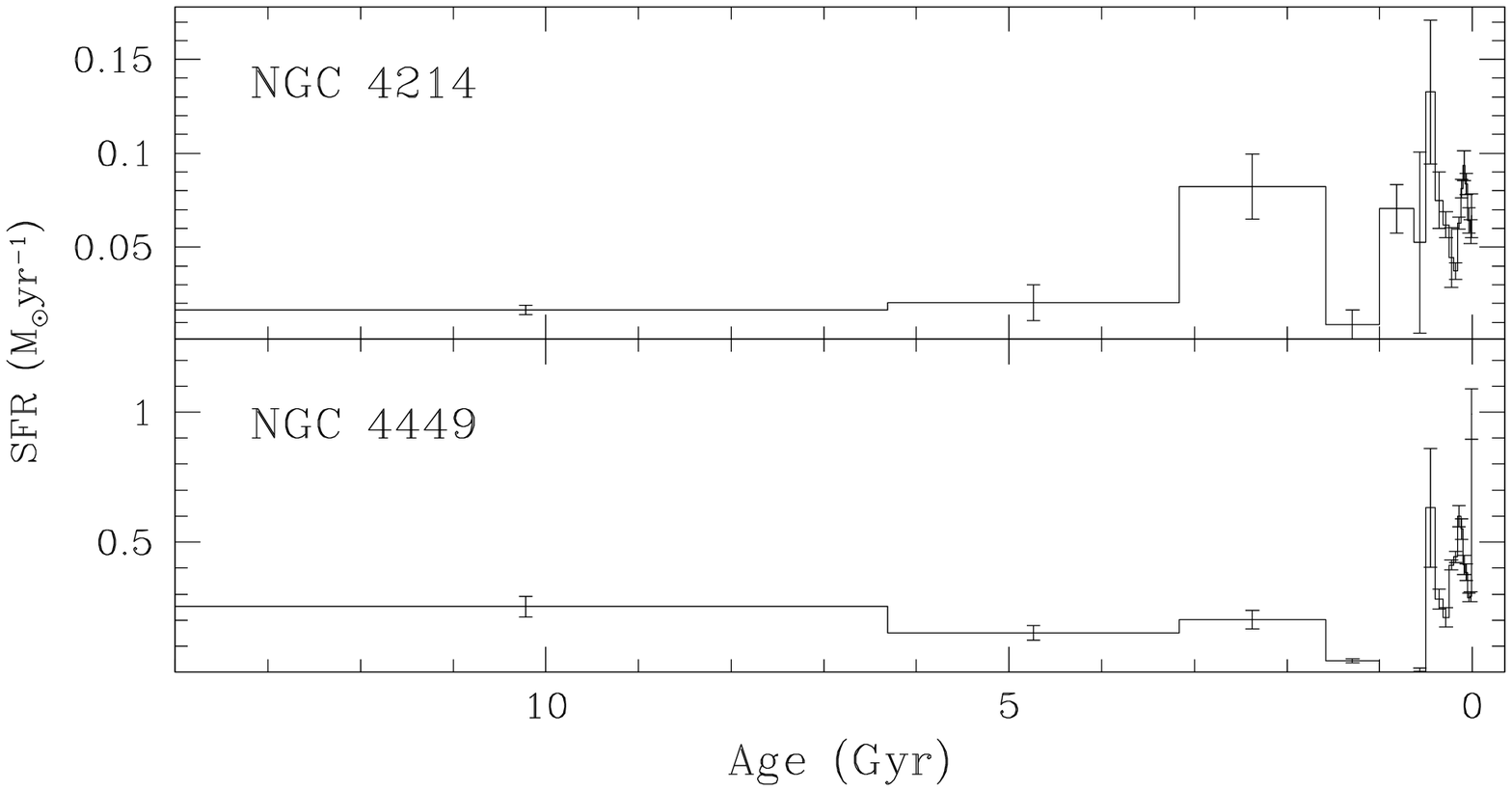}
\caption{\textit{Lifetime SFHs continued: NGC~4214, NGC~4449}}
\end{figure}

\clearpage

\clearpage
\begin{figure}
\figurenum{\ref{fig:sfh_1gyr}}
\plotone{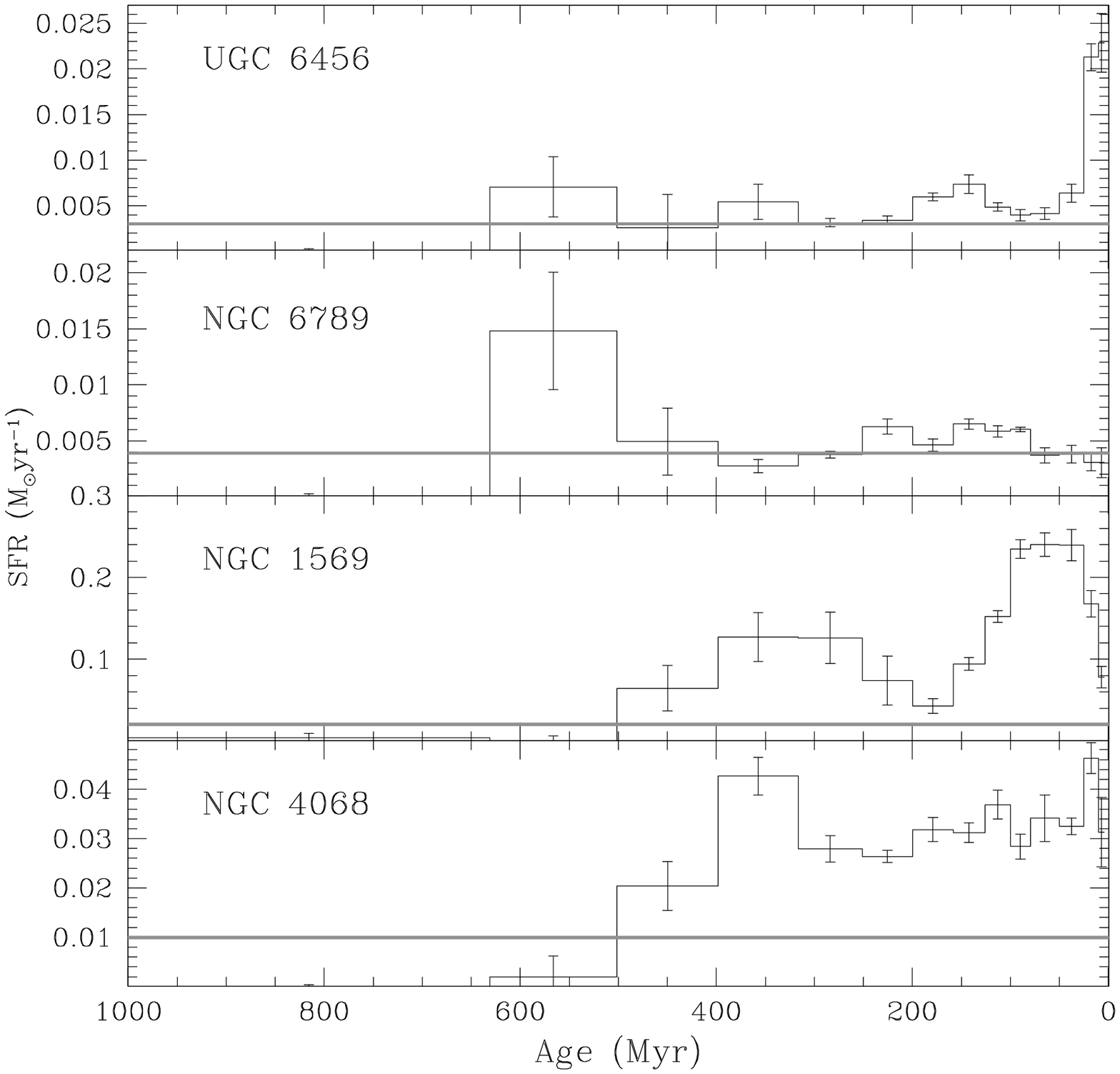}
\caption{\textit{Recent SFHs continued: UGC~6456, NGC~6789, NGC~1569, NGC~4068}}
\end{figure}

\clearpage
\begin{figure}
\figurenum{\ref{fig:sfh_1gyr}}
\plotone{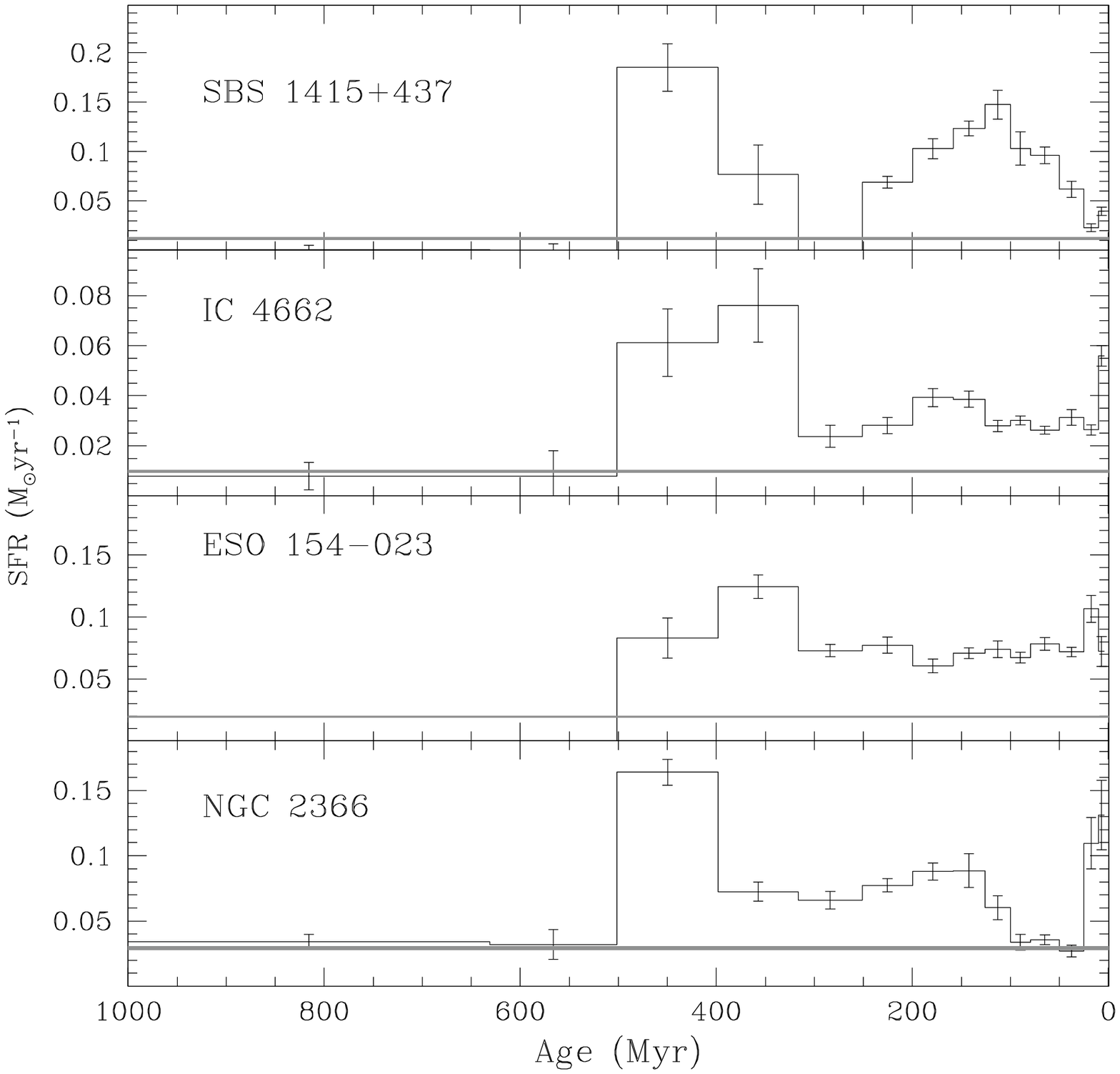}
\caption{\textit{Recent SFHs continued: SBS1415$+$437, IC~4662, ESO154$-$023, NGC~2366}}
\end{figure}

\clearpage
\begin{figure}
\figurenum{\ref{fig:sfh_1gyr}}
\plotone{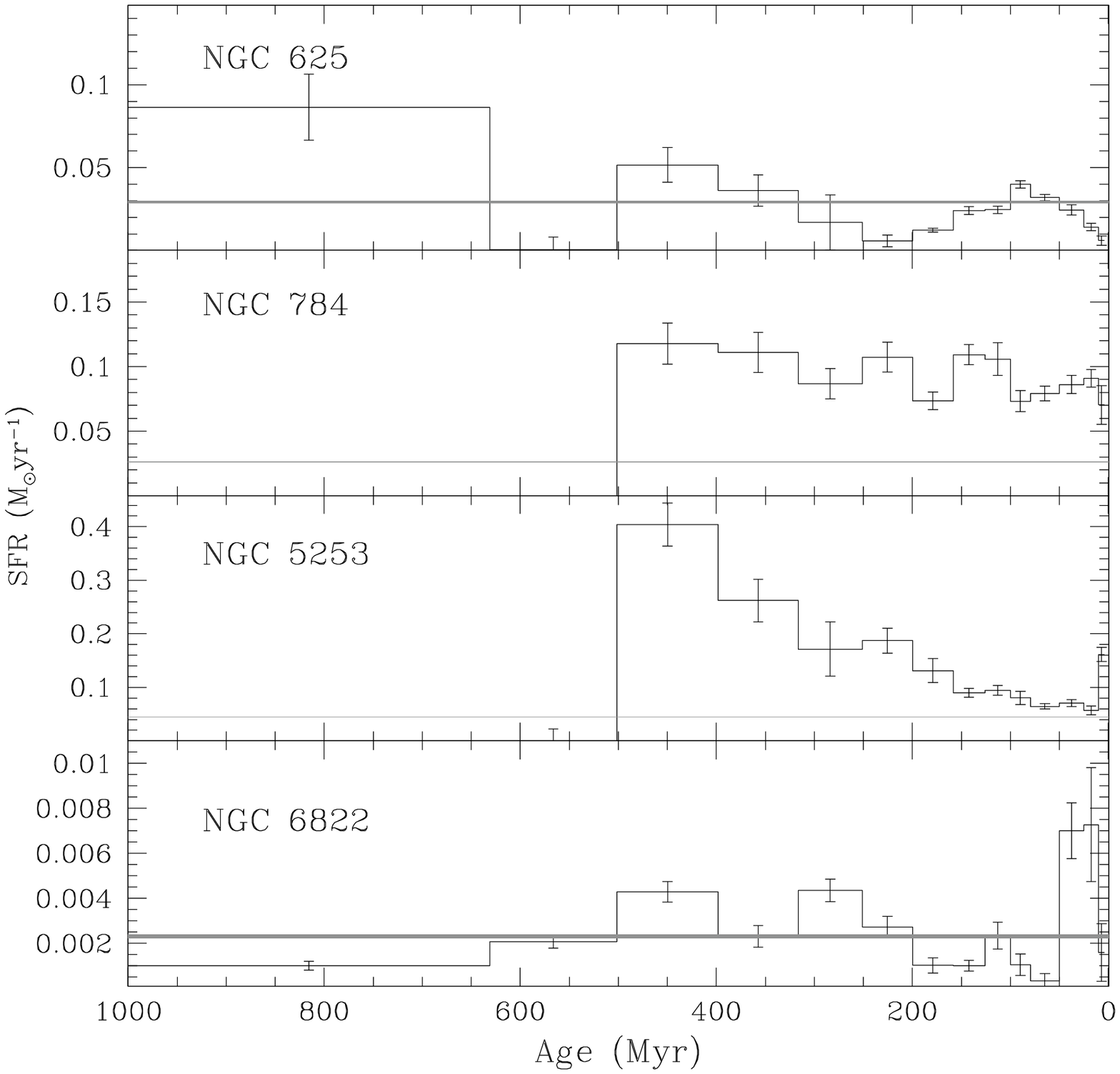}
\caption{\textit{Recent SFHs continued: NGC~625, NGC~784, NGC~5253, NGC~6822}}
\end{figure}

\clearpage
\begin{figure}
\figurenum{\ref{fig:sfh_1gyr}}
\plotone{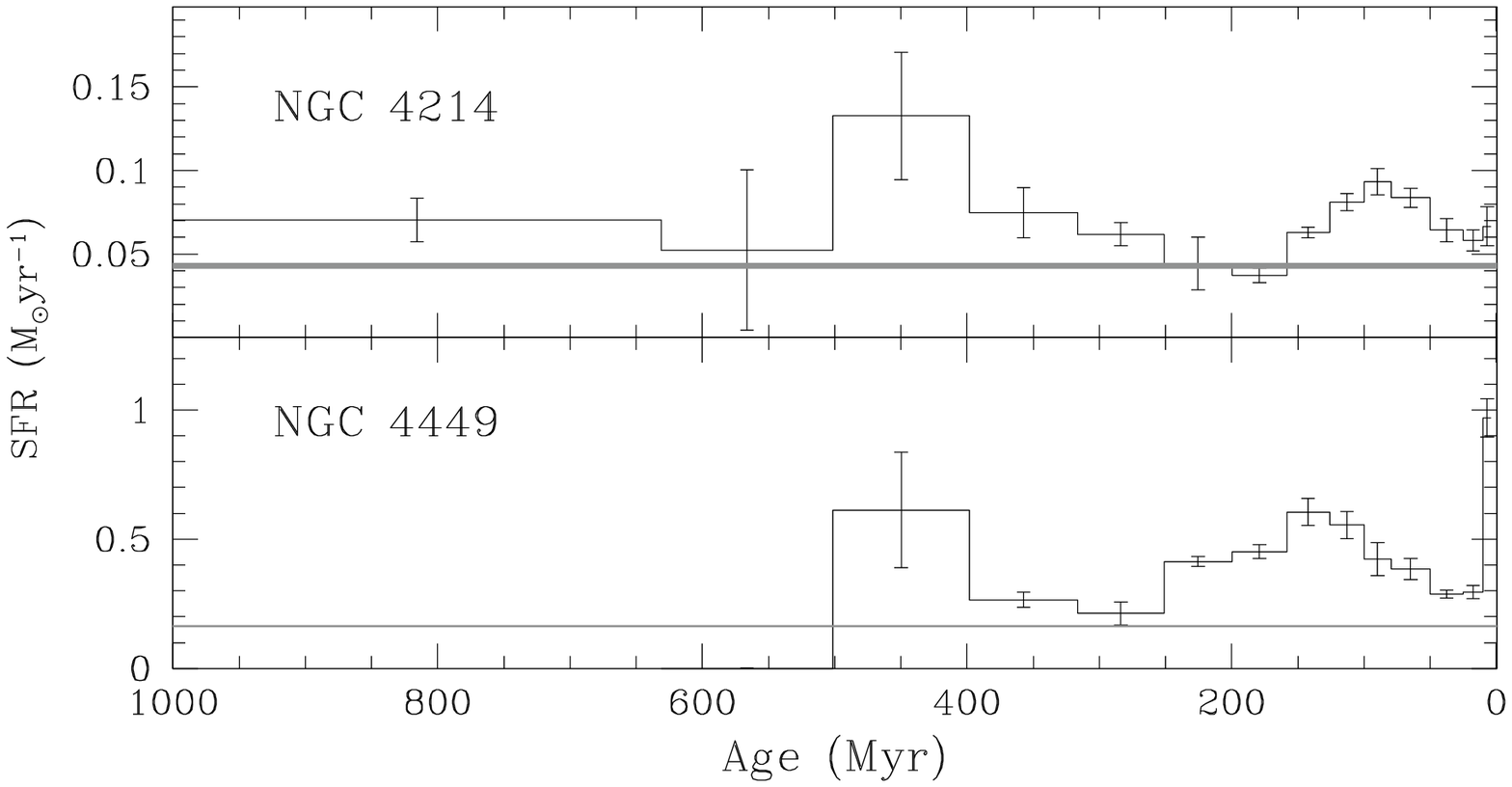}
\caption{\textit{Recent SFHs continued: NGC~4214, NGC~4449}}
\end{figure}

\clearpage

\clearpage


\begin{thebibliography}{}
\bibitem[Aloisi et al.(2005)]{Aloisi2005} Aloisi, A., van der 
Marel, R.~P., Mack, J., Leitherer, C., Sirianni, M., \& Tosi, M.\ 2005, \apjl, 631, L45 
\bibitem[Angeretti et al.(2005)]{Angeretti2005} Angeretti, L., Tosi, 
M., Greggio, L., Sabbi, E., Aloisi, A., \& Leitherer, C.\ 2005, \aj, 129, 2203 
\bibitem[Annibali et al.(2008)]{Annibali2008} Annibali, F., Aloisi, 
A., Mack, J., Tosi, M., van der Marel, R.~P., Angeretti, L., Leitherer, C., 
\& Sirianni, M.\ 2008, \aj, 135, 1900 
\bibitem[Aparicio \& Hidalgo(2009)]{Aparicio2009} Aparicio, A., \& Hidalgo, S.~L.\ 2009, \aj, 138, 558 
\bibitem[Babul \& Ferguson(1996)]{Babul1996} Babul, A., \& Ferguson, H.~C.\ 1996, \apj, 458, 100 
\bibitem[Barnes \& de Blok(2001)]{Barnes2001} Barnes, D.~G., \& de Blok, W.~J.~G.\ 2001, \aj, 122, 825 
\bibitem[Barone et al.(2000)]{Barone2000} Barone, L.~T., 
Heithausen, A., H{\"u}ttemeister, S., Fritz, T., \& Klein, U.\ 2000, \mnras, 317, 649 
\bibitem[Begum et al.(2008)]{Begum2008} Begum, A., Chengalur, 
J.~N., Karachentsev, I.~D., Sharina, M.~E., 
\& Kaisin, S.~S.\ 2008, \mnras, 386, 1667 
\bibitem[Bertelli et al.(1994)]{Bertelli1994} Bertelli, G., Bressan, A., Chiosi, C., Fagotto, F., \& Nasi, E.\ 1994, \aaps, 106, 275 
\bibitem[Bolatto et al.(2008)]{Bolatto2008} Bolatto, A.~D., Leroy, 
A.~K., Rosolowsky, E., Walter, F., \& Blitz, L.\ 2008, \apj, 686, 948 
\bibitem[Bonnarel et al.(2000)]{Bonnarel2000} Bonnarel, F., et al.\ 2000, \aaps, 143, 33 
\bibitem[Bottinelli et al.(1990)]{Bottinelli1990} Bottinelli, L., Gouguenheim, L., Fouque, P., \& Paturel, G.\ 1990, \aaps, 82, 391 
\bibitem[Brinchmann et al.(2004)]{Brinchmann2004} Brinchmann, J., 
Charlot, S., White, S.~D.~M., Tremonti, C., Kauffmann, G., Heckman, T., 
\& Brinkmann, J.\ 2004, \mnras, 351, 1151 
\bibitem[Bruzual \& Charlot(2003)]{Bruzual2003} Bruzual, G., \& Charlot, S.\ 2003, \mnras, 344, 1000 
\bibitem[Calzetti et al.(1997)]{Calzetti1997} Calzetti, D., Meurer, 
G.~R., Bohlin, R.~C., Garnett, D.~R., Kinney, A.~L., Leitherer, C., 
\& Storchi-Bergmann, T.\ 1997, \aj, 114, 1834 
\bibitem[Cannon et al.(2003)]{Cannon2003} Cannon, J.~M., Dohm-Palmer, R.~C., Skillman, E.~D., Bomans, D.~J., C{\^o}t{\'e}, S., \& Miller, B.~W.\ 2003, \aj, 126, 2806 
\bibitem[Cannon \& Skillman(2004)]{Cannon2004} Cannon, J.~M., \& Skillman, E.~D.\ 2004, \apj, 610, 772 
\bibitem[Cole et al.(2007)]{Cole2007} Cole, A.~A., et al.\ 2007, 
\apjl, 659, L17 
\bibitem[Colless et al.(1994)]{Colless1994} Colless, M., Schade, 
D., Broadhurst, T.~J., \& Ellis, R.~S.\ 1994, \mnras, 267, 1108 
\bibitem[Condon et al.(1982)]{Condon1982} Condon, J.~J., Condon, 
M.~A., Gisler, G., \& Puschell, J.~J.\ 1982, \apj, 252, 102 
al.\ 2009, \apjs, 183, 67 
\bibitem[de Blok \& Walter(2000)]{deBlok2000} de Blok, W.~J.~G., \& Walter, F.\ 2000, \apjl, 537, L95 
\bibitem[de Vaucouleurs et al.(1974)]{deVaucouleurs1974} de Vaucouleurs, 
G., de Vaucouleurs, A., \& Pence, W.\ 1974, \apjl, 194, L119 
\bibitem[de Vaucouleurs et al.(1991)]{deVaucouleurs1991} de Vaucouleurs, 
G., de Vaucouleurs, A., Corwin, H.~G., Jr., Buta, R.~J., Paturel, G., 
\& Fouqu{\'e}, P.\ 1991, Third Reference Catalogue of Bright Galaxies, Springer, New York, NY (USA)
\bibitem[Dohm-Palmer \& Skillman(2002)]{Dohm-Palmer2002} Dohm-Palmer, R.~C., \& Skillman, E.~D.\ 2002, \aj, 123, 1433 
\bibitem[Dolphin(2000)]{Dolphin2000} Dolphin, A.~E.\ 2000, \pasp, 
112, 1383 
\bibitem[Dolphin et al.(2001)]{Dolphin2001} Dolphin, A.~E., et al.\ 2001, \mnras, 324, 249 
\bibitem[Dolphin(2002)]{Dolphin2002} Dolphin, A.~E., 2002, \mnras, 332, 91
\bibitem[Drozdovsky et al.(2001)]{Drozdovsky2001} Drozdovsky, I.~O., 
Schulte-Ladbeck, R.~E., Hopp, U., Crone, M.~M., \& Greggio, L.\ 2001, \apjl, 551, L135 
\bibitem[Drozdovsky et al.(2002)]{Drozdovsky2002} Drozdovsky, I.~O., 
Schulte-Ladbeck, R.~E., Hopp, U., Greggio, L., \& Crone, M.~M.\ 2002, \aj, 124, 811 
\bibitem[Ferguson \& Babul(1998)]{Ferguson1998} Ferguson, H.~C., \& Babul, A.\ 1998, \mnras, 296, 585 
\bibitem[Gallagher \& Hunter(1986)]{Gallagher1986} Gallagher, J.~S., III, \& Hunter, D.~A.\ 1986, \aj, 92, 557 
\bibitem[Gallart et al.(1996)]{Gallart1996} Gallart, C., Aparicio, A., \& Vilchez, J.~M.\ 1996, \aj, 112, 1928 
\bibitem[Gieren et al.(2006)]{Gieren2006} Gieren, W., Pietrzy{\'n}ski, G., Nalewajko, K., Soszy{\'n}ski, I., Bresolin, F., 
Kudritzki, R.-P., Minniti, D., \& Romanowsky, A.\ 2006, \apj, 647, 1056 
\bibitem[Gil de Paz et al.(2003)]{GildePaz2003} Gil de Paz, A., 
Madore, B.~F., \& Pevunova, O.\ 2003, \apjs, 147, 29 
\bibitem[Gogarten et al.(2009)]{Gogarten2009} Gogarten, S.~M., et 
al.\ 2009, \apj, 691, 115 
\bibitem[Grocholski et al.(2008)]{Grocholski2008} Grocholski, A.~J., et al.\ 2008, \apjl, 686, L79 
\bibitem[Harris \& Zaritsky(2001)]{Harris2001} Harris, J., \& Zaritsky, D.\ 2001, \apjs, 136, 25 
\bibitem[Harris et al.(2004)]{Harris2004} Harris, J., Calzetti, D., Gallagher, J.~S., Smith, D.~A. \& Conselice, C.~J., 2004, \apj, 603, 503
\bibitem[Heckman et al.(1998)]{Heckman1998} Heckman, T.~M., Robert, 
C., Leitherer, C., Garnett, D.~R., \& van der Rydt, F.\ 1998, \apj, 503, 646 
\bibitem[Heckman et al.(2005)]{Heckman2005} Heckman, T.~M., et al.\ 
2005, \apjl, 619, L35 
\bibitem[Heydari-Malayeri et al.(1990)]{Heydari-Malayeri1990} Heydari-Malayeri, M., Melnick, J., \& Martin, J.-M.\ 1990, \aap, 234, 99 
\bibitem[Hodge(1980)]{Hodge1980} Hodge, P.~W.\ 1980, \apj, 241, 125
\bibitem[Holtzman et al.(1999)]{Holtzman1999} Holtzman, J.~A., et 
al.\ 1999, \aj, 118, 2262 
\bibitem[Holtzman et al.(2006)]{Holtzman2006} Holtzman, J.~A., 
Afonso, C., \& Dolphin, A.\ 2006, \apjs, 166, 534 
\bibitem[Huchra et al.(1983)]{Huchra1983} Huchra, J.~P., Geller, 
M.~J., Gallagher, J., Hunter, D., Hartmann, L., Fabbiano, G., 
\& Aaronson, M.\ 1983, \apj, 274, 125 
\bibitem[Huchtmeier et al.(1981)]{Huchtmeier1981} Huchtmeier, W.~K., Seiradakis, J.~H., \& Materne, J.\ 1981, \aap, 102, 134 
\bibitem[Huchtmeier et al.(2005)]{Huchtmeier2005} Huchtmeier, W.~K., 
Gopal, K., \& Petrosian, A.\ 2005, VizieR Online Data Catalog, 343, 40887 
\bibitem[Hunter et al.(1982)]{Hunter1982} Hunter, D.~A., 
Gallagher, J.~S., \& Rautenkranz, D.\ 1982, \apjs, 49, 53 
\bibitem[Israel et al.(1995)]{Israel1995} Israel, F.~P., Tacconi, L.~J., \& Baas, F.\ 1995, \aap, 295, 599 
\bibitem[Israel(1997)]{Israel1997} Israel, F.~P.\ 1997, \aap, 328, 471
\bibitem[Karachentsev et al.(2002)]{Karachentsev2002} Karachentsev, I.~D., et al.\ 2002, \aap, 383, 125 
\bibitem[Karachentsev et al.(2003)]{Karachentsev2003} Karachentsev, I.~D., et al.\ 2003, \aap, 398, 467 
\bibitem[Karachentsev et al.(2006)]{Karachentsev2006} Karachentsev, I.~D., et al.\ 2006, \aj, 131, 1361 
\bibitem[Kauffmann et al.(2003)]{Kauffmann2003} Kauffmann, G., et 
al.\ 2003, \mnras, 341, 33 
\bibitem[Kaviraj et al.(2007)]{Kaviraj2007} Kaviraj, S., Kirkby, 
L.~A., Silk, J., \& Sarzi, M.\ 2007, \mnras, 382, 960 
\bibitem[Kennicutt et al.(1987)]{Kennicutt1987} Kennicutt, R.~C., 
Jr., Roettiger, K.~A., Keel, W.~C., van der Hulst, J.~M., 
\& Hummel, E.\ 1987, \aj, 93, 1011 
\bibitem[Kennicutt et al.(1994)]{Kennicutt1994} Kennicutt, R.~C.,
Jr., Tamblyn, P., \& Congdon, C.~E.\ 1994, \apj, 435, 22
\bibitem[Kennicutt(1998)]{Kennicutt1998} Kennicutt, R.~C., Jr., 1998, \araa, 36, 189 
\bibitem[Kennicutt \& Skillman(2001)]{Kennicutt2001} Kennicutt, R.~C., Jr., \& Skillman, E.~D.\ 2001, \aj, 121, 1461 
\bibitem[Kennicutt et al.(2005)]{Kennicutt2005} Kennicutt, R.~C., Jr., Lee, J.~C., Funes, J.~G., Sakai, S., \& Akiyama, S., 2005, ASSL Volume 329, Starbursts from 30 Doradus to Lyman Break Galaxies, eds., De Gris, R. \& Gonz\'{a}lez Delgado, R.~M., Springer (the Netherlands), 187
\bibitem[Lee et al.(2008)]{Lee2008} Lee, J.~C., et al.\ 2008, 
Astronomical Society of the Pacific Conference Series, 396, 151 
\bibitem[Lee et al.(2009)]{Lee2009} Lee, J.~C., Kennicutt, 
R.~C., Jos{\'e} G.~Funes, S.~J., Sakai, S., \& Akiyama, S.\ 2009, \apj, 692, 1305 
\bibitem[Leroy et al.(2005)]{Leroy2005} Leroy, A., Bolatto, 
A.~D., Simon, J.~D., \& Blitz, L.\ 2005, \apj, 625, 763 
\bibitem[Loose \& Thuan(1986)]{Loose1986} Loose, H.-H., \& Thuan, T.~X.\ 1986, \apj, 309, 59 \bibitem[L{\'o}pez-S{\'a}nchez et al.(2008)]{Lopez2008}L{\'o}pez-S{\'a}nchez, {\'A}.~R., Koribalski, B., Esteban, C., Garc{\'{\i}}a-Rojas, J.\ 2008, ASSP, eds. Koribalski, B.S.; Jerjen, H., Galaxies in the Local Volume, Springer (The Netherlands). p.299
\bibitem[L{\'o}pez-S{\'a}nchez et al.(2009)]{Lopez2009}L{\'o}pez-S{\'a}nchez, {\'A}.~R., Koribalski, B., Esteban, C., Popping, A., van Eymeren J., \& Hibbard, J.\ in press, Star Forming Galaxies Workshop Proceedings
\bibitem[Madau et al.(1996)]{Madau1996} Madau, P., Ferguson, 
H.~C., Dickinson, M.~E., Giavalisco, M., Steidel, C.~C., 
\& Fruchter, A.\ 1996, \mnras, 283, 1388 
\bibitem[Marigo \& Girardi(2007)]{Marigo07} Marigo, P., \& Girardi, L.\ 2007, \aap, 469, 239
\bibitem[Marlowe et al.(1997)]{Marlowe1997} Marlowe, A.~T., Meurer, 
G.~R., Heckman, T.~M., \& Schommer, R.\ 1997, \apjs, 112, 285 
\bibitem[Marlowe et al.(1999)]{Marlowe1999} Marlowe, A.~T., Meurer, 
G.~R., \& Heckman, T.~M.\ 1999, \apj, 522, 183 
\bibitem[Mas-Hesse \& Kunth(1999)]{Mas-Hesse1999}Mas-Hesse, J. M. \& Kunth, D., 1999, A\&A, 349, 765
\bibitem[Mayer et al.(2001)]{Mayer2001} Mayer, L., Governato, F.,
Colpi, M., Moore, B., Quinn, T., Wadsley, J., Stadel, J., \& Lake, G.\
2001, \apj, 559, 754
\bibitem[McQuinn et al.(2009)]{McQuinn2009} McQuinn, K.~B.~W., 
Skillman, E.~D., Cannon, J.~M., Dalcanton, J.~J., Dolphin, A., Stark, D., 
\& Weisz, D.\ 2009, \apj, 695, 561 
\bibitem[McQuinn et al.(submitted)]{McQuinn2010} McQuinn, K.~B.~W., et al., 2010, ApJ, submitted, Paper II
\bibitem[Meurer et al.(1997)]{Meurer1997} Meurer, G.~R., Heckman, 
T.~M., Lehnert, M.~D., Leitherer, C., \& Lowenthal, J.\ 1997, \aj, 114, 54 
\bibitem[Meurer(2000)]{Meurer2000} Meurer, G.~R.\ 2000, Massive 
Stellar Clusters, 211, 81 
\bibitem[M{\'e}ndez et al.(2002)]{Mendez2002} M{\'e}ndez, B., 
Davis, M., Moustakas, J., Newman, J., Madore, B.~F., \& Freedman, W.~L.\ 2002, \aj, 124, 213 
\bibitem[Nishi \& Tashiro(2000)]{Nishi2000} Nishi, R., \& Tashiro, M.\ 2000, \apj, 537, 50 
\bibitem[O'Connell(2005)]{OConnell2005}O'Connell, R.~W., 2005, ASSL Volume 329, Starbursts from 30 Doradus to Lyman Break Galaxies, eds. De Grijs, R. \& Gonz\'{a}lez Delgado, R.~M., Springer (The Netherlands), 333
\bibitem[Oey et al.(2007)]{Oey2007} Oey, M.~S., et al.\ 2007, 
\apj, 661, 801 
\bibitem[Omukai \& Nishi(1999)]{Omukai1999} Omukai, K., \& Nishi, R.\ 1999, \apj, 518, 64 
\bibitem[Pasetto et al.(2003)]{Pasetto2003} Pasetto, S., Chiosi, C., \& Carraro, G.\ 2003, \aap, 405, 931 
\bibitem[Paturel et al.(2003)]{Paturel2003} Paturel G., Petit C., Prugniel P., Theureau G., Rousseau J., Brouty M., Dubois P., Cambr{\'e}sy L., 2003, A\&A, 412, 45
\bibitem[Pelupessy et al.(2004)]{Pelupessy2004} Pelupessy, F.~I., van der Werf, P.~P., \& Icke, V.\ 2004, \aap, 422, 55 
\bibitem[Romano et al.(2006)]{Romano2006}Romano, D., Tosi, M. \& Matteucci, F., 2006, \mnras, 365, 759
\bibitem[Reynolds(1984)]{Reynolds1984} Reynolds, R.~J.\ 1984, \apj, 
282, 191 
\bibitem[Roberts(1963)]{Roberts1963} Roberts, M.~S.\ 1963, \araa, 1, 149 
\bibitem[Sakai et al.(2004)]{Sakai2004} Sakai, S., Ferrarese, L., 
Kennicutt, R.~C., Jr., \& Saha, A.\ 2004, \apj, 608, 42 
\bibitem[Salpeter(1955)]{Salpeter1955} Salpeter, E.~E.\ 1955, \apj, 
121, 161
\bibitem[Scalo(1986)]{Scalo1986} Scalo, J.~M.\ 1986, Fundamentals 
of Cosmic Physics, Volume 11, 1
\bibitem[Schaerer, Contini, \& Kunth(1999)]{Schaerer1999} Schaerer, D., Contini, T., \& Kunth, D., 1999, A\&A, 341, 399
\bibitem[Schlegel et al.(1998)]{Schlegel1998} Schlegel, D.~J., 
Finkbeiner, D.~P., \& Davis, M.\ 1998, \apj, 500, 525 
\bibitem[Searle \& Sargent(1972)]{Searle1972} Searle, L., \& 
Sargent, W.~L.~W.\ 1972, \apj, 173, 25
\bibitem[Searle et al.(1973)]{Searle1973} Searle, L., Sargent, 
W.~L.~W., \& Bagnuolo, W.~G.\ 1973, \apj, 179, 427
\bibitem[Skillman et al.(1988)]{Skillman1988} Skillman, E.~D., Terlevich, R., Teuben, P.~J., \& van Woerden, H.\ 1988, \aap, 198, 33 
\bibitem[Skillman(1996)]{Skillman1996} Skillman, E.~D.\ 1996, The 
Minnesota Lectures on Extragalactic Neutral Hydrogen, 106, 208
\bibitem[Skillman(1997)]{Skillman1997} Skillman, E.~D.\ 1997, 
Revista Mexicana de Astronomia y Astrofisica Conference Series, 6, 36 
\bibitem[Skillman \& Gallart(2002)]{Skillman2002} Skillman, E.~D., \& Gallart, C.\ 2002, Observed HR Diagrams and Stellar Evolution, 274, 535 
\bibitem[Skillman et al.(2003)]{Skillman2003} Skillman, E.~D., 
Tolstoy, E., Cole, A.~A., Dolphin, A.~E., Saha, A., Gallagher, J.~S., 
Dohm-Palmer, R.~C., \& Mateo, M.\ 2003, \apj, 596, 253 
\bibitem[Skillman(2005)]{Skillman2005} Skillman, E.~D.\ 2005, New 
Astronomy Review, 49, 453 
\bibitem[Spaans \& Norman(1997)]{Spaans1997} Spaans, M. \& Norman, C.~A., 1997, \apj, 483, 87
\bibitem[Stinson et al.(2007)]{Stinson2007} Stinson, G.~S., 
Dalcanton, J.~J., Quinn, T., Kaufmann, T., 
\& Wadsley, J.\ 2007, \apj, 667, 170 
\bibitem[Swaters \& Balcells(2002)]{Swaters2002} Swaters, R.~A., \& Balcells, M.\ 2002, \aap, 390, 863 
\bibitem[Taylor et al.(1998)]{Taylor1998} Taylor, C.~L., 
Kobulnicky, H.~A., \& Skillman, E.~D.\ 1998, \aj, 116, 2746 
\bibitem[Telesco et al.(1988)]{Telesco1988} Telesco, C.~M., 
Wolstencroft, R.~D., \& Done, C.\ 1988, \apj, 329, 174 
\bibitem[Thornley et al.(2000)]{Thornley2000} Thornley, M.~D., Schreiber, N.~M.~F., Lutz, D., Genzel, R., Spoon, H.~W.~W. \& Kunze, D., 2000, \apj, 539, 641
\bibitem[Thuan \& Seitzer(1979)]{Thuan1979} Thuan, T.~X., \& Seitzer, P.~O.\ 1979, \apj, 231, 327 
\bibitem[Thuan \& Martin(1981)]{Thuan1981} Thuan, T.~X., \& Martin, G.~E.\ 1981, \apj, 247, 823 
\bibitem[Tolstoy \& Saha(1996)]{Tolstoy1996} Tolstoy, E., \& Saha, A.\ 1996, \apj, 462, 672 
\bibitem[Tolstoy et al.(2009)]{Tolstoy2009} Tolstoy, E., Hill, V., \& Tosi, M.\ 2009, \araa, 47, 371 
\bibitem[Tosi et al.(1989)]{Tosi1989} Tosi, M., Greggio, L., \& Focardi, P.\ 1989, \apss, 156, 295 
\bibitem[Tremonti et al.(2001)] {Tremonti2001} Tremonti, C.~A., Calzetti, D., Leitherer, C. \& Heckman, T.~M., 2001, \apj, 555, 322 
\bibitem[Tully et al.(2006)]{Tully2006} Tully, R.~B., et al.\ 2006, \aj, 132, 729 
\bibitem[van Eymeren et al.(2009)]{vanEymeren2009} van Eymeren, J., Marcelin, M., Koribalski, B., Dettmar, R.-J., Bomans, D.~J., Gach, J.-L., \& Balard, P.\ 2009, \aap, 493, 511 
\bibitem[van Zee(2001a)]{vanZee2001} van Zee, L.\ 2001, \aj, 121, 2003 
\bibitem[van Zee et al.(2001b)]{vanZee2001b} van Zee, L., Salzer, 
J.~J., \& Skillman, E.~D.\ 2001, \aj, 122, 121 
\bibitem[Walter et al.(2001)]{Walter2001} Walter, F., Taylor, 
C.~L., H{\"u}ttemeister, S., Scoville, N., 
\& McIntyre, V.\ 2001, \aj, 121, 727 
\bibitem[Walter et al.(2008)]{Walter2008} Walter, F., Brinks, E., 
de Blok, W.~J.~G., Bigiel, F., Kennicutt, R.~C., Thornley, M.~D., 
\& Leroy, A.\ 2008, \aj, 136, 2563 
\bibitem[Weisz et al.(2008)]{Weisz2008} Weisz, D.~R., Skillman, 
E.~D., Cannon, J.~M., Dolphin, A.~E., Kennicutt, R.~C., Jr., Lee, J., 
\& Walter, F.\ 2008, \apj, 689, 160 
\bibitem[Whiting et al.(1997)]{Whiting1997} Whiting, A.~B., Irwin, 
M.~J., \& Hau, G.~K.~T.\ 1997, \aj, 114, 996 
\bibitem[Williams et al.(2009)]{Williams2009} Williams, B.~F., et 
al.\ 2009, arXiv:0911.4121 
\bibitem[Williams et al.(2010)]{Williams2010} Williams, B.~F., et 
al.\ 2010, \apj, 709, 135 
\bibitem[Wilson(1995)]{Wilson1995} Wilson, C.~D.\ 1995, \apjl, 
448, L97 
\end{thebibliography}
\end{document}